\documentclass[preprint,12pt]{elsarticle}

\usepackage{amssymb}
\usepackage{amsmath}
\usepackage{graphicx}
\usepackage{color}
\usepackage{bbold}
\usepackage[makeroom]{cancel}
\usepackage{epsfig}
\usepackage{epsf}
\usepackage{dcolumn}
\usepackage{bm}
\usepackage{hhline}
\usepackage{multirow}
\usepackage{dsfont}
\usepackage{braket}
\usepackage{slashed}
\usepackage{amssymb}
\usepackage{tikz,xcolor}
\usepackage[CJKbookmarks=true, colorlinks=true, linkcolor=blue, urlcolor=blue,citecolor=blue]{hyperref}
\usepackage[utf8]{inputenc}
\usepackage{amssymb}
\pdfoutput=1 % if your are submitting a pdflatex (i.e. if you have
\usepackage[T1]{fontenc} % if needed
\usepackage{amsmath}
\usepackage{mathrsfs}
\usepackage{color}
\usepackage{comment}
\usepackage{fancyvrb}
\usepackage{graphicx}
\usepackage{amsmath}
\usepackage{amssymb}
\usepackage{bm}
\usepackage{amsfonts}
\usepackage{times}
\usepackage{epsfig}
\usepackage{dcolumn}
\usepackage{bm}
\usepackage{color}
\usepackage{longtable}
\usepackage{array}
\usepackage{multirow}
\usepackage{verbatim}
\usepackage[utf8]{inputenc}
\usepackage{scalerel}
\usepackage{graphicx}
\usepackage[normalem]{ulem}
\usepackage{overpic}
\usepackage{graphicx}
\usepackage{bbm,bm}
\usepackage{multirow}
\usepackage{lineno}
\biboptions{sort&compress}

\definecolor{mygreen}{rgb}{0, 0.7, 0.2}

\newcommand{\dd}{\text{d}}
\renewcommand{\Im}{\textrm{Im }}
\renewcommand{\Re}{\textrm{Re }}

\usepackage{color}
\usepackage{braket}
\usepackage{dsfont}
\usepackage{multicol}

\journal{PHYSICS REPORTS}

\def\be{\begin{eqnarray}}
\def\en{\end{eqnarray}}

\def\gsim{ {\ \lower-1.2pt\vbox{\hbox{\rlap{$>$}\lower5pt\vbox{\hbox{$\sim$}
}}}\ } }

\newcommand{\SM}{\text{SM}}
\newcommand{\NP}{\text{NP}}
\newcommand{\CKM}{\text{CKM}}
\newcommand{\STCF}{\text{STCF}}
\newcommand{\ie}{\it i.e.}
\newcommand{\eg}{\it e.g.}

\newcommand{\ee}{e^+ e^-}

\newcommand{\pipi}{\pi^+\pi^-}

\newcommand{\jpsi}{J/\psi}
\newcommand{\pip}{\pi^+}
\newcommand{\pim}{\pi^-}
\newcommand{\piz}{\pi^0}
\newcommand{\KS}{K_{S}^0}
\newcommand{\KL}{K_{L}^0}
\newcommand{\Kone}{K_{1}}
\newcommand{\Ktwo}{K_{2}}
\newcommand{\Kp}{K^{+}}
\newcommand{\Km}{K^{-}}
\newcommand{\Kz}{K^{0}}
\newcommand{\Kzbar}{\bar{K}^{0}}

\newcommand{\rt}{\rightarrow}

\newcommand{\BR}{{\cal B}}

\newcommand{\CP}{C\!P}
\newcommand{\CPT}{C\!P\!T}

\newcommand{\Ham}{{\mathcal H}}
\newcommand{\Mass}{{\mathcal M}}
\newcommand{\Gmm}{{\boldsymbol \Gamma}}

\newcommand{\Str}{${\mathcal S}$}

\newcommand{\eps}{\varepsilon}
\newcommand{\epsp}{\varepsilon^{\prime}}

\newcommand{\smallonehalf}{{\scriptstyle \frac{1}{2}}}

\newcommand{\rootionehalf}{{\textstyle \frac{i}{\sqrt{2}}}}
\newcommand{\smallrootonehalf}{{\scriptstyle \sqrt{\frac{1}{2}}}}

\newcommand{\cp}{CP~}

\newcommand{\psip}{\psi(3686)}
\newcommand{\DDzbar}{D^{0}\bar{D}^{0}}
\newcommand{\DpDm}{D^{+}D^{-}}
\newcommand{\Dsp}{D_{s}^{+}}
\newcommand{\Dsm}{D_{s}^{-}}

\begin{document}
%\begin{sloppypar}
    
\begin{frontmatter}

\title{$\CP$ violation studies at Super Tau-Charm Facility}

\author[inst0]{Hai-Yang~Cheng}
\author[inst1]{Zhi-Hui~Guo}
\author[inst2]{Xiao-Gang~He}
\author[inst3]{Yingrui~Hou}
\author[inst4]{Xian-Wei~Kang}
\author[inst5,inst6]{Andrzej~Kupsc}
\author[inst7,inst13]{Ying-Ying~Li}
\author[inst7]{Liang~Liu}
\author[inst3]{Xiao-Rui~Lyu}
\author[inst17,inst8]{Jian-Ping~Ma}
\author[inst9,inst10]{Stephen~Lars~Olsen}
\author[inst7]{Haiping~Peng}
\author[inst16]{Qin~Qin}
\author[inst11,inst12]{Pablo~Roig% Garcés
}
\author[inst13]{Zhi-Zhong~Xing}
\author[inst14]{Fu-Sheng~Yu}
\author[inst15]{Yu~Zhang}
\author[inst3]{Jianyu~Zhang}
\author[inst7]{Xiaorong~Zhou}

\affiliation[inst0]{organization={Institute of Physics, Academia Sinica},%Department and Organization
            city={Taipei},
            postcode={11529}, 
            country={China}}
\affiliation[inst1]{organization={Hebei Normal University},%Department and Organization
            city={Shijiazhuang},
            postcode={050024}, 
            country={China}}
\affiliation[inst2]{organization={
Tsung-Dao Lee Institute and School of Physics and Astronomy,
Shanghai Jiao Tong University},%Department and Organization
            city={Shanghai},
            postcode={200240}, 
            country={China}}
\affiliation[inst3]{organization={University of Chinese Academy of Sciences},%Department and Organization
            city={Beijing},
            postcode={100049}, 
            country={China}}
\affiliation[inst4]{organization={Beijing Normal University},%Department and Organization
            city={Beijing},
            postcode={100875}, 
            country={China}}
\affiliation[inst5]{organization={National Centre for Nuclear Research},%Department and Organization
            city={Warsaw},
            postcode={02-093}, 
            country={Poland}}
\affiliation[inst6]{organization={Uppsala University},%Department and Organization
            city={Uppsala},
            postcode={SE-75120}, 
            country={Sweden}}
\affiliation[inst7]{organization={University of Science and Technology of China},%Department and Organization
            addressline={Address One}, 
            postcode={230026}, 
            country={China}}
\affiliation[inst17]{organization={Henan Normal University},%Department and Organization
            city={Xinxiang},
            postcode={453007}, 
            country={China}}
\affiliation[inst8]{organization={Institute of Theoretical Physics, Chinese Academy of Sciences},%Department and Organization
            city={Beijing},
            postcode={100190}, 
            country={China}}
\affiliation[inst9]{organization={High Energy Physics Center, Chung-Ang University},%Department and Organization
            city={Seoul},
            postcode={06974}, 
            country={Korea}}
\affiliation[inst10]{organization={Particle and Nuclear Physics Institute, Institute for Basic Science},%Department and Organization
            city={Daejeon},
            postcode={34126}, 
            country={Korea}}
\affiliation[inst16]{organization={Huazhong University of Science and Technology},%Department and Organization
            city={Wuhan},
            postcode={430074}, 
            country={China}}
\affiliation[inst11]{organization={Departamento de F\'isica, Centro de Investigaci\'on y de Estudios Avanzados del Instituto
Polit\'ecnico Nacional},%Department and Organization
            city={Mexico City},
            postcode={AP 14740, CP 07000}, 
            country={Mexico}}
\affiliation[inst12]{organization={IFIC, Universitat de Val\`encia – CSIC%, Catedr\´atico Jos\´e Beltr\´an 2
},%Department and Organization
            city={Paterna},
            postcode={E-46980}, 
            country={Spain}}
\affiliation[inst13]{organization={Institute of High Energy Physics, Chinese Academy of Sciences},%Department and Organization
            city={Beijing},
            postcode={100049}, 
            country={China}}
\affiliation[inst14]{organization={Lanzhou University},%Department and Organization
            city={Lanzhou},
            postcode={730000}, 
            country={China}} 
\affiliation[inst15]{organization={University of South China},%Department and Organization
            city={Hengyang},
            postcode={421001}, 
            country={China}}

\begin{abstract}

Charge-parity ($C\!P$) violation in the tau-charm energy region is a promising area for sensitive tests of Standard Model~(SM) predictions and searches for new, beyond the SM physics. A future Tau-Charm Facility that operates at center-of-mass energies between 2.0 and 7.0~GeV, with a peak luminosity of $0.5\times10^{35}$~cm$^{-2}$s$^{-1}$, would provide huge numbers of hadrons and tau ($\tau$) leptons that are produced in low-background environments and with well understood kinematic properties.  In this report, prospects for unique studies of $C\!P$ violation in the decay of charmed hadrons, and in the production and decay of hyperons and $\tau$ leptons at a next-generation tau-charm facility are discussed. In addition, opportunities for improved tests of $\CP T$ invariance test in $K^{0}-\bar{K}^{0}$ mixing are presented.

\end{abstract}

% \begin{keyword}
% %% keywords here, in the form: keyword \sep keyword
% keyword one \sep keyword two
% %% PACS codes here, in the form: \PACS code \sep code
% %% MSC codes here, in the form: \MSC code \sep code
% %% or \MSC[2008] code \sep code (2000 is the default)
% \end{keyword}

\end{frontmatter}

\newpage

\tableofcontents

\newpage

%\linenumbers

\section{Introduction}
\label{sec:intro}
%\documentclass[11pt]{article}
%\usepackage{graphicx}
%\usepackage{bbm,bm}
%\usepackage{besphysics}
%\usepackage{multirow}

%\usepackage{epsfig}

%\input colordvi
%\input blackdvi

%\setlength{\textwidth}{175mm}
%\setlength{\textheight}{220mm}
%\setlength{\oddsidemargin}{-5mm}
%\setlength{\topmargin}{-6mm}

%\renewcommand{\baselinestretch}{1}

%\newcommand{\hatp}{\hat{\bm p}}
%\newcommand{\hatk}{\hat{\bm k}}
%\newcommand{\hatn}{\hat{\bm n}}
%\newcommand{\hatl}{\hat{\bm l}}

%\newcommand{\stcflum}{0.5\times 10^{35} ~\rm cm^{-2} c^{-1}}
%\newcommand{\stcfene}{2\sim7~\rm GeV}

%\pagestyle{plain}

%\begin{document}

%\def\bfej{\mbox{\boldmath$\varepsilon$}}
%\newcommand {\sla}[1]{ #1 \!\!\!/}
%\newcommand{\jgmm}{\jp\to \gamm MM}
%\newcommand{\st}[1]{|#1\rangle}
%\newcommand{\jgpp}{\jp\to\gamma\phi\phi}
%\def \qqbar {q\bar q}
%\newcommand{\str}[1]{|#1\rangle}
%\newcommand{\elmnt}[3]{\langle #1|#2|#3\rangle}
%\newcommand{\cp}{CP~}
%\newcommand{\cpv}{CP violation~}

%\begin{center}
%{\bf\large Introduction   }
%\end{center}
%\par
%\vskip20pt

 The future Super Tau-Charm Facility (STCF) will operate at center-of-mass energies between 2.0 and 7.0~GeV with a peak luminosity of $0.5\times10^{35}$~cm$^{-2}$s$^{-1}$, and produce huge numbers of hadrons and tau ($\tau$) leptons with experimentally favorable kinematic conditions and
low-backgrounds in a state-of-the-art $4\pi$-solid-angle particle detector. It will support a rich physics program that addresses issues involving 
non-perturbative Quantum Chromodynamics (QCD) and exotic hadron spectroscopy, precision measurements of electroweak interactions, as well as searches for
the new, beyond the Standard Model physics (NP)\,\cite{Achasov:2023gey}. In this report, we review the prospects for using the large STCF  data samples
for studies of $\CP$ violations in charmed particle decays, searches for $\CP$ violations in the production and decays of hyperons and $\tau$ leptons,
and tests of $\CP T$ invariance with large samples of strangeness-tagged neutral kaon decays. 
 
The study of $\CP$ violations is a subject of central  importance to particle physics and cosmology.  The universe is predominantly composed of matter, with
trace amounts of anti-matter particles that have only fleeting existences.  The Standard Model (SM) has no explanation for  this famous Baryon Asymmetry of the Universe (BAU), and whatever the (still unknown) new physics  processes that are responsible for this asymmetry might be, they must violate the $CP$ symmetry. 

In the SM of elementary particles~\cite{Weinberg:2018apv}, the source of $\CP$-violation is an irreducible complex phase in the Cabibbo\,\cite{Cabibbo:1963yz}
Kobayashi-Maskawa\,\cite{Kobayashi:1973fv} quark-flavor  mixing matrix , $V_{CKM}$, that modifies the coupling of $W$-bosons to the three-generation quark  currents,
the so-called Kobayashi-Maskawa (KM) mechanism. In this scheme, the interaction Lagrangian is given by
\begin{eqnarray}
{\cal L} = - {g\over \sqrt{2}}\bar U \gamma^\mu W^+_\mu V_{CKM} D + h.c.,
\end{eqnarray}
where $U = (u, c, t)$ and $D = (d, s, b)$ represent the known quarks. $V_{CKM}$ is a $3\times 3$ unitary matrix that has three mixing angles $\theta_{12}$, $\theta_{23}$, $\theta_{13}$ and a complex phase $\delta$ and is usually parameterized as
\begin{eqnarray}
\label{mixpar}
    V_{KM} &=& \left ( 
    \begin{array} {ccc}
    V_{ud}&V_{us}&V_{ub}\\
    V_{cd}&V_{cs}&V_{cb}\\
    V_{td}&V_{ts}&V_{tb}
    \end{array} 
    \right ) \nonumber\\
    &=& \left ( 
    \begin{array} {ccc}
    c_{12} c_{13}&s_{12}c_{13}&s_{13} e^{-i\delta}\\
    -s_{12}c_{23} - c_{12}s_{23}s_{13} e^{i\delta}&c_{12}c_{23} - s_{12}s_{23}s_{13} e^{i\delta}&s_{23}c_{13}\\
    s_{12}s_{23}-c_{12}c_{23}s_{13}e^{i\delta}&-c_{12} s_{23} - s_{12} c_{23}s_{13} e^{i\delta}&c_{23}c_{13}
    \end{array} 
    \right ) ,\;\;
\end{eqnarray}
where $s_{ij} = \sin\theta_{ij}$ and $c_{ij} = \cos\theta_{ij}$. The phase $\delta$ is the source of $\CP$-violation.

%{\color{red} Some where near here, should we consider introduce KM matrix here and summarize the values?
%Because we keep referring numbers for KM matrix elements and the KM phase}
 
$\CP$ violation in the neutral $K$ meson system was discovered in particle-physics experiments prior to the emergence of SM  and the KM mechanism. Since
then, $\CP$ violations
have also been observed in the  $B$- and $D$-meson systems.  Unlike violations of $P$, or $C$, which are maximal, $\CP$ violations in the $K$ and $D$ meson
systems  are small effects, but they can still be consistently explained by the KM phase in SM. All of the elements in the $V_{CKM}$ matrix, including the
$\CP$ phase $\delta$ have been experimental measured with reasonably good precision, and are summarized in the following widely used Wolfenstein parameterization \cite{Wolfenstein:1983yz},
\begin{eqnarray}\label{eq_Wolfenstein}
    V_{CKM} &=& \left ( 
    \begin{array} {ccc}
    1-\lambda^2/2&\lambda&A\lambda^3(\rho - i\eta)\\
    -\lambda&1-\lambda^2/2&A\lambda^2\\
    A\lambda^3(1-\rho -i\eta)&-A\lambda^2&1
    \end{array} 
    \right )  + {\cal O}(\lambda^4),
\end{eqnarray}
where $\lambda = 0.22501\pm0.00068$, $A = 0.826^{+0.016}_{-0.015}$, $\bar \rho = \rho(1-\lambda^2/2) = 0.1591\pm 0.0094 $ and
$\bar \eta = \eta(1-\lambda^2/2) = 0.3523^{+0.0073}_{-0.0071}$.
In terms of the quark mixing angles and $\CP$ phase, the parametrization of Eq.~(\ref{mixpar}) translates  to numerical values of $s_{12}=0.22501\pm0.00068$,
$s_{13}=0.003732^{+0.000090}_{-0.000085}$, $s_{23}=0.04183^{+0.00079}_{-0.00069}$ and $\delta=1.147\pm0.026$~\cite{ParticleDataGroup:2024cfk}. 

In the SM, the single complex phase $\delta$ accounts for all of the $\CP$-violating phenomena that have been observed in laboratory-based experiments, but
it cannot explain the baryon asymmetry in the universe. This is characterized by the ratio of the baryon and photon densities in the universe, which is
measured with about 10\% precision to be $\eta\approx 6\times 10^{-10}$.  However, calculations of $\eta$ that are based on the phase $\delta$
in the CKM matrix, find values of $\eta$ that are nine orders of magnitude smaller than its measured value. This large discrepancy strongly motivates
experimental searches for $\CP$ violations in laboratory experiments that cannot be explained by the SM's KM mechanism and is, therefore, evidence for
new, non-SM physics interactions.

\par 
Although all observations of $\CP$ violations in $K$-, $D$-, and $B$-meson systems that have been reported to date can be explained by the
SM's $\delta$ phase, there is still a need to establish that the KM-mechanism is correct in a broader scope of interactions. The nonzero
phase in the mixing matrix originates from the mass matrix that couples quarks to the scalar sector of the SM that involves the Higgs field and Electro-weak spontaneous symmetry  breaking. Unlike the other sectors of the SM, the Higgs sector is still poorly understood. It is not clear why there are three generations of quarks with the observed inter-generation mixing pattern between generations that produces the $\CP$ violation that
is observed.  Although the SM has been very successful, there are both observational and theoretical reasons that indicate that NP beyond SM should exist. On the observational side, in addition to the BAU problem mentioned above, the are no SM explanations for the non-vanishing neutrino masses that have been established by the discovery of neutrino oscillations, QCD's strong
$\CP$ problem, or for the universe's dark matter and dark energy. Among the theoretical arguments that call for NP are the many free parameters in the SM, the poorly understood Higgs sector physics,  and its related hierarchy problem. In this review, we focus on possible NP sources of $CP$ violation. Most
proposed NP scenarios have additional $\CP$-violating phases. If a $\CP$ violation is found in a process for which  the SM prediction is zero or unmeasurably small, it will  be an unquestionable evidence for NP. Therefore, searches for $\CP$ violation will play an important role for a better understanding of the
SM as well as in the quest for NP.  

\par
\begin{table}[htbp]
\begin{center}
\caption{Some typical event numbers expected per year at STCF.}
\label{tablelumi}
\begin{tabular}{c|c|c|c|c}
     \hline \hline
     c.m.~energy               & Lumi              &  ~~~~~~Samples~~~~~~ & $ ~~\sigma~~$  & ~Numbers~ \\ 
     ({\rm GeV})              &  ${\rm ab}^{-1}$)   &                      & $({\rm nb})$  & ~of Events~ \\ \hline
       \multirow{2}{*}{3.097}                  & \multirow{2}{*}{1}                  &  $\jpsi$                &  3400 & $3.4\times 10^{12}$   \\
                                &               &  $J/\psi\to\Lambda\bar{\Lambda}$  &   & $7.0\times10^{9}$  \\  
                               &               &  $J/\psi\to\Xi\bar{\Xi}$  &   & $2.3\times10^{9}$  \\ 
                               &               &  $J/\psi\to K^\mp\pi^\pm K^0/\bar{K}^0$  &   & $3.4\times10^{9}$  \\ \hline
       3.670                  & 1                  &  $\tau^+\tau^-$         &  2.4  & $2.4\times 10^{9} $   \\ \hline
       \multirow{2}{*}{3.686} & \multirow{2}{*}{1}                   &  $\tau^+\tau^-$         &  2.5  & $2.5\times 10^{9} $   \\
                              &                    &  $\psip\to \tau^+\tau^-$&       & $2.0\times 10^{9} $   \\ \hline
       \multirow{3}{*}{3.770} & \multirow{3}{*}{1} &  $\psi(3770)\to\DDzbar $             &  3.6  & $3.6\times 10^{9} $   \\
                              &                    &  $\psi(3770)\to \DpDm   $             &  2.8  & $2.8\times 10^{9} $   \\
                              &                    &  $\psi(3770)\to \tau^+\tau^-$         & 2.9   & $2.9\times 10^{9} $\\  \hline
       \multirow{2}{*}{4.040} &  \multirow{2}{*}{1}             &  $\psi(4040)\to\Dsp\Dsm$             & 0.20  & $2.0\times 10^{8} $       \\
                              &                    &  $\psi(4040)\to \tau^+\tau^-$         & 3.5   & $3.5\times 10^{9} $ \\  \hline
       \multirow{2}{*}{4.630} & \multirow{2}{*}{1}                     &  $\psi(4630)\to \Lambda_c\bar\Lambda_c$& 0.20 & $2.0\times 10^{8} $  \\
                              &                    &  $\psi(4630)\to \tau^+\tau^-$          & 3.4  & $3.4\times 10^{9} $   \\ 
   \hline\end{tabular}
\end{center}
\end{table}

Since the observational effects of $\CP$ violation are, in general, expected to be small, tests of $\CP$ symmetry and searches for new examples of
$\CP$-violating processes will require large statistical  samples of high quality data that would be available at the proposed STCF~\cite{Achasov:2023gey}. 
In the energy range covered by the STCF, a number of $J^{PC}=1^{--}$ charmonium states can be copiously produced via $s$-channel $e^+e^-$ annihilations at
c.m. energies corresponding to the resonance peaks.   The numbers of quantum-correlated $\tau^+\tau^-$, or
meson-antimeson. or baryon-antibaryon pairs that can be produced via decays of these resonance states are shown for some example processes in
Table~\ref{tablelumi}. (More detailed compilations of expected event numbers can be found in Ref.~\cite{Achasov:2023gey}.) To the extent that $\CP$ is
conserved in the $e^+e^-$ annihilation and charmonium decay processes, these particle-antiparticle
pairs will be produced in a pure $\CP$-odd eigenstate.  These conditions, $\ie$ the particle-antiparticle quantum correlations and their pure $\CP$
quantum number, make
charmonium$\to$particle-antiparticle pairs unique, and very powerful reactions for $\CP$ violation studies.  Quantum correlations are essential for
tests of $\CP$ symmetry in hyperon systems.

Improved searches for $\CP$ violations in non-leptonic decays of strange hyperon, especially $\Lambda\to N\pi$ and $\Xi\to\Lambda\pi$, are of
particular interest because the level of SM-induced $\CP$ violations in these decays\,\cite{Tandean:2002vy} are two orders  of magnitude below the current
levels of experimental sensitivity.
%@article{Tandean:2002vy,
%    author = "Tandean, Jusak and Valencia, G.",
%    title = "{CP violation in hyperon nonleptonic decays within the standard model}",
%    archivePrefix = "arXiv",
%    doi = "10.1103/PhysRevD.67.056001",
%    journal = "Phys. Rev. D",
%   volume = "67",
%    pages = "056001",
%   year = "2003"
%} 
 The effectiveness of using the $J/\psi\to$hyperon antihyperon reaction for $\CP$ studies has been demonstrated by the BESIII experiment that has
reported the world's best limits on  NP $\CP$ violating processes in strange hyperon decays using 3.2\,M $J/\psi\to\Lambda\bar{\Lambda}$ and 0.9\,M 
$J/\psi\to\Xi\bar{\Xi}$ events\,\cite{BESIII:2021ypr,BESIII:2022qax,BESIII:2023drj}.  These measurements have $\mathcal{O}(1\%)$ backgrounds and small
$\mathcal{O}(0.2\%)$ systematic errors. At the STCF, the $J/\psi \to \Lambda\bar\Lambda$ and $\Xi\bar{\Xi}$ event samples would be more than three orders of magnitude larger than the existing BESIII data samples, and could support searches for NP $\CP$ violations with a factor of $\sim$40 higher sensitivities, which
would nearly reach the SM-model-predicted levels. 

Included in Table~\ref{tablelumi} is a projection of 3.4\,B detected strangeness-tagged $K^0$ and $\bar{K}^0$ events produced via $J/\psi\to K^\mp\pi^\pm K^0(\bar{K}^0$ decays that can be used to improve the precision of the magnitude and phases
of the $\eta_{+-}$ and $\eta_{00}$ $\CP$-violating parameters of the neutral kaon system, and extend the sensitivity of tests of $\CP T$ invariance in
$K^0$-$\bar{K}^0$ mixing. The current state of the art sensitivities for neutral kaon $\CP$ parameter measurements and $\CP T$ tests are thirty
year old CPLEAR measurements that are based on $\sim$70\,M strangeness-tagged neutral kaon decays\,\cite{Apostolakis:1999zw,CPLEAR:1998ohr}. With fifty
times more strangeness-tagged decays in a more sophisticated detector and a cleaner event environment,  STCF would provide an order of magnitude improved sensitivity.

In addition to studies of $\CP$ violation in the decays of the above-mentioned particles, searches for $\CP$ violation in the charmonium decay processes
or in their production via the $e^+e^-$ annihilation at non-resonant energy points that could be performed are discussed below.  This can provide additional
information about $\CP$ symmetry.

 Since hyperon-antihyperon, $\tau^+\tau^-$ and charmed-anticharmed hadron pair data samples at STCF will be especially large, this review is mainly focus on
$\CP$ tests with these systems.  In addition we include a study of the possibility of an improved test of the combined $C$, $P$ and time reversal $T$ symmetries,
 $\ie$  $CPT$ invariance. $\CP T$ invariance is expected to be exact from our current understanding of space-time and interactions, at last in the framework of local quantum field theory, but its validity has to be experimentally tested. In Chaps.~\ref{sec:hyperon},~\ref{sec:tau} and~\ref{sec:charm}, we discuss current theoretical and experimental progress, as well as the prospected potential at STCF, on relevant processes for hyperon-antihyperon, $\tau^+\tau^-$ and
charmed-anticharmed hadron systems, respectively. In Chap.~\ref{sec:kaon}, we discuss how $CPT$ invariance could be tested with the large data sample of 
strangeness-tagged neutral $K$-mesons that will be produced at STCF. In Chap.~\ref{sec:sum}, we provide our summaries.

\newpage

%\par\vskip40pt
%\begin{thebibliography}{99}
%\bibitem{STCF}  M. Achasov, {it et. al. }, Front. Phys. (Beijing) 19 (2024) 1, 14701,  e-Print: 2303.15790 [hep-ex]. 

%\end{thebibliography}

%\end{document}

%\section{Probe of CP violation in tau-charm energy region}
%\label{sec:taucharm}
%\input{sec2_taucharm}

%\section{The STCF project}
%\label{sec:stcf}
%\input_sec7_stcf}

\section{$\CP$ violation in hyperon sector}
\label{sec:hyperon}
\subsection{Direct $\CP$ violation in strange quark systems}
\label{sec3:intro}

Direct $\CP$ violation requires that amplitudes for a given process and its $\CP$-transformed process
to have different complex phases. For example, if the amplitude for a particle decay is ${\cal A}=|{\cal A}|\exp(i\xi+i\delta)$, then the amplitude for the $\CP$-transformed process is $\bar {\cal A}=|{\cal A}|\exp(-i\xi+i\delta)$, where $\xi$ is the  $\CP$-odd phase that changes sign after the $\CP$ transformation,  while $\delta$ is the $\CP$-even strong phase due to final state re-scattering, that does not change sign. In order to observe an effect due to the $\CP$-odd phases, an interference with other amplitudes contributing to the process is needed. 
A usually studied CP violating observable is the decay rate asymmetry, $A_{CP}$, defined as
\begin{eqnarray}\label{eq:Acp-direct}
A_{CP} &=& {|{\cal A}|^2 - |\bar {\cal A}|^2\over |{\cal A}|^2 + |\bar {\cal A}|^2} \nonumber\\
    &=& - {2\sum_{j\neq k}|{\cal A}_j| |{\cal A}_k| \sin(\xi_j - \xi_k) \sin(\delta_j - \delta_k)
\over \sum_{j} |{\cal A}_j|^2 + {2\sum_{j\neq k}|{\cal A}_j| |{\cal A}_k|} \cos(\xi_j - \xi_k) \cos(\delta_j - \delta_k)}\;.
\end{eqnarray}
It is clear that, in order to have a non-zero $A_{CP}$, there must exist at least two amplitudes with different CP conserving and violating phases. That is, ${\cal A}=\sum_j |{\cal A}_j|\exp(i\xi_j+i\delta_j)$, with $j$ larger than 1.
Since in the $\SM$ the $\CP$-odd phases $\xi_i$ are small, one has $\sin(\xi_1 - \xi_2) \approx \xi_1 - \xi_2$ for the weak transitions of the strange quarks. Therefore, a significant contribution to the $\CP$ violation might be due to a NP effect.

One process where the direct $\CP$ violation was first observed is the combined weak and strong decay of
$K^0$ meson into a pion pair  $K^0\to\pi^+\pi^-$. 
There are two final states that have distinct properties with respect to the strong interactions by having different values of the isospin quantum number $I=0$ and $I=2$ for the two-pion system. The initial weak transitions can proceed to one of the two states and are represented by the amplitudes ${\cal A}_0$ and ${\cal A}_2$. One parameterizes the decay amplitudes as %\textcolor{violet}{Pablo: Given the first eq. below, shouldn't there the coefficients of the second one be multiplied by $\sqrt{2}$?}
\begin{eqnarray}
&&A(K^0\to \pi^+\pi^-) = \sqrt{{2\over 3}} A_0 e^{i(-\xi_0 + \delta_0)} + \sqrt{1\over 3} A_2 e^{i(-\xi_2 + \delta_2)}\;,\nonumber\\
&&A(K^0\to \pi^0\pi^0) = \sqrt{{1\over 3}} A_0 e^{i(-\xi_0 + \delta_0)} - \sqrt{2\over 3} A_2 e^{i(-\xi_2 + \delta_2)}\;,
\end{eqnarray}
where, experimentally, ${\cal A}_2/{\cal A}_0\approx 0.05$.
%The strong interaction of the pions 
%is described by the strong phases
%$\delta_0$ and $\delta_2$ that are different for each isospin state. The complete amplitude for the %$K^0\to\pi^+\pi^-$ process can then be written as ${\cal A}=|{\cal A}_0|\exp(i\xi_0+i\delta_0)+|{\cal %A}_2|\exp(i\xi_2+i\delta_2)$. The amplitude of the $\CP$-transformed process $\bar K^0\to\pi^-\pi^+$, $\bar{\cal %A}$, differs only by the signs of the $\CP$-odd phases $\xi_0$ and $\xi_2$. 
%The decay rates $|{\cal A}|^2$ and $|\bar{\cal A}|^2$ can be different if $\CP$ is violated. 

The direct $\CP$ violation observable $A_{CP}$ is expressed as~\cite{Bigi:2000yz,Sozzi:2004py}: 
\begin{align}
    A(K^0\to \pi^+\pi^-)_{CP} \approx \sqrt{2} (\xi_0-\xi_2)\sin(\delta_0-\delta_2)\frac{{\cal A}_2}{{\cal A}_0}\;.
\end{align}
This quantity is directly related to the well-know experimentally measured quantity $\text{Re}(\epsilon') \approx(\xi_0-\xi_2)\sin(\delta_0-\delta_2)({\cal A}_2/{\cal A}_0)$, with
\begin{eqnarray}
Re(\epsilon') = {1\over 3}Re \left  ({A(K_L \to \pi^+\pi^-)\over A(K_S \to \pi^+ \pi^-)} - {A(K_L \to \pi^0\pi^0)\over A(K_S \to \pi^0 \pi^0)}\right ) \approx (\xi_0-\xi_2)\frac{{\cal A}_2}{{\cal A}_0}\;.
\end{eqnarray}

The measured value of $\text{Re}(\epsilon')$ is $(3.7\pm0.5)\times10^{-6}$~\cite{ParticleDataGroup:2022pth}, corresponding to the weak-phase difference $\xi_0-\xi_2\approx10^{-4}$ radian.
\begin{figure}
    \centering
   \includegraphics[width=0.95\textwidth]{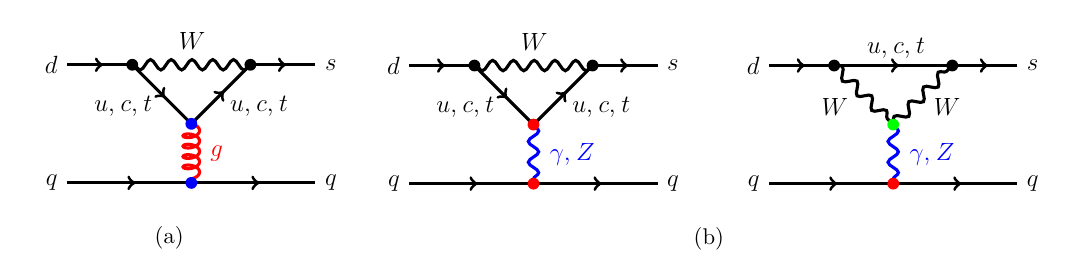}    
\caption{Quark level diagrams for the weak processes involving kaons or hyperons that contribute to the $\CP$-odd phases in $\SM$: (a)  QCD-penguin operators  and (b) electroweak penguin operators. This figure was created using a modified script from Ref.~\cite{Gisbert:2017vvj}. % need confirm, was Ref.~[7].
}
\label{fig:CPV}
\end{figure}

In the $\SM$, the contributions to the $\CP$-odd phases $\xi_{0,2}$ are given by the loops that involve all three quark generations, as shown in the diagrams in Fig.~\ref{fig:CPV}. An order of magnitude estimate for these contributions is determined from the $\CKM$ matrix~\cite{Cabibbo:1963yz,Kobayashi:1973fv}
and can be expressed by the product of the Wolfenstein parameters~\cite{Wolfenstein:1983yz} as $\lambda^4A^2\eta\approx %-
6\times10^{-4}$~\cite{ParticleDataGroup:2024cfk,ParticleDataGroup:2022pth}.%\textcolor{violet}{PR: The sign does not come from the expression in terms of Wolfenstein parameters.}
Recent $\SM$ predictions for the $\CP$ violation in kaon decays that include hadronic effects are given in the framework of the low-energy effective field theory~\cite{Gisbert:2017vvj} and the lattice QCD~\cite{RBC:2015gro,RBC:2020kdj}. The agreement with the $\text{Re}(\epsilon')$ parameter measurement is satisfactory but -due to the large uncertainties- there is room for a $\NP$ contribution.

%%%%%%%%%%%%%%%%%%%%%%%%%%%%%%%%%%%%%%%%%%%%%%%%%%%%%%%%%%%%%%%%%%%%%%%%%%%%%%%%%%%%%%%%%%%%%%%%%%%%%%%%%%%%%%%%%%%%%%%%%%%%%%%%%%%%%%%%%%%%%%%%%%%%%%%%%%%%%%%%%%%%%%%%

\subsection{Hyperon two-body hadronic weak decays }
% The main decay modes of the ground-state hyperons are weak-$|\Delta S|\!=\!1$ transitions into a baryon and a pseudoscalar meson. 
The  direct $\CP$ violation is also possible in the main decays of the ground-state baryons with one or more strange quarks --- hyperons.
The spin of  a baryon provides an additional degree of freedom that can be used to construct independent $\CP$ tests that can be very sensitive. A quantum state of a spin-$1/2$ baryon  $B$ is described by Pauli matrices:
\begin{equation}
    \ket{B}=\sigma_0+{\bf P}_B\cdot {\boldsymbol{\sigma}}\ ,
\end{equation}
where $\sigma_0$ is the $2\times 2$ unit matrix and $\boldsymbol{\sigma}:=(\sigma_1,\sigma_2,\sigma_3)$ represents spin $x,y,z$ projections in the baryon rest frame. The polarization vector ${\bf P}_B$ describes the preferred spin direction for an ensemble of the $B$-baryons.
Consider the $B \to b \pi$ transition between two spin-1/2 baryons with positive internal parities where a negative parity pion is emitted. Examples of such processes are the decays $\Xi^-\to \Lambda\pi^-$ or $\Lambda\to p\pi^-$. The  baryon $b$ can have the same or opposite direction of spin as the baryon $B$, and the angular momentum conservation requires $s$- or $p$-wave for the $b$--$\pi$ system.  The parity of the final state is negative for the $s$-wave and positive for the $p$-wave. Both are allowed, since parity is not conserved in weak decays. The decay amplitude is given by the transition operator
\begin{equation}
   {\cal A}(B\to b\pi)\propto {\cal S}\sigma_0 +{\cal P}{\boldsymbol{\sigma}}\cdot {\bf\hat{n}}\ , \label{eq:Ampli}
   \end{equation}
where  ${\bf\hat{n}}$ is the direction of the $b$-baryon in the $B$-baryon rest frame. The complex parameters   ${\cal S}$ and $ {\cal P}$ can be represented as
${\cal S}=|{\cal S}|\exp(i\xi_S+i\delta_S)$ and  ${\cal P}=|{\cal P}|\exp(i\xi_P+i\delta_P)$. 
The strong interaction of the $b$--$\pi$ system is given by the $\delta_P$ and $\delta_S$ phases and the  $\CP$-odd phases of weak interaction are $\xi_P$ and $\xi_S$. The amplitude ratios $|{\cal P}/{\cal S}|$ for $\Lambda\to p\pi^-$ and $\Xi^-\to \Lambda\pi^-$ are $0.442(4)$  and $0.188(2)$, respectively~\cite{Salone:2022lpt}. Since only the angular distributions will be discussed, the probability of the decay $\Gamma\propto|{\cal S}|^2+|{\cal P}|^2$ is set to  $|{\cal S}|^2+|{\cal P}|^2=1$. The real and imaginary parts of the amplitude interference term  are
\begin{align}
 \alpha&={2\text{Re} ({\cal S}^*{\cal P})}\propto2|{\cal S}||{\cal P}| \cos(\xi_P+\delta_P-\xi_S-\delta_S),\\ 
 \beta&={2\text{Im} ({\cal S}^*{\cal P})}\propto2|{\cal S}||{\cal P}| \sin(\xi_P+\delta_P-\xi_S-\delta_S)\ .
\end{align}
The real part,  given by the $\alpha$ parameter, can be determined from the angular distribution of the baryon $b$ when the  baryon $B$ has known non-zero polarization, or by measuring the polarization of the daughter baryon. 
For example, the proton angular distribution in the $\Lambda(\Lambda\to p\pi^-)$ decay is given as 
\begin{equation}
    \frac{1}{\Gamma}\frac{\dd\Gamma}{\dd\Omega}=\frac{1}{4\pi}\left(1+\alpha_\Lambda {\bf P}_\Lambda\cdot{{\bf\hat{n}}}\right)\ ,
\end{equation}
where ${\bf\hat{n}}$ is the proton momentum direction and ${\bf P}_\Lambda$ is the $\Lambda$ polarization vector. 

The imaginary part of the interference term given by the parameter $\beta$  is a $T$-odd variable. 
Measurement of $\beta$ requires that the polarization of both baryon $B$ and the daughter baryon is determined.  For the decay $\Xi^-\!\to\Lambda\pi^-$, where the cascade is polarized, the $\beta_\Xi$ parameter can be determined using the subsequent $\Lambda\to p\pi^-$ decay that acts as $\Lambda$ polarimeter. { The polarization ${\bf P}_\Lambda$ of the final baryon $\Lambda$ is given by the Lee--Yang formula \cite{Lee:1957qs}
\begin{eqnarray}
    {\bf P}_\Lambda = { (\alpha_\Xi + {\bf P}_\Xi\cdot {{\bf\hat{n}}}){{\bf\hat{n}}} + \beta_\Xi  {\bf P}_\Xi \times {{\bf\hat{n}}} + \gamma_\Xi {{\bf\hat{n}}}\times ({\bf P}_\Xi \times {{\bf\hat{n}}}) \over 1 +\alpha_\Xi {\bf P}_\Xi\cdot {{\bf\hat{n}}}}\;,
\end{eqnarray}
where ${\bf\hat{n}}$ is, now, the $\Lambda$ momentum direction and $\gamma=|{\cal S}|^2-|{\cal P}|^2$.
%The relation between the initial $\Xi^-$ polarization ${\bf P}_\Xi$ and the daughter $\Lambda$
%polarization ${\bf P}_\Lambda$ is given by the Lee--Yang formula \cite{Lee:1957qs}. 
If the ${\bf \hat z}$ axis in the $\Xi^-$ rest frame is defined  along the ${\bf\hat n}$ direction, then the relation between the polarization vectors is
% \begin{equation}
%    \left[\begin{array}{c}
%     P_\Lambda^x\\
%     P_\Lambda^y\\
%     P_\Lambda^z    
%     \end{array}\right]= 
%     \frac{1}{1+\alpha_{\Xi}P_\Xi^z}\left(
% \begin{array}{c}
%   \gamma_{\Xi} P_\Xi^x \red{+}\beta_{\Xi}P_\Xi^y \\
%   \gamma_{\Xi}P_\Xi^y \red{-}\beta_{\Xi} P_\Xi^x \\
%  \alpha _{\Xi} +P_\Xi^z \\
% \end{array}
% \right)
%     \label{eq:LY}\ .
% \end{equation}
%% old 
\begin{equation}
   \left[\begin{array}{c}
    P_\Lambda^x\\
    P_\Lambda^y\\
    P_\Lambda^z    
    \end{array}\right]=
    \frac{1}{1+\alpha_{\Xi}P_\Xi^z}\left(
\begin{array}{ccc}
  \gamma_{\Xi} & \beta_\Xi & 0 \\
  -\beta_\Xi & \gamma_\Xi & 0 \\
   0 & 0 & 1 \\
\end{array}
\right)  
\left[\begin{array}{c}
    P_\Xi^x\\
    P_\Xi^y\\
    \alpha_\Xi+P_\Xi^z    
    \end{array}\right]
    \label{eq:LY}\ .
\end{equation}
%where the parameter $\gamma=|{\cal S}|^2-|{\cal P}|^2$.  
This equation implies that $\Lambda$ in the decay of unpolarized $\Xi^-$ has the longitudinal polarization vector component $P_\Lambda^z=\alpha _{\Xi}$. Using a polar angle parameterization such that $\beta=\sqrt{1-\alpha^2}\sin\phi$ and $\gamma=\sqrt{1-\alpha^2}\cos\phi$, the $\phi$ parameter has the interpretation of the rotation angle between  the $\Xi$ and $\Lambda$ polarization vectors in the transversal $x$--$y$ plane. It is convenient to rewrite the Eq.~\eqref{eq:LY} for a  general two-body decay $D\equiv B \to b \pi$, where $\pi$ represents a pseudoscalar meson, in terms of a $4\times4$ matrix,
$a^D_{\mu\nu}$, describing the transformations between the Pauli matrices $ \sigma^{B}_\mu$ and  $\sigma_\nu^{b}$  defined in the $B$
and $b$ baryon helicity frames, respectively~\cite{Perotti:2018wxm}:
\begin{equation}
\sigma^{B}_\mu\to\sum_{\nu=0}^3a_{\mu\nu}^{D}\sigma_\nu^{b}\ . \label{eq:decay}
\end{equation}
The explicit form of the matrix is
\begin{equation}
    a^D_{\mu\nu}= 
      \left(
\begin{array}{cccc}
 1 & 0 & 0 & \alpha_D \\
 0 & \gamma_D & -\beta_D & 0 \\
 0 & \beta_D & \gamma_D & 0 \\
 \alpha_D & 0 & 0 & 1 \\
\end{array}
\right)
     \label{eq:LY4}\ .
\end{equation}
Here $\alpha_D$, $\beta_D$ and $\gamma_D$ correspond to the Lee-Yang polarization parameter $-\alpha_B$, $\beta_B$ and $\gamma_B$, respectively.     

We consider the case in which the direction of the outgoing baryon $b$ in the $B$ rest frame is given by an arbitrary (not aligned with the $B$ spin direction) $\theta_b$ and $\varphi_b$ polar and azimuth angles, respectively. The polarization transformation matrix should be multiplied by the following four-dimensional rotation matrix:
\begin{equation}
{\cal R}_{\mu\nu}^{(4)}(\Omega)=   \left(
\begin{array}{cccc}
 1 & 0 & 0 & 0 \\
 0 & \cos \theta    \cos \phi  &  -  \sin \phi  & \sin \theta \cos \phi  \\
 0 & \cos \theta  \sin \phi   & \cos \phi  & \sin \theta \sin \phi  \\
 0 & -\sin \theta   & 0 & \cos \theta  \\
\end{array}
\right)\ ,\label{eq:R4D}
\end{equation}
which is the 4D rotation where the spatial part ${\cal R}_{jk}(\Omega)$ corresponds to the product of the following three axial rotations:
\begin{equation}
\begin{split}
    {\cal R}_{jk}(\Omega)&=R_z(\phi)R_y(\theta)\\
    &=\left(
\begin{array}{ccc}
 \cos \phi  & -\sin \phi  & 0 \\
 \sin \phi  & \cos \phi  & 0 \\
 0 & 0 & 1 \\
\end{array}
\right)\left(
\begin{array}{ccc}
 \cos \theta  & 0 & \sin \theta  \\
 0 & 1 & 0 \\
 -\sin \theta  & 0 & \cos \theta  \\
\end{array}
\right) .
\end{split}
\end{equation}
We call the resulting matrix, the {\it decay matrix} $a_{\mu\nu}(\theta_b,\varphi_b;\alpha_D,\beta_D)$: 
\begin{equation}
a_{\mu\nu}
:=   \left(
\begin{array}{cccc}
 1 & 0 & 0 & 0 \\
 0 & \cos \theta_b    \cos \varphi_b  &  -  \sin \varphi_b  & \sin \theta_b \cos \varphi_b  \\
 0 & \cos \theta_b  \sin \varphi_b   & \cos \varphi_b  & \sin \theta_b \sin \varphi_b  \\
 0 & -\sin \theta_b   & 0 & \cos \theta_b  \\
\end{array}
\right)
\left(
\begin{array}{cccc}
 1 & 0 & 0 & \alpha_D \\
 0 & \gamma_D & -\beta_D & 0 \\
 0 & \beta_D & \gamma_D & 0 \\
 \alpha_D & 0 & 0 & 1 \\
\end{array}
\right)\ .\label{eq:decayM}
\end{equation}
If the polarization of the final baryon $b$ is not measured, as in the case of $\Lambda\to p\pi^-$ decay, only $a_{\mu0}$
elements of the matrix contribute. Contrary, for the unpolarized baryon $B$, the $b$-baryon polarization is given by the $a_{0\mu}$ elements. In the case of two body weak decays, this polarization is aligned along the $b$-momentum in ${\mathbb R}_B$ reference frame. The remaining elements of the matrix show direction between the $B$ and $b$ polarization vectors.

Some of the hyperon hadronic processes are listed in Table~\ref{tab:decayproperties}, where the Particle Data Group (PDG) averages  before 2018~\cite{ParticleDataGroup:2018ovx} and the most recent results are reported.
\begin{table}[ht!]
\begin{center}
 \caption{{\bf Properties of two-body hadronic decays of the ground-state hyperons.} The  recent BESIII results, \textbf{in bold},  are compared to the world averages before 2018 (reported in PDG18~\cite{ParticleDataGroup:2018ovx}). In addition BESIII~\cite{BESIII:2021ypr} and \cite{BESIII:2023jhj} measured $B_{\rm CP}=-0.005(14)(3)$ and $-0.003(8)(7)$ for  $\Xi^-\to \Lambda\pi^-$, $B_{\rm CP}=-0.0001(69)(9)$ for $\Xi^{0}\to\Lambda\pi^{0}$~\cite{BESIII:2023drj}.  Branching fractions ${\cal B}$ of the decays are rounded within $0.5\%$ accuracy.  }
 \small
  \begin{tabular}{lrllll}
    \hline\hline
    &${\cal B}$&$\braket{\alpha}$&$\braket{\phi}$&$A_{\rm CP}$&Comment\\
    \hline
    $\Lambda\to p\pi^-$&$64\%$&$\phantom{-}0.642(13)^*$&$-0.113(61)^*$&$\phantom{-}0.006(21)$&PDG18 \cite{ParticleDataGroup:2018ovx}\\
    &&{\boldmath$\phantom{-}0.754(3)(2)$}&--&{\boldmath$-0.006(12)(7)$}&BESIII \cite{BESIII:2018cnd}\\
    &&$\phantom{-}0.721(6)(5)^*$&--&--&CLAS \cite{Ireland:2019uja}\\
    &&{\boldmath$\phantom{-}0.760(6)(3)$}&--&{\boldmath $-0.004(12)(9)$}&BESIII \cite{BESIII:2021ypr}\\ 
    &&{\boldmath$\phantom{-}0.7542(10)(24)$}&--&{\boldmath$-0.0025(46)(12)$}&BESIII \cite{BESIII:2022qax}\\
    $\Lambda\to n\pi^0$&$36\%$&$\phantom{-}0.65(4)^*$&--&--&PDG18 \cite{ParticleDataGroup:2018ovx}\\
    &&{\boldmath$-0.692(17)^{**}$}&--&--&BESIII \cite{BESIII:2018cnd}\\
    &&{\boldmath$\phantom{-}0.670(9)(9)$}&--&{\boldmath$\phantom{-}0.001(9)(3)$}& BESIII \cite{BESIII:2023jhj} \\
    $\Sigma^+\to p\pi^0$&$52\%$&$-0.980(23)^*$&$\phantom{-}0.628(59)^*$&--&PDG18 \cite{ParticleDataGroup:2018ovx}\\
   &&{\boldmath$-0.994(4)(2)$}&-- 
   & {\boldmath$-0.004(37)(1)$}&BESIII \cite{BESIII:2020fqg}\\
       $\Sigma^+\to n\pi^+$&$48\%$&$\phantom{-}0.068(13)^*$&$\phantom{-}2.91(35)^*$&--&PDG18 \cite{ParticleDataGroup:2018ovx}\\
     && {\boldmath$\phantom{-}0.0506(26)(19)$}&--&{\boldmath$-0.080(52)(28)$}& BESIII \cite{BESIII:2023sgt} \\ 
    $\Sigma^-\to n\pi^-$&$100\%$&$-0.068(8)^*$&$\phantom{-}0.174(26)^*$&--&PDG18 \cite{ParticleDataGroup:2018ovx}\\
    $\Xi^0\to \Lambda\pi^0$&$100\%$&$-0.406(13)^*$&$\phantom{-}0.36(21)^*$&--&PDG18 \cite{ParticleDataGroup:2018ovx}\\
    && {\boldmath$-0.3770(24)(14)$}&--&{\boldmath$\phantom{-}0.0069(58)(18)$}& BESIII \cite{BESIII:2023drj} \\ 
    $\Xi^-\to \Lambda\pi^-$&$100\%$&$-0.458(12)^{*,\circ}$&$-0.042(16)^*$&--&PDG18 \cite{ParticleDataGroup:2018ovx,HyperCP:2004not} \\
    &&{\boldmath $-0.373(5)(2)$}&{\boldmath$\phantom{-}0.016(14)(7)$} &{\boldmath$\phantom{-}0.006(13)(6)$}&BESIII \cite{BESIII:2021ypr}\\
    && {\boldmath$-0.367(4)(3)$}& {\boldmath$-0.013(12)(8)$} &{\boldmath$-0.009(8)(7)$}& BESIII \cite{BESIII:2023jhj} \\ 
    \hline
    $\Omega^-\to \Lambda K^-$&$68\%$&{$\phantom{-}0.0181(22)^{*}$}&--&$ -0.02(13)$& PDG18 \cite{ParticleDataGroup:2018ovx}\\
    &&{\boldmath$-0.04(3)$}&{\boldmath$\phantom{-}1.08(51)$}&--&BESIII \cite{BESIII:2020lkm}\\
 $\Omega^-\to \Xi^0 \pi^-$&$24\%$&{$\phantom{-}0.09(14)$}&--&--& PDG18 \cite{ParticleDataGroup:2018ovx}\\
 $\Omega^-\to \Xi^- \pi^0$&$8\%$&{$\phantom{-}0.05(21)$}&--&--& PDG18 \cite{ParticleDataGroup:2018ovx}\\
    \hline\hline
   \multicolumn{6}{l}{${}^{\phantom{*}*}$ -- result only available for hyperons}\\
   \multicolumn{6}{l}{${}^{**}$ -- result only available for antihyperons}\\
  \multicolumn{6}{l}{${}^{\phantom{*}\circ}$ -- measured as a product with the $\alpha_\Lambda$ value}
  \end{tabular}
    \label{tab:decayproperties}
\end{center}
\end{table}

Several processes can be related by isospin 
relations. Using the isospin decomposition, in the notation similar to Ref.~\cite{Overseth:1969bxc}, of the $L=S,P$ amplitudes for $\Lambda\to N\pi$ into the two possible weak transitions  $\Delta I=1/2,3/2$:
\begin{align}
    L_{[\Lambda p]}&=-\sqrt{\frac{2}{3}}L_{1,1}\exp\!{(i\xi^{L}_{1,1}+i\delta_{1}^L)} +
    \sqrt{\frac{1}{3}}L_{3,3}\exp\!{(i\xi^{L}_{3,3}+i\delta_{3}^L)} \ , \nonumber \\
    L_{[\Lambda n]}&=\phantom{-}\sqrt{\frac{1}{3}}L_{1,1}\exp\!{(i\xi^{L}_{1,1}+i\delta_{1}^L)} +
    \sqrt{\frac{2}{3}}L_{3,3}\exp\!{(i\xi^{L}_{3,3}+i\delta_{3}^L)} \ , \label{eq:Lisospin}
\end{align}
where in the $L=P$ case $P_{[\Lambda p]}$ and $P_{[\Lambda n]}$ on the left-hand sides are to be replaced by
$(1+\Delta_{[\Lambda p]})P_{[\Lambda p]}$ and $(1+\Delta_{[\Lambda n]})P_{[\Lambda n]}$, respectively, as per the discussion in the previous paragraph.
Analogously, for the $\Xi\to\Lambda\pi$ channels one has
\begin{align}
    L_{[\Xi -]}&=L_{1,2}\exp\!{(i\xi^{L}_{1,2}+i\delta_{2}^L)} + \frac{1}{2}L_{3,2}\exp\!{(i\xi^{L}_{3,2}+i\delta_{2}^L)} \ , \nonumber \\
    L_{[\Xi 0]} & = \frac{1}{\sqrt2} L_{1,2}\exp\!{(i\xi^{L}_{1,2}+i\delta_{2}^L)} -  \frac{1}{\sqrt2} L_{3,2}\exp\!{(i\xi^{L}_{3,2}+i\delta_{2}^L)} \ . \label{eq:Xisospin}
\end{align}
The isospin decomposition for the processes $\Sigma\to N\pi$ is more complicated, since the initial stage has $I=1$ and the final states with the isospins $1/2$ and $3/2$ can be reached by weak transitions with $\Delta I=1/2,3/2$ and $5/2$. Each weak transition can include a weak $\CP$-odd phase. However, the $\Delta I=1/2$ transitions dominate, and the $\CP$ violation experiments can be analysed neglecting other contributions. 

%\section{CP violation in hyperon decays}
For the charge conjugation-transformed baryon decay process $\bar B\to \bar b\bar \pi$, the amplitude is:
\begin{equation}
\bar {\cal A}(\bar B\to \bar b\bar \pi)\propto \bar {\cal S}\sigma_0 -\bar {\cal P}{\boldsymbol{\sigma}}\cdot {\bf\hat{n}}\ ,
   \end{equation}
where the complex parameters $\bar {\cal S}$ and $\bar {\cal P}$ are obtained from  $ {\cal S}$ and $ {\cal P}$ by reversing the  sign for the weak $\CP$-odd phases $\xi_S$ and $\xi_P$.
Since the product of the $P$-odd and  $P$-even terms changes sign, the decay parameters are:
\begin{align}
 \bar\alpha&\propto-2|{\cal S}||{\cal P}| \cos(-\xi_P+\delta_P+\xi_S-\delta_S) ,\\    
 \bar\beta&\propto-2|{\cal S}||{\cal P}| \sin(-\xi_P+\delta_P+\xi_S-\delta_S)\ .
\end{align}
The weak phase difference $\xi_P-\xi_S$ can be determined using two independent experimental observables:
\begin{align}
    A_{CP}=\frac{\alpha+\bar\alpha}{\alpha-\bar\alpha}&={(\xi_P-\xi_S)\tan(\delta_P-\delta_S)}, \ 
   B_{CP}=\frac{\beta+\bar\beta}{\alpha-\bar\alpha}={(\xi_P-\xi_S)}\ .
\end{align}
Here we have assumed $\xi_P-\xi_S$ to be small, so that $\sin(\xi_P-\xi_S) \approx \xi_P-\xi_S$. 

The goal of the $\CP$ tests in hadronic decays of hyperons is to determine the weak-phase differences between $P$ and $S$ final states  for the two decays.

The BESIII results based on $1.3\times10^{9}$ and $10^{10}$ $J/\psi$ data for $(\xi_P-\xi_S)^\Xi$~\cite{BESIII:2021ypr} and $(\xi_P-\xi_S)^\Lambda$~\cite{BESIII:2022qax}, respectively,  are presented in Fig.~\ref{fig:results} by the blue rectangle. The red band represents the HyperCP experimental result on $A^\Lambda_{CP}+A^\Xi_{CP}$~\cite{HyperCP:2004zvh}. Interpretation of this measurement requires the value of $\tan(\delta_P-\delta_S)_\Xi$, that is poorly known.
Projection for the statistical uncertainties of the analysis using full BESIII data set for $e^+e^-\to J/\psi\to \Xi\overline\Xi$ is given by the brown rectangle.  The results should be compared to $\SM$ predictions of $(\xi_P-\xi_S)^\Lambda=(-0.2\pm2.2)\times10^{-4}$ and $(\xi_P-\xi_S)^\Xi=(-2.1\pm1.7)\times10^{-4}$~\cite{Donoghue:1985ww,Donoghue:1986hh}. 
%{\color{blue} cite: Donoghue and Pakvasa PRL Phys.Rev.Lett. 55 (1985) 162; Donoghue, He and Pakvasa, Phys.Rev.D 34 (1986) 833and etc. }

In case of hyperon decays, the $\SM$ contribution is dominated by QCD-penguin diagrams shown in Fig.~\ref{fig:CPV}(a). The results on the weak phases in kaon and in the hyperon decays can be combined in order to search for $\NP$. 
The present kaon data imply the limits $|\xi_P-\xi_S|_{\rm NP}^{\Lambda}\leq5.3\times10^{-3}$ and $|\xi_P-\xi_S|_{\rm NP}^{\Xi}\leq3.7\times10^{-3}$~\cite{He:1999bv}. 
Clearly, the hyperon $\CP$-violation measurements with much improved precision will provide an independent constraint on the NP contributions in the strange quark sector. 
However, a lot also remains to be done on the theory side, as the present predictions suffer from considerable uncertainties. It is hoped that the lattice analyses~\cite{Beane:2003yx} could help solve this problem in the future. 

%\section{Weak phases}
\begin{figure}[htbp!]
    \centering
\includegraphics[width=0.6\textwidth]{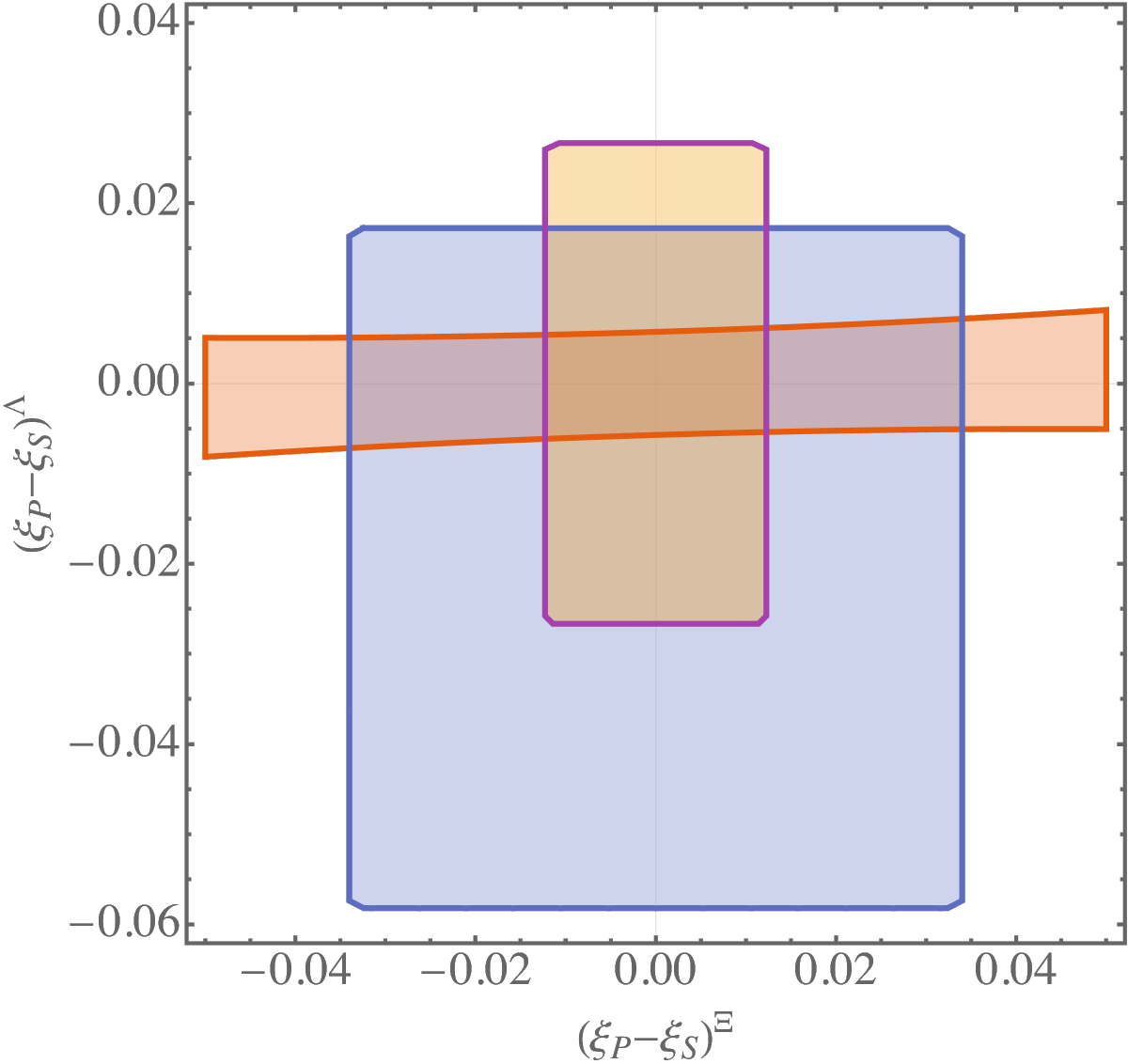}
    \caption{Summary of the results for the weak phases  $(\xi_P-\xi_S)^\Xi$ and $(\xi_P-\xi_S)^\Lambda$. The present BESIII results are given by the blue rectangle~\cite{BESIII:2021ypr,BESIII:2022qax}. The red horizontal band is the HyperCP result~\cite{HyperCP:2004zvh}. The brown rectangle is the projection of the statistical uncertainties for the complete BESIII data set.}
    \label{fig:results}
\end{figure}

%%%%%%%%%%%%%%%%%%%%%%%%%%%%%%%%%%%%%%%%%%%%%%%%%%%%%%%%%%%%%%%%%%%%%%%%%%%%%%%%%%%%%%%%%%%%%%%%%%%%%%%%%%%%%%%%%%%%%%%%%%%%%%%%%%%%%%%%%%%

\subsection{Spin entangled baryon--antibaryon systems}
STCF provides a novel method to search for $\CP$ violation in baryon decays using entangled baryon--antibaryon pairs. For stable spin $1/2$ hyperons, copiously produced pairs from the $J/\psi$ resonance decays can be used. In general, the spin state of a baryon--antibaryon system can be represented as
\begin{equation}
    \ket{B\bar B}=\sum_{\mu,\nu=0}^3C_{\mu\nu}^{B\bar B}\sigma_\mu^{(B)}\otimes\sigma_\nu^{(\bar B)}\ ,
\end{equation}
where the Pauli operators $\sigma_\mu^{(B)}$  and $\sigma_\nu^{(\bar B)}$ act in the rest frames of the baryon and antibaryon, respectively.
The coefficients of the spin correlation--polarization matrix, $C^{B\bar B}_{\mu\nu}$, depend on the baryon-pair production process. At STCF, the pair can be produced in a single photon annihilation process of an electron--positron pair. 

The vertex function $\Gamma_\mu $
%$\bar u(p_2,\lambda_2) \, \Gamma_\mu \, v(p_1,\lambda_1)$ 
for the single photon annihilation in the $e^+e^-\to B\overline B$ process with a particle-antiparticle pair of spin 1/2 and mass $m_B$ is
\cite{Korner:1976hv}
\begin{eqnarray}
  \label{eq:elasticFF12Gam}
  \Gamma_\mu := F_1(q^2) \, \gamma_\mu + F_2(q^2) \, \frac{i \sigma_{\mu\nu} q^\nu}{2m_B},
\end{eqnarray}
where $q=p_1+p_2$ denotes the momentum of the virtual photon.
These form factors are related to the helicity amplitudes by
\begin{eqnarray}
 G_E\propto A_{+1/2,+1/2} & = & 2 m_B \left(F_1 + \tau \, F_2 \right)\,, \nonumber \\
  G_M\propto A_{+1/2,-1/2} & = & \sqrt{2q^2} \, \left(F_1 + F_2\right)\,,
  \label{eq:relelFF12helamp}  
\end{eqnarray}
where $\tau=q^2/{4m^2_B}$. The expressions are formally the same for the process that proceeds via a decay of  a neutral vector meson, such as $J/\psi$ or $\psi(2S)$. If the meson decays via a single photon annihilation, the electromagnetic form factors $G_E$ and $G_M$ at the resonance are directly related to the continuum form factors via hadronic vacuum polarization.
However, if $\eg$ the decay process is strong, the interpretation of the form factors is different and relates to the properties of the resonance. For example, for intermediate vector charmonia resonances, they are often called {\it psonic} form factors~\cite{Faldt:2017kgy} to differentiate them from the electromagnetic form factors. Therefore, the single photon annihilation is fully described by two complex scalars. If we ignore the unmeasurable global phase, there are three real parameters that fully define the properties of the process. One of the parameters defines the overall normalization and the total cross section of the reaction.
In the standard setting, the collider beams are unpolarized and the elements of the $C^{B\bar B}_{\mu\nu}$ matrix are known functions of the baryon $B$ production angle, $\theta$. The functions for a given c.m.~energy depend on the two remaining parameters that need to be determined from data~\cite{Faldt:2017kgy,Perotti:2018wxm}. 
%All elements of the $C^{B\bar B}_{\mu\nu}= $ matrix depend on the two global parameters that are related to the amplitudes  describing spin flip and non-flip contributions to the annihilation process. 
The $G_M$ and $G_E$ form factors are functions of the centre-of-mass energy squared, $s$. The  distribution of the production angle $\theta$ is given as 
\begin{equation}
    \frac{1}{\sigma} \frac{\dd\sigma}{\dd\cos\theta}=\frac32\frac{1+\alpha_P\cos^2\theta}{3+\alpha_P}\ ,
\end{equation}
where 
\begin{equation}
    \alpha_P=\frac{\tau|G_M|^2-|G_E|^2}{\tau|G_M|^2+|G_E|^2}=\frac{\tau-R^2}{\tau+R^2}
\end{equation}
with $R:=|G_E|/|G_M|$.
In addition, if the relative complex phase between the amplitudes $\Delta\Phi:={\rm Arg}(G_E/G_M)$ is not zero, the baryons spin is polarized in the direction perpendicular to the reaction plane,  given by the $\hat{\bf y}$ unit vector. The polarization vector component  $P_y(\theta)$ and the spin correlation terms $C_{ij}(\theta)$, $i,j=x,y,z$ are:
\begin{equation}
C^{B\bar B}_{\mu\nu}= \frac{1}{\sigma}\frac{\dd\sigma}{\dd\cos\theta}\left(
\begin{array}{cccc}
1&0&P_y&0\\
0&C_{xx}&0&C_{xz}\\
-P_y&0&C_{yy}&0\\
0&-C_{xz}&0&C_{zz}\\
\end{array}
\right)\ ,\label{eqn:cxx}
\end{equation}
where the orientations of the coordinate systems in the rest frames of $B$ and $\bar B$ are related as $(\hat{\bf x},\hat{\bf y},\hat{\bf z})_B=(\hat{\bf x},-\hat{\bf y},-\hat{\bf z})_{\bar B}$. 
The explicit expressions for the spin correlation matrix terms are:
    \begin{align*}
    P_y&=\sqrt{1-\alpha_P^2}\sin(\Delta\Phi)\sin\theta\cos\theta:=\beta_P\sin\theta\cos\theta,\\
    C_{xy}&=\sqrt{1-\alpha_P^2}\cos(\Delta\Phi)\sin\theta\cos\theta:=\gamma_P\sin\theta\cos\theta,\\
    C_{xx}&=\sin^2\theta,\\
     C_{yy}&= \alpha_P\sin^2\theta,\\    
     C_{zz}&=\alpha_P+\cos^2\theta \ .    
\end{align*}
The above expressions were derived for unpolarized electron and positron beams. At STCF longitudinally polarized electron beams will be available. The corresponding spin correlation matrix is given in Ref.~\cite{Salone:2022lpt}.

For hyperon decay studies at STCF, reactions of special interest are $J/\psi$ decays. This is due to the huge yield of hyperon--antihyperon pairs from strong $J/\psi$ decays. These are manifested by the large branching fractions of $1$--$2$ per mil. The branching ratios and the form factor parameters $\alpha_\psi$ and $\Delta\Phi_\psi$ for the relevant $e^+e^-\to J/\psi\to B\overline B$ processes are given in Table~\ref{tab:Prod}. Due to the non-zero phase between the form factors, the hyperons are polarized in the direction perpendicular to the production plane in the case of unpolarized beams. This polarization changes with the production angle as shown in Fig.~\ref{fig:polar} by the solid lines. The impact of the electron beam polarization for the absolute value of the hyperon polarization is shown by dashed and dotted lines for the beam polarization of 80\% and 100\%, respectively. For the polarized beam, the direction of the hyperon polarization vector is not restricted to the direction orthogonal to the production plane. 
\begin{table}
\begin{center}
  \caption{Properties of the $e^+e^-\to J/\psi\to B\overline B$ decays to the pairs of ground-state octet hyperons. \label{tab:Prod}}
%\begin{ruledtabular}
\begin{tabular}{lllll}
\hline\hline
  Final state&${\cal B}(\times 10^{-4})$& $\alpha_\psi$&$\Delta\Phi_\psi (\text{rad})$&Comment\\ \hline
$\Lambda\overline{\Lambda}$&$19.43(3)$&$\phantom{-}0.4748(22)(31)$&$\phantom{-}0.7521(42)(66)$&\cite{BESIII:2018cnd,BESIII:2017kqw,BESIII:2022qax}\\
$\Sigma^+\overline\Sigma\vphantom{X}^-$&$15.0(24)$&$-0.508(7)$&$-0.270(15)$ &\cite{BES:2008hwe, BESIII:2020fqg}\\
$\Sigma^-\overline\Sigma\vphantom{X}^+$&\multicolumn{4}{c}{--- no data ---}\\
$\Sigma^0\overline\Sigma^0$&$11.64(4)$&$-0.4133(85)$&$-0.0828(76)$&\cite{BESIII:2024nif}\\
$\Xi^0\overline\Xi\vphantom{X}^0$&$11.65(43)$&$\phantom{-}0.514(6)(15)$& $\phantom{-}1.168(19)(18)$&\cite{BESIII:2016nix,BESIII:2023drj}\\
$\Xi^-\overline\Xi\vphantom{X}^+$&$\phantom{0}9.7(8)$&$\phantom{-}0.611(7)(13)$&$\phantom{-}1.30(3)(3)$&\cite{ParticleDataGroup:2020ssz,BESIII:2021ypr,BESIII:2023jhj}\\
\hline\hline 
 \end{tabular}
\end{center}
%\end{ruledtabular}
\end{table}

\begin{figure}
\centering
\includegraphics[width=\textwidth]{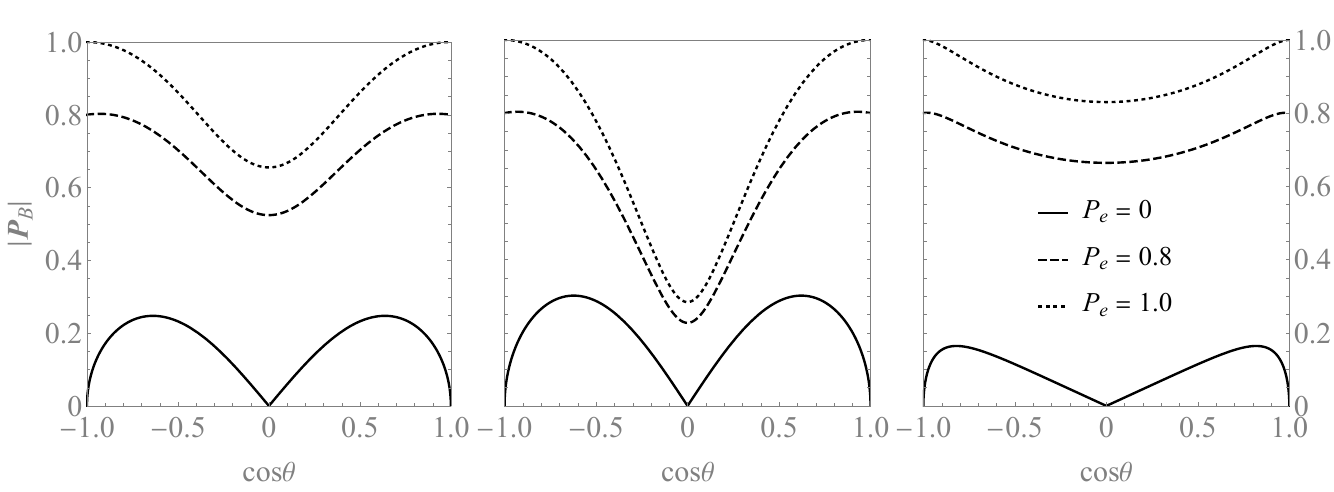}
\caption[]{ Magnitudes of the hyperon polarization as a function of the production angle for:
(a) $\Lambda$, (b) $\Xi^-$ and (c) $\Sigma^+$ for the electron beam polarizations $P_e=0$, $0.8$, $1$ (solid, dashed and dotted lines, respectively).
  The  $\alpha_\psi$ and $\Delta\Phi_\psi$ values are taken from Table~\ref{tab:Prod}.
  \label{fig:polar}}
\end{figure}

Equation~\eqref{eq:Ampli} implies that in the parity-violating weak decays of the hyperons, the average spin direction is reflected by the angular distributions of the decay products. For example, in the case of the $e^+e^-\to \Lambda\overline\Lambda$ reaction with one-step decays $\Lambda(\bar\Lambda)$  into $p\pi^-(\bar p\pi^+)$ the complete five-fold angular distribution %for  
is
\begin{align}
   &{\cal W}(\theta,{\bf n}_{\Lambda},{\bf n}_{\bar\Lambda} )\propto\\
   & \left(1+\alpha_P\cos^2\theta\right)\left[1+\left(\alpha_\Lambda n_{\Lambda,y}+\alpha_
    {\bar\Lambda} n_{\bar\Lambda,y}\right)P_y
    +\alpha_\Lambda\alpha_
    {\bar\Lambda}\sum_{i,j=x,y,z}C_{ij}^{\Lambda\bar\Lambda}  n_{\Lambda,i} n_{\bar\Lambda,j}\ 
    \right],
\end{align}
where $n_{\Lambda,i}$ and $n_{\bar\Lambda,j}$ are directional cosines of the proton (antiproton) in the $\Lambda$ and $\bar\Lambda$ reference frames, respectively. 
It can be rewritten in the modular form as 
\begin{equation}
\label{eqn:jointL}
    {{\cal{W}}}(\boldsymbol{\xi};\boldsymbol{\omega}_\Lambda) = \sum_{\mu,\nu = 0}^{3}C^{\Lambda\bar\Lambda}_{\mu\nu} a_{\mu0}^{\Lambda} a_{\nu0}^{\overline\Lambda}\ .  
\end{equation}

For multi-step processes there are more kinematic variables %degrees 
and the expression for the probability density function is much more complicated. They can be conveniently expressed in modular form, as derived in Ref.~\cite{Perotti:2018wxm}. In addition to the $\alpha_B$ parameter also the $\beta_B$ can be involved. 
For example, in the $e^+e^-\to \Xi\overline\Xi$ process with two-step decay sequence $\Xi\to\Lambda\pi$ and $\Lambda\to p\pi^-$, one can determine separately $\alpha_\Xi$, $\beta_\Xi$ and  $\alpha_\Xi$ parameters.
The production and the two-step decays in the $e^+e^-\to\Xi^-\overline\Xi\vphantom{\Xi}^+$ case are described by a nine-dimensional vector $\boldsymbol{\xi}$ of the helicity angles. The structure of the nine-dimensional angular distribution is determined by eight global parameters $\boldsymbol{\omega}_\Xi:=(\alpha_{\psi},\Delta\Phi,\alpha_{\Xi},\phi_{\Xi},\overline{\alpha}_{\Xi},\overline{\phi}_{\Xi},\alpha_{\Lambda},\overline{\alpha}_{\Lambda})$ and in the modular form gives \cite{Perotti:2018wxm}:
\begin{equation}
\label{eqn:jointXi}
    {{\cal{W}}}(\boldsymbol{\xi};\boldsymbol{\omega}_\Xi) = \sum_{\mu,\nu = 0}^{3}C^{\Xi\overline\Xi}_{\mu\nu}\sum_{\mu'\nu' = 0}^{3} a_{\mu\mu'}^{\Xi} a_{\nu\nu'}^{\overline{\Xi}} a_{\mu'0}^{\Lambda} a_{\nu'0}^{\overline\Lambda}\ .  
\end{equation}

The BESIII experiment has studied two reactions that provide the best sensitivity for the $\CP$ tests in hyperon decays:
 $e^+e^-\to J/\psi\to \Lambda\overline\Lambda$~\cite{BESIII:2018cnd,BESIII:2022qax} and $e^+e^-\to J/\psi\to \Xi\overline\Xi$~\cite{BESIII:2021ypr}.
At the BESIII experimental setting -that should be similar to STCF- for every million of the produced $J/\psi$ resonances, 320 and 56 of fully reconstructed $\Lambda\overline \Lambda$ and $\Xi\overline \Xi$ pairs are selected, respectively.

With $10^{10}$ $J/\psi$ mesons available at BESIII, precision measurements of the hyperon decay parameters are possible.
The average baryon polarizations $|P_y|$ in the $e^+e^-\to J/\psi\to \Lambda\overline\Lambda$ and $e^+e^-\to J/\psi\to \Xi\overline\Xi$ reactions are 18\% and 23\%, respectively.  The two production reaction parameters, the decay parameters $\alpha_\Lambda$, $\alpha_\Xi$ and $\beta_\Xi$, as well as the $\CP$-violating variables $A_{CP}^\Lambda$, $A_{CP}^\Xi$ and $B_{CP}^\Xi$, can be determined using unbinned maximum likelihood fit to the measured angular distributions. 
The statistical uncertainty of the observable $A_{CP}^\Lambda$ in the process $e^+e^-\to J/\psi\to \Lambda\overline\Lambda$ is inversely proportional to the $\Lambda$ polarization~\cite{Salone:2022lpt}.
The weak phase difference for the $\Xi$ baryon decay is directly given by the measurement of the  $B_{CP}^\Xi$ observable in the process $e^+e^-\to J/\psi\to \Xi\overline\Xi$. The sensitivity of such measurement depends on the average squared of the $\Xi$ polarization and the $\Xi\overline\Xi$ spin correlation terms~\cite{Salone:2022lpt}.

%%%%%%%%%%%%%%%%%%%%%%%%%%%%%%%%%%%%%%%%%%%%%%%%%%%%%%%%%%%%%%%%%%%%%%%%%%%%%%%%%%%%%%%%%%%%%%%%%%%%%%%%%%%%%%

\subsection{Radiative and semileptonic decays}
Hyperon radiative decays $B_1\to B_2\gamma$ have branching fractions of the order of per mil. The angular distribution of the daughter baryon in a radiative decay of a polarized hyperon $B_1$ has the same form as in the the case of weak hadronic decays. 
It is described by a decay asymmetry parameter $\alpha_\gamma$. The branching fractions and the  values of decay parameters are summarized in Table~\ref{tab:radiative}. However, polarization of the baryon $B_2$ is given by
\begin{table}[htbp!]
    \centering
        \caption{Hyperon radiative decays: branching fractions, ${\cal B}$, and decay asymmetries, $\alpha_\gamma$.}
    \begin{tabular}{c|llc}
    \hline\hline
    Deacay&${\cal B}\times10^3$&$\alpha_\gamma$&Comment\\
    \hline
             $\Sigma^+\to p\gamma$&1.04(6) & $-0.76(8)$& \cite{ParticleDataGroup:2024cfk}\\
         $\Sigma^+\to p\gamma$&$0.996(28)$ & $-0.652(56)$& \cite{BESIII:2023fhs}\\
         $\Lambda\to n\gamma$&0.832(66) & $-0.16(11)$&\cite{ParticleDataGroup:2024cfk}\\
        $\Xi^0\to\Sigma^0\gamma$&3.33(10)  &$-0.69(6)$&\cite{ParticleDataGroup:2024cfk}\\
         $\Xi^0\to\Lambda\gamma$&$1.17(7)$&$-0.70(7)$ &\cite{ParticleDataGroup:2024cfk}\\
         $\Xi^{0}\to\Lambda\gamma$&$1.347(85)$&$-0.741(65)$ & \cite{BESIII:2024lio}\\
      $\Xi^-\to\Sigma^-\gamma$&$0.13(2)$& --&\cite{ParticleDataGroup:2024cfk}\\
    \hline\hline
    \end{tabular}
    \label{tab:radiative}
\end{table}
%such as $\Lambda\to n\gamma$ or $\Sigma^+\to p\gamma$ 
the following aligned decay matrix corresponding to the situation where the radiative photon helicities are unmeasured% is
~\cite{Batozskaya:2023rek}: 
\begin{equation}
    b_{\mu\nu}^\gamma\propto\left(
\begin{array}{rrrr}
 1 & \phantom{-}0 & \phantom{-}0 & \alpha_\gamma \\
 0 & 0 & 0 & 0 \\
 0 & 0 & 0 & 0 \\
 -\alpha_\gamma & 0 & 0 & -1 \\
\end{array}
\right)\ .
\end{equation}
Therefore, even if the final  baryon polarization is measured, only the $\alpha_\gamma$ parameter is accessible. The $\CP$ symmetry can be studied by comparing the branching fractions and the decay parameters for a hyperon decay to the charge conjugated process of the antihyperon. Studies with the hyperon--antihyperon pairs produced at an electron--positron collider provide these quantities simultaneously. For example, for $\Sigma^+\to p\gamma$, BESIII has determined $\Delta_{CP}=({{\cal B}_+-{\cal B}_-})/({{\cal B}_++{\cal B}_-})=0.006(12)$ and $A_{CP}=({\alpha_++\alpha_-})/({\alpha_+-\alpha_-})=0.095(89)$~\cite{BESIII:2023fhs}.

The weak radiative processes and the hadronic weak decays are coupled channels that can mix due to parity-conserving electromagnetic interactions.
For example, according to the BESIII data~\cite{BESIII:2022rgl}, the branching fraction for decay $\Lambda\to n\gamma$ is completely saturated by the unitarity lower limit of $0.85\times10^{-3}$ ~\cite{Farrar:1971xi}, where the two intermediate weak processes $\Lambda\to p\pi^-$ and $\Lambda\to n\pi^0$ combine with the inverse 
photoproduction processes $p\pi^-\to n\gamma$ 
and $n\pi^0\to n\gamma$% are considered
. On the other hand, the amplitude of $\Sigma^+\to p\gamma$ requires a significant real part of the amplitude, due to short-range contributions, since the unitarity lower limit is only $0.69\times10^{-5}$, $\ie$ more than two orders of magnitude lower the the observed branching fraction.
Closely related are decays where the produced radiative photon converts externally into an electron--positron pair. It is possible to define additional $\CP$ tests (similar as for semileptonic~(SL) processes discussed below), however, they are suppressed by $\alpha_\text{em}$, what limits the sensitivity. An example of such process is $\Xi^0\to\Lambda e^+e^-$, with branching fraction $7.6(6)\times10^{-6}$, and the decay asymmetry $\alpha_{\gamma^*}=-0.8(2)$, measured by the NA48 collaboration~\cite{NA48:2007smd}.
Other interesting conversion decay is $\Sigma^+\to p\mu^+\mu^-$ , where the first observation at HyperCP~\cite{HyperCP:2005mvo} suggested an anomalously large branching fraction of $\left(8.6^{+6.6}_{-5.5}\pm 5.5\right)\times10^{-8}$. Moreover, the $\mu^+\mu^-$ invariant masses for the three observed events were within 1 MeV, suggesting a narrow resonance contribution. 
The experiment was repeated by the LHCb collaboration, determining the branching fraction  $\left(2.2^{+0.9+1.5}_{-0.8-1.1}\right)\times10^{-8}$~\cite{LHCb:2017rdd}, consistent with the SM predictions~\cite{He:2005yn,He:2018yzu}. It is important to mention the first attempts to calculate this decay from first principles using Lattice QCD~\cite{Erben:2022tdu}.

In addition, in the electromagnetic decay $\Sigma^0\to\Lambda\gamma$, flavour-diagonal  $\CP$ violation can be searched for, by first determining a non-zero value for $\alpha_\gamma$ which would indicate edm contribution and then comparing it with the $\bar\alpha_\gamma$ value~\cite{Pospelov:2005pr, Nair:2018mwa}. 

SL decays $B_1\to B_2\ell\bar\nu_\ell$ proceed via external emission of the intermediate $W$ boson.  At STCF exclusive measurements of many SL decays can be performed using a tagging method, that allows to determine the complete kinematic information of the processes with the neutrino escaping the detector. Therefore, in principle, all decay form factors can be determined.

\begin{table}[ht]
    \centering
        \caption{Branching fractions of semileptonic decays of hyperons~\cite{ParticleDataGroup:2024cfk}.}
    \begin{tabular}{l|c c}
    \hline\hline
        Transition & \multicolumn{2}{c}{$\cal B$} \\
        &$e^-\bar\nu_e$ &$\mu^-\bar\nu_\mu$\\
        \hline
       $\Lambda\to pW^-$ & $8.34(14)\times10^{-4}$ &
       $1.51(19)\times10^{-4}$\\
       $\Sigma^-\to nW^-$ &
       $1.007(34)\times10^{-3}$ &
       $4.5(4)\times10^{-4}$\\
       $\Sigma^+\to \Lambda W^+$ & $2.3(4)\times10^{-5}$ & $-$ \\
       $\Sigma^+\to nW^+$ & $<5\times10^{-6}$& $<3.0\times10^{-5}$ \\
       $\Xi^0\to\Sigma^+W^-$ & $2.52(8)\times10^{-4}$& $2.33(35)\times10^{-6}$\\
       $\Xi^{0}\to\Sigma^{-}W^{+}$ & $<1.6\times10^{-4}$ & $<9\times10^{-4}$ \\
       $\Xi^{0}\to p W^{-}$ & $<1.3\times10^{-3}$ & $<1.3\times10^{-3}$ \\
        $\Xi^-\to\Xi^0 W^-$ & $<1.9\times10^{-5}$ & $-$\\
       $\Xi^-\to\Sigma^0W^-$ & $8.7(17)\times10^{-5}$ & $<8\times10^{-4}$ \\
       $\Xi^-\to\Lambda W^-$ & $1.27(23)\times10^{-4}$&  $3.5(35)\times10^{-4}$\\
    \hline\hline
    \end{tabular}
    \label{tab:SL}
\end{table}

The amplitude for a SL decay of a 1/2$^+$ hyperon $B_1$ into a 1/2$^+$ baryon $B_2$ and an off-shell $W^-$-boson decaying to the lepton pair $l^-\bar\nu_l$ with the momenta and masses denoted as $B_1(p_1,M_1)\to B_2(p_2,M_2)+l^-(p_l,m_l)+\bar\nu_l(p_{\nu},0)$ due to the vector $J^V_{\mu}$ and axial-vector $J^A_{\mu}$ currents is~\cite{Kadeer:2005aq}
\begin{equation}\label{eq:matrixelem_semil}
 \begin{aligned}
 \langle B_2|J^{V}_{\mu}+J^{A}_{\mu}|B_1\rangle 
& = \bar{u}(p_2)\left[\gamma_{\mu}\left(F^V_1(q^2)+F^A_1(q^2)\gamma_5\right)+\frac{i\sigma_{\mu\nu}q^{\nu}}{M_1}\left(F^V_2(q^2)+F^A_2(q^2)\gamma_5\right)\right.\\
&+\left.\frac{q_{\mu}}{M_1}\left(F^V_3(q^2)+F^A_3(q^2)\gamma_5\right)\right]u(p_1) \ ,
 \end{aligned}
\end{equation}
where $q_{\mu}:=(p_1-p_2)_{\mu}=(p_l+p_\nu)_\mu$ is the four-momentum transfer. The four-momentum squared $q^2$ ranges from $m_l^2$ to $(M_1-M_2)^2$. The form factors $F_{1,2,3}^{V,A}(q^2)$ are complex functions of $q^2$ that describe hadronic effects in the transition.  Neglecting  possible $\CP$-odd weak phases, the  corresponding form factors are the same for the $(l^-,\bar{\nu}_l)$ and $(l^+,\nu_l)$ transitions. 
To fully determine the hadronic part of a SL decay, the six involved form factors should be extracted as a function of $q^2$. The form factors  are usually parameterized by the axial-vector to vector $g^D_{av}$ coupling, the weak-magnetism  $g^D_w$ coupling and the pseudoscalar $g^D_{av3}$ coupling. They are obtained by normalizing to $F_1^V(0)$:
\begin{equation}\label{eq:g_q2}
 g^D_{av}(q^2)=\frac{F_1^A(q^2)}{F_1^V(0)}\ , \qquad g^D_w(q^2)=\frac{F_2^V(q^2)}{F_1^V(0)}\ , \qquad
 g^D_{av3}(q^2)=\frac{F_3^A(q^2)}{F_1^V(0)}\ .
\end{equation}
Since the hyperon decays have  limited range of $q^2$, the $q^2$-dependence, in the first approximation, can be completely neglected by using the values at the $q^2=0$ point. An improved approximation includes an effective-range parameter $r_i$ that represents a  linear dependence on $q^2$:
\begin{equation}
    F_i(q^2)=F_i(0)\left[1+r_iq^2+...\right]\label{eq:g_q2r}\ .
\end{equation}
In addition, since there are no hadronic intermediate states in this limited $q^2$ range, the form factors are real-valued functions. The  couplings and the effective range parameters can be determined from the multidimensional distributions observed in experiments. The optimal approach for such parametric estimation is the maximum likelihood method using multidimensional unbinned data. The experiments do not measure spin of the leptons but if a hyperon is produced, its spin can be determined from a sub-sequential weak decay. 
This information about a SL decay can be summarized in terms of the following aligned decay matrix~\cite{Batozskaya:2023rek}: 
\begin{equation}
    b_{\mu\nu}=\left(
\begin{array}{cccc}
b_{00} &{b}_{01} & 0 & {b}_{03} \\
 {b}_{10}&b_{11} & 0& {b}_{13}\\
 0& 0&b_{22} & 0 \\
 {b}_{30} & {b}_{31} & 0& b_{33} \\
\end{array}\label{eq:Bmatrix}
\right) \ .   
\end{equation}
In the $B_1$ reference frame,
$\mathbb{R}_1$, the four-momentum vector of the off-shell $W^-$ is 
$q_{\mu}=\left(q_0,p\sin\theta_W\cos\phi_W ,p\sin\theta_W \sin\phi_W,p\cos\theta_W\right)$. The energy $q_0$ and the magnitude of the  three-momentum $p$ of the off-shell $W^-$ boson are the following functions of $q^2$
\begin{equation}\label{eq:q0}
 q_0(q^2) = \frac{1}{2M_1}(M^2_1-M^2_2+q^2)
\end{equation}
and
\begin{equation}\label{eq:momentum_b2}
 p(q^2)=|{\bf p}_2|=\frac{1}{2M_1}\sqrt{Q_+Q_-},
\end{equation}
where
\begin{equation}\label{eq:Qpm}
 Q_{\pm}=(M_1\pm M_2)^2-q^2.
\end{equation}

The decay $W^-\to l^-\bar\nu_l$ is described in the $\mathbb{R}_W$ reference frame, where the emission angles of the $l^-$ lepton are $\theta_l$ and $\phi_l$. The value of the lepton momentum in this frame is 
\begin{equation}\label{eq:lepmom}
    |{\bf p}_l|=\frac{q^2-m_l^2}{2\sqrt{q^2}}\ .
\end{equation}

Contrary to the decay matrices for two body decays, each element is a function of $q^2$ and $\cos\theta_l$. Moreover, the alignment of the decay plane for a three-body decay requires all three Euler angles.
The differential decay rate is obtained from $b_{00}$ by multiplying by the kinematic and spinor normalization factors that depend on $q^2$
\begin{align}\label{eq:dg}
    {\dd\Gamma}&= \frac{G_F^2}{(2\pi)^5}|V_{us}|^2\frac{|{\bf p}_l| |{\bf p}_2|}{16 M_1^2} (q^2-m_l^2)b_{00}{\dd q \dd\Omega_2 \dd\Omega_l}\\
    &= {G_F^2}|V_{us}|^2{V_{Ph}(q^2)} (q^2-m_l^2)b_{00}{\dd q \dd\Omega_2 \dd\Omega_l}\ ,
\end{align}
where $V_{Ph}(q^2)= {(2\pi)^{-5}}{(4 M_1)^{-2}}{|{\bf p}_l| |{\bf p}_2|}$ is the three-body phase space density factor~\cite{ParticleDataGroup:2022pth}. The momenta $|{\bf p}_2|$ and $|{\bf p}_l|$ of the baryon $B_2$ and the lepton are given in Eqs.~\eqref{eq:momentum_b2} and~\eqref{eq:lepmom}, respectively.
The virtue of hyperon SL decays is that they can be used as complementary determination of the $|V_{us}|$ element of the $\CKM$ matrix~\cite{Cabibbo:2003cu}. The necessary form factors are two-point functions which might be soon accessible in Lattice QCD calculations~\cite{Sasaki:2012ne,Cooke:2014tpy}.

%%%%%%%%%%%%%%%%%%%%%%%%%%%%%%%%%%%%%%%%%%%%%%%%%%%%%%%%%%%%%%%%

\subsection{$\CP$ violation in production via EDM}
%\color{red}comments of Pablo:  shouldn't one discuss G-parity, CVC, and other (approximate) symmetries explaining hierarchy among big and small FF contributions? Perhaps also Ademollo-Gatto, symm. breaking? 

$\Lambda$ hyperon is the only one in the hyperon family with a measured EDM upper limit of $1.5\times 10^{-16}$ $e$ cm, at Fermilab~\cite{Pondrom:1981gu}. The $\Lambda$ EDM absolute value, indirectly estimated from the neutron EDM upper limit, is $< 4.4\times 10^{-26}$ $e$ cm~\cite{Guo:2012vf,Atwood:1992fb,Pich:1991fq,Borasoy:2000pq}. No indirect predictions exist for hyperons with two or three strange valence quarks. Challenges in direct EDM measurements for hyperons include their limited lifetimes and difficulties in preparing diverse hyperon sources for a single fixed-target experiment.

STCF can efficiently produce and reconstruct massive entangled pairs of $\Lambda$, $\Sigma$, and $\Xi$ hyperon-antihyperon from charmonium $J/\psi$ decays, which allows us to study minute violations of conservation laws. The entangled production of hyperon-antihyperon pairs incorporates the electric dipole form factor in the $P$- and $\CP$-violating components of the Lorentz invariant amplitude, providing a unique opportunity to indirectly extract hyperon EDM. 
If we assume a $J=1$ object in the $s$-channel for the annihilation of electron--positron into a baryon--antibaryon pair but without imposing $P$ and $C$ conservation, there are four complex form factors to describe the process~\cite{He:2022jjc}. 
%The helicity amplitude of $J/\psi \rightarrow B\bar{B}$ decay with four complex form factors can be written as 
To make direct link to the EDM of baryon, we write the helicity amplitude of $J/\psi\to B\bar{B}$ decay with four
complex form factors following Ref.~\cite{He:2022jjc} as:
\begin{equation}
    \begin{aligned}
    \mathcal{M}_{\lambda_{1},\lambda_{2}}=\epsilon_{\mu}(\lambda_1-\lambda_2)\bar{u}(\lambda_{1},p_1) \Big[F_{V}(q^{2})\gamma^{\mu}+\frac{i}{2m}\sigma^{\mu\nu}q_{\nu}H_{\sigma}(q^{2})\\
    +\gamma^{\mu}\gamma^{5}F_{A}(q^{2})+\sigma^{\mu\nu}\gamma^{5}q_{\nu}H_{T}(q^{2}) \Big]v(\lambda_{2},p_2),
    \end{aligned}
\end{equation}
where $q^{2}=M^{2}_{J/\psi}$, $m$ is $B$ hyperon mass, and $p_{1}$ and $p_{2}$ are the four-momenta of hyperon $B$ and anti-hyperon $\overline{B}$, respectively.

Processes involving a flavor-diagonal $\CP$-violating vertex contribute to the electric dipole form factor $H_T$, connecting hyperon EDM to fundamental theories. Various extensions of the SM lead to unique contributions to these operators, impacting hyperon EDM differently. Hyperon EDM measurements offer a direct way to assess the contributions of quark EDM and quark chromoelectric dipole moment~\cite{Du:2024jfc}, as the effects of high-dimensional operators are suppressed and neutron EDM measurements constrain the QCD $\theta$ term. NP, like supersymmetry, could significantly enhance hyperon EDM, which would indicate a special coupling between strange quarks and NP. By establishing a connection between the $\CP$ violation form factor $H_{T}$ and the hyperon EDM contribution, one finds~\cite{He:1992ng, He:1993ar, He:2022jjc}: 
\begin{equation}
	\begin{aligned}
        H_{T}(q^{2})=\frac{2e}{3M^{2}_{J/\psi}}g_{V}d_{B}(q^{2}).
	\end{aligned}
\end{equation} 
The EDM form factor $d_{B}(q^{2})$ is typically a complex number for nonzero timelike momentum transfer, equal to the real EDM of hyperon $B$ in the vanishing $q^{2}$ limit. This form factor can effectively be treated as an EDM under the assumption that the momentum transfer dependence is negligible, by considering an unknown extension to the zero region. 
The contribution from the exotic form factors could be accommodated in 
the spin-density matrix $C^{B\bar B}_{\mu\nu}$ corresponding to eq.~\eqref{eqn:cxx} by adding the components related to the CPV contribution
\begin{align}
   C_{0x}^\text{CPV}&=\phantom{-}m_B^2\sin2\theta\left(
   {\tau}\Im a-\Re b\right)\nonumber\\ 
   C_{x0}^\text{CPV}&=-m_B^2\sin2\theta\left(
   {\tau}\Im a+\Re b\right)\nonumber\\ 
C_{0z}^\text{CPV}&={-}\sqrt{\tau}\left[
   (\Re c- 2\Im d)\cos^2\theta+\Re c+ 2\Im d\right]\nonumber\\  
   C_{0z}^\text{CPV}&=\phantom{-}\sqrt{\tau}\left[
   (\Re c+ 2\Im d)\cos^2\theta+\Re c- 2\Im d\right]\\ 
   C_{xy}^\text{CPV}&={-}\sqrt{\tau}\sin^2\theta\left(\Im c- 2\Re d\right)\nonumber\\
    C_{yx}^\text{CPV}&=\phantom{-}\sqrt{\tau}\sin^2\theta\left(\Im c+ 2\Re d\right)\nonumber\\
       C_{yz}^\text{CPV}&=\phantom{-}m_B^2\sin2\theta\left(
   {\tau}\Im a-\Re b\right)\nonumber\\ 
   C_{zy}^\text{CPV}&=-m_B^2\sin2\theta\left(
   {\tau}\Im a+\Re b\right)\nonumber
\end{align}
% \begin{equation}
% \resizebox{\textwidth}{!}{% 
%     $ C_{\mu\nu}^\text{CPV}=\frac12\sin\theta\left(
%     \begin{array}{cccc}
%      0 & 0 & 0 & 0 \\
%      0 & 0 & - \sin\theta \left(a+2 M_\Lambda b\right) & 0 \\
%      0 &  \sin\theta \left(a-2 M_\Lambda b\right) & 0 & -\frac{1}{4} \cos\theta \left((s-4) a-2 \sqrt{s} \left(M_\Lambda c-2 d\right)\right) \\
%      0 & 0 & -\frac14  \cos\theta \left((s-4) a-2 \sqrt{s} \left(M_\Lambda c+2 d\right)\right) & 0 \\
%     \end{array}\right) $}
% \end{equation}
and the ones due to the EDM contribution
\begin{equation}
    C_{\mu\nu}^P=-2m_B^2d_J|G_M|^2\frac{\tau}{1+\alpha_P}\left(
    \begin{array}{cccc}
     0 & \gamma_P\sin\theta & 0 & 2\tau\zeta_P\cos\theta \\
     \gamma_P\sin\theta & 0 & 0 & 0 \\
     0 & 0 & 0 & \gamma_P\sin\theta  \\
     -2\tau\zeta_P\cos\theta & 0 & \gamma_P\sin\theta & 0 \\
    \end{array}\right)
\end{equation}
with
\begin{gather}
\zeta_P:=\frac{1+\alpha_P+\gamma_P}{1+\alpha_P}\\
    a:=F_A G_M^* \\
    b:=\frac{1}{2m_B}H_T G_E^* \\
    c:=\frac{1}{m_B}F_A G_E^* \\ 
    d:=\frac{1}{m_B^2}H_T G_M^*\ .
\end{gather}
The combined production matrix is then the sum $C^{B\bar B}_{\mu\nu}+C_{\mu\nu}^\text{CP}+C_{\mu\nu}^P$ and it can be applied for a modular description of the joined angular distributions.

\begin{figure*}[!htbp]
    \centering
    \includegraphics[width=0.8\linewidth]{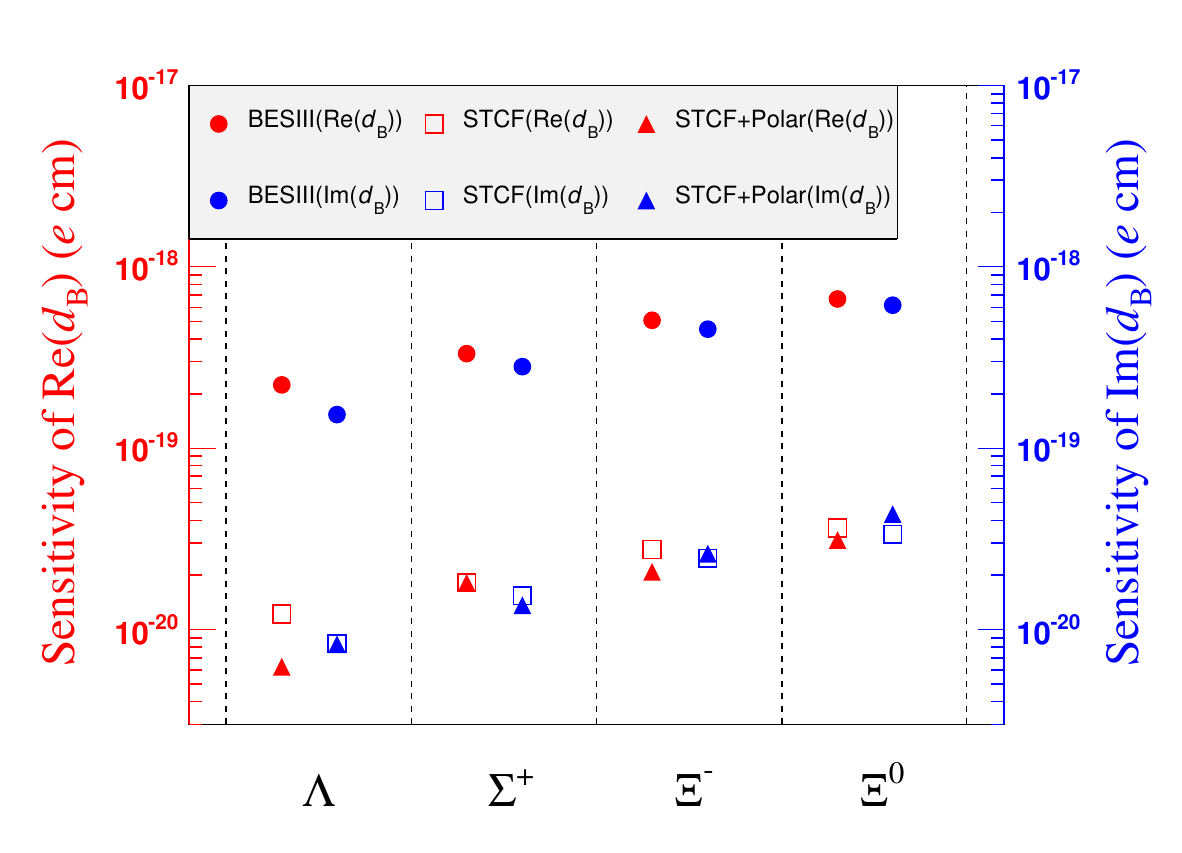}
    \caption{\label{fig:sensitivity}
	Sensitivity of EDM for different hyperons. The markers for hyperons $\Lambda$, $\Sigma^{+}$, $\Xi^{-}$, and $\Xi^{0}$ are located within dashed regions, respectively. The red and blue markers correspond to the sensitivity of the physical quantities represented by the red title (left Y axis) and blue title (right Y axis) in the graph. The full circle, open square, and full triangle up correspond to the estimated sensitivities for the BESIII experiment, the STCF experiment with an unpolarized beam, and the STCF experiment with $80\%$ polarized electron beam, respectively.}
\end{figure*}

To evaluate the EDM measurement sensitivity for various hyperon systems at STCF, we generated 500 pseudo-data sets using a cascade decay probability density function that incorporates hyperon production and decay processes~\cite{He:2022jjc, Fu:2023ose}. Factors such as branching fractions, detection efficiencies, and the effect of a longitudinally polarized electron beam were taken into account. These pseudo-data sets were analyzed through a maximum likelihood fit, simultaneously extracting all hyperon-related parameters and estimating the expected experimental sensitivities. The results are presented in Fig~\ref{fig:sensitivity}~\cite{Fu:2023ose}. The sensitivity for $\Lambda$ EDM reaches $10^{-21}$ $e$ cm (red full triangle), marking a significant 5-orders-of-magnitude enhancement compared to the existing measurement at Fermilab with similar statistics. Additionally, the STCF experiment maintains high sensitivities of $10^{-20}$ $e$ cm for the $\Sigma^+$, $\Xi^-$, and $\Xi^0$ hyperons~\cite{Fu:2023ose,Du:2024jfc}. Hyperon EDM measurements will serve as a crucial milestone and a rigorous test for NP theories like SUSY and the left-right symmetrical model.

At $J/\psi$ one can use hadronic vacuum polarization to increase sensitivity for the EDM effect. Contribution of the two photon processes is negligible.

%%%%%%%%%%%%%%%%%%%%%%%%%%%%%%%%%%%%%%%%%%%%%%%%%%%%%%%%%%%%%%%%%%%%%%

\subsection{Prospect of hyperon $\CP$ violation study at STCF}

The advantage of studying hyperon $\CP$ violation at Tau-Charm factory is obvious. The polarization of hyperons can be disentangled, and the measured precision of the hyperon $\CP$ violation can match that of fixed target experiment with three orders of magnitude less statistics. The $\CP$ violation test of all the SU(3) hyperons can be performed with pair production from $J/\psi$ decay. Thus more $\CP$ violation observables can be constructed.
Since 2019, the BESIII experiment has shown its great potential in the search of hyperon $\CP$ violation via
	the decay of 10 billion $J/\psi$ events~\cite{BESIII:2018cnd}. 
With the large branching fraction of $J/\psi\to B\bar{B}$ with $B\bar{B}$ denoting hyperon pairs, ten million of hyperons and
anti-hyperons can be obtained at BESIII. Moreover, the hyperons are naturally polarized transversely 
due to the non-zero phase of two helicity amplitudes
\begin{equation}
    P_{y}=-\frac{\sqrt{1-\alpha_\psi^{2}}\sin\theta\cos\theta}{1+\alpha_{\psi}\cos^{2}\theta}\sin\Delta\Phi_{\psi},
   \label{equation_py}
\end{equation}
where $\alpha_{\psi}$ and $\Delta\Phi$ are the parameters that describe $J/\psi\to B\bar{B}$.
%Table~\ref{tab:CPhyperon} summarizes the current related measurements of hyperon pair production from $J/\psi$ decay. 
Moreover, by the cascade decays, the weak phase can be determined for the first time~\cite{BESIII:2021ypr}. It is quite promising to discover the
hyperon $\CP$ violation with more statistics at STCF. 
With the current designed energy spread, more than 3.4 trillion $J/\psi$ events will be collected at STCF with $1$~ab$^{-1}$ integrated luminosity.  
Following the statistical sensitivity of the $\CP$ violation test will be discussed. 

\begin{comment}    
\begin{table*}[htbp!]
%    \centering
    \caption{The branching fraction, the observable $A_{CP}$, the production parameters $\alpha_{\psi}$ and $\Delta\Phi$, and the maximum polarization $P_{y}^{\rm max}$ of the hyperons produced in $J/\psi$ decay.}
    \small
    \label{tab:CPhyperon}
    \begin{tabular}{c|c|c|c|c|c}
      \hline
      \hline
       $Y\bar{Y}$ mode  &  $\mathcal{B}(\times10^{-3})$ & $\alpha_{\psi}$ & $\Delta\Phi$ & $P_{y}^{\rm max}$ & $A_{CP}$\\
       \hline
       $\Lambda\bar{\Lambda}$  &  $1.89\pm0.09$ & $0.475\pm0.003$ & $0.752\pm0.008$  & $25\%$ & $-0.003\pm0.005$~\cite{BESIII:2022qax}\\
       $\Sigma^{+}\bar{\Sigma}^{-}$ & $1.07\pm0.04$ & $-0.508\pm0.007$ & $-0.27\pm0.02$  & $16\%$ & $-0.004\pm0.038$~\cite{BESIII:2020fqg}\\
       $\Xi^{0}\bar{\Xi}^{0}$  &  $1.17\pm0.04$   &  $0.66\pm0.06$ & $1.16\pm0.02$ & $27\%$ & $-0.005\pm0.007$~\cite{BESIII:2023drj}\\
       $\Xi^{-}\bar{\Xi}^{+}$   & $0.97\pm0.08$   &  $0.59\pm0.02$ & $1.21\pm0.05$ & $30\%$ & $0.006\pm0.014$~\cite{BESIII:2021ypr}\\
    \hline
    \hline
    \end{tabular}
\end{table*}
\end{comment}

%\subsubsection{Event selection}
Charged tracks are selected based on the criteria in the fast simulation. Efficiency loss occurs due to the acceptance requirement $|\cos\theta|<0.93$, where $\theta$ is set in reference to the beam direction, and the requirements on the $\Lambda$ mass and decay vertex.
The decay process of $J/\psi \rightarrow \Lambda\bar{\Lambda}$ can be described as follows: the $\Lambda$ decays into a $p$ and $\pi^{-}$, while the $\bar{\Lambda}$ decays into a $\bar{p}$ and $\pi^{+}$. Therefore, the candidate event must have at least four charged tracks.
Charged tracks are divided into two categories, where positively charged tracks are $p$ and $\pi^{+}$, and negatively charged tracks are $\bar{p}$ and $\pi^{-}$.
Based on the momentum of tracks, the momentum of the particle track can subsequently be identified as $p (\bar{p})$ or $\pi^{+} (\pi^{-})$. Specifically, particles with a momentum exceeding 500~MeV/$c$ are identified as $p (\bar{p})$, while those with a momentum less than 500~MeV/$c$ are classified as $\pi^{+} (\pi^{-})$. For selected events, multiple charged tracks must be present for $p$, $\bar{p}$, $\pi^{+}$ and $\pi^{-}$, respectively.
A second vertex fit is performed by looping over all combinations of positively and
negatively charged tracks. The selected $p\pi^{-}~(\bar{p}\pi^{+})$ pairs must decay from the same vertex. The invariant mass of $\Lambda$($\bar{\Lambda}$) must fall within the range of $1.111<M_{p \pi^{-}/\bar{p} \pi^{+}}<1.121$~GeV/$c^{2}$. The event selection efficiency is finally determined to be 38.2\%.   
The influence of the polarization of the electron beam on event selection efficiency has also been studied, and the selection efficiency is found to be unaffected by beam polarization.

%\subsubsection{Sensitivity to $\CP$ violation in hyperon decay}
Based on the joint angular distribution, a maximum likelihood fit is performed with four free parameters ($\alpha_{\psi}$, $\alpha_{-}$, $\alpha_{+}$,  $\Delta\Phi_{\psi}$). The joint likelihood function, as shown in Eq.~(\ref{eq9}) , is used for this purpose.
\begin{footnotesize}
	\begin{align}
		\mathcal{L}=\prod\limits_{i=1}^{N}\mathcal{P}(\xi^{i}, \alpha_{\psi}, \alpha_{-}, \alpha_{+}, \Delta\Phi)=\prod\limits_{i=1}^{N}\mathcal{C}\mathcal{W}(\xi^{i}, \alpha_{\psi},\alpha_{-}, \alpha_{+}, \Delta\Phi)\epsilon(\xi^{i}),
		\label{eq9}
	\end{align}
\end{footnotesize}\\
The probability density function of the kinematic variable~$\xi^{i}$~for event $i$ denoted as $\mathcal{P}(\xi^{i},~\alpha_{\psi},~\alpha_{-},~\alpha_{+},~\Delta\Phi)$ is used in a maximum likelihood fit. The detection efficiency is represented by $\epsilon(\xi^{i})$ and $N$ denotes the total number of events. The normalization factor, denoted as $\mathcal{C}^{-1} =\frac{1}{N_{MC}}\sum\limits^{N_{MC}}_{j=1}\mathcal{W}(\xi^{j},~\alpha_{\psi},~\alpha_{-}, ~\alpha_{+},~\Delta\Phi)\epsilon(\xi^{j})$, is estimated the $N_{\rm MC}$ events generated with the phase space model, which is about ten times the size of mDIY MC. Usually, the minimization of $-{\rm ln}\mathcal{L}$ is performed by using {\sc MINUIT}:
\begin{align}
	-{\rm ln}\mathcal{L}=-\sum\limits^{N}_{i=1}{\rm ln}\mathcal{C}\mathcal{W}(\xi^{i},\alpha_{\psi},\alpha_{-},\alpha_{+},\Delta\Phi)\epsilon(\xi^{i})
	\label{eq10}.
\end{align} 
In this analysis, we extrapolate the sensitivity of $\CP$ violation for a large number of $J/\psi$ events generated at future STCF, taking into account the effects of excessive storage pressure. This extrapolation is based on the relationship between the sensitivity of 
$\CP$ violation and generated 0.1 trillion $J/\psi$ events. We investigate this relation using a sample size ranging from 0.01 to 0.1 trillion 
$J/\psi$, with a step size of 0.01 trillion 
$J/\psi$. The sensitivity analysis is presented in Fig.~\ref{fig:hyperoncp}(a), where we examine the impact of event statistics on the sensitivity of 
$\CP$ violation. The sensitivity can be described using the following formula: 
\begin{equation}
	\sigma_{A_{CP}}\times\sqrt{N_{\rm fin}}=k.
	\label{ex}
\end{equation}
The variable $N_{\rm fin}$ represents the number of events that pass the final selection criteria, while $k$ is a constant with a value of 7.82.
As shown in Fig.~\ref{fig:hyperoncp}(b), the sensitivity improves proportionally with the polarization. This observation provides a foundation for extrapolating 
$\CP$ violation sensitivity from the size of the data sample.
\par Five different beam polarizations were utilized to generate a sample of 0.1 trillion MC events, with the specific aim of investigating the quantity 
$\sigma_{A_{CP}}$. The resulting five sets of data points were fit using Eq.~(\ref{eq11}) \cite{Salone:2022lpt}: 
\begin{equation}
	\sigma_{A_{CP}}\approx \sqrt{\frac{3}{2}} \frac{1}{\alpha_{-}\sqrt{N_{sig}}\sqrt{\left \langle P_{\Lambda}^{2} \right \rangle}}.
	\label{eq11}
\end{equation}

Under the same sample size, the obtained results align with those in Ref.~\cite{Bondar:2019zgm}, maintaining consistency at the order of magnitude level.
By extrapolating the number of $J/\psi$ events based on the 3.4 trillion events expected to be generated annually by future STCF, the statistical sensitivity of 
$\CP$ violation will reach the order of $\mathcal O$ $(10^{-4})$,
and a beam polarization of 80\% will further improve the sensitivity by
a factor of 3.
%at a beam polarization of 80\%.

\begin{figure}[htbp!]
\begin{center}
\begin{overpic}[width=6cm,angle=0]{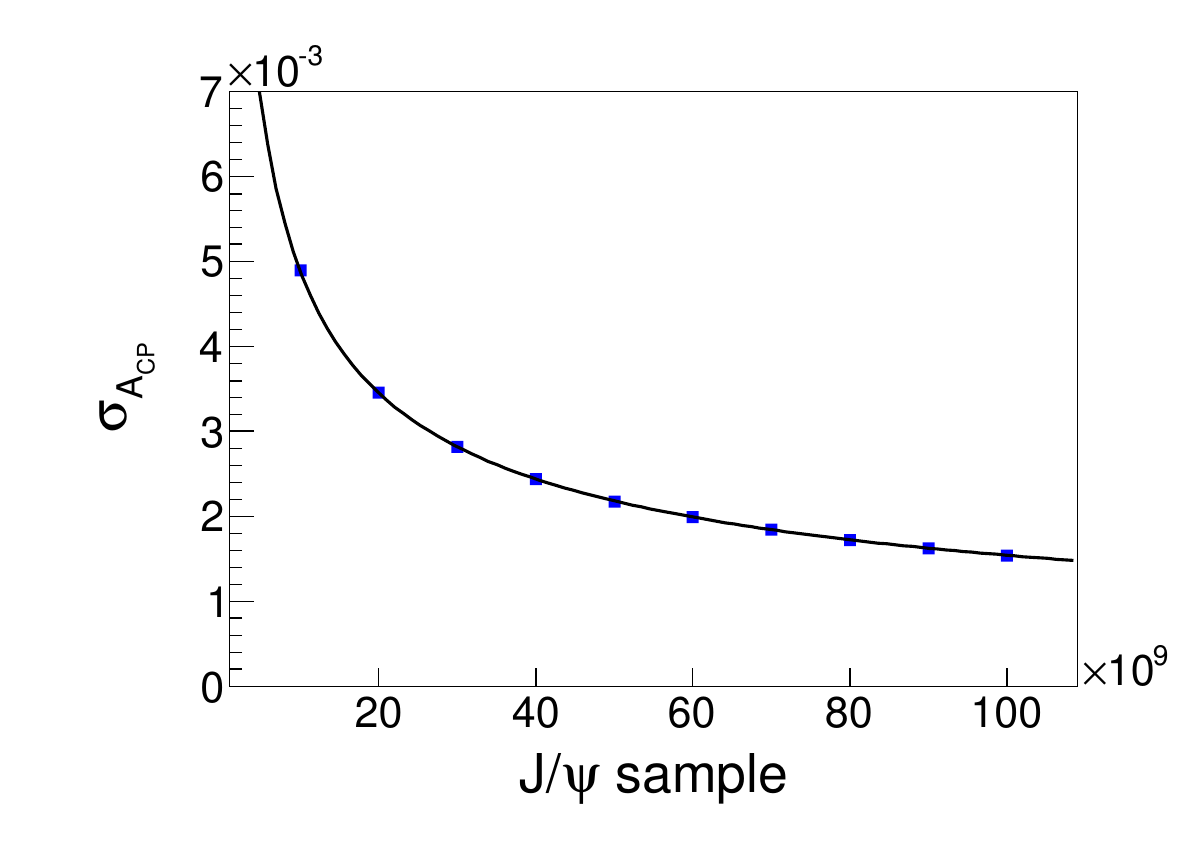}
\end{overpic}
\begin{overpic}[width=6cm,angle=0]{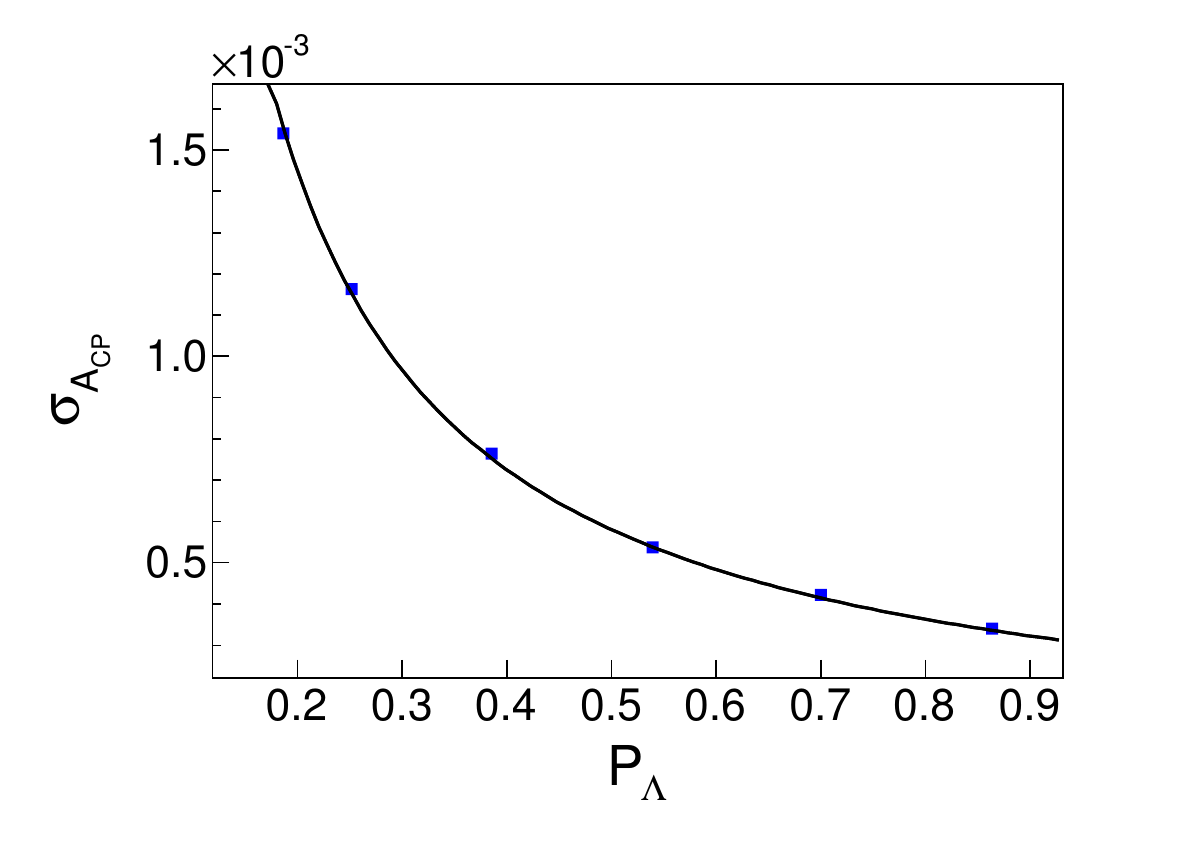}
\end{overpic}
\end{center}
\vspace{-0.5cm}
\caption{The blue dots represent the statistical errors of $\CP$ violation, while the black line is fitted using Eq.~(\ref{ex}).
The blue dots represent the values of $\sigma_{A_{CP}}$ under different $\Lambda$ polarization.}
\label{fig:hyperoncp}
\end{figure}

%\subsection{Comparison of hyperon $\CP$ sensitivity among different experiments}
A comparison of the hyperon CP sensitivity at STCF with various experiments, namely the LHCb and BelleII experiments, is presented as follows.
\begin{itemize}
    \item \textbf{Production mechanism of hyperons} \\
    The LHCb experiment produces hyperons mainly via promote production, and the decay of $B$ meson or charmed/bottom baryons. 
With the data collected in run2, about 13 million of $\Xi^{-}$ and 5 million of $\Omega^{-}$ hyperons have been reconstructed inclusively. 
The Hyperons are not polarized from promote production, and
may have polarization from decays, $\eg$ from charmed baryon. 
It is therefore essential to distinguish the hyperons from promote production or from decay. 
With the run3 data, five times more hyperons can be produced than that of run2. 
As for run4 and run5, the integrated luminosities of 40~fb$^{-1}$ and 300~fb$^{-1}$ are expected, respectively.

The Belle and Belle~II experiments produce hyperons depending on the running strategy, mainly from decays of $B$ meson and charmed baryons (and also $\Upsilon(nS)$~($n=1,\,2,\,3$) at Belle), and $e^{+}e^{-}\to q\bar{q}$ ($q=u,\,d,\,s$). 
The cross sections of $e^{+}e^{-}\to b\bar{b}$ and $c\bar{c}$ are about 1.1~nb and 1.3~nb, respectively. 
Using the $\Lambda_{c}^+\to\Lambda\pi^+$ decay where the $\Lambda_{c}^+$ is produced from $e^{+}e^{-}\to c\bar{c}$ process with 1~ab$^{-1}$ integrated luminoistiy at Belle experiment, more than $10^{5}$ $\Lambda$ events are reconstructed with the precision of decay asymmetry parameters 
determined to be at the level of $\mathcal{O}(10^{-3})$. 
The $\Lambda_{c}$ is supposed to be not polarized via the process 
$e^{+}e^{-}\to c\bar{c}$ and the final $\Lambda$ is polarized due to the weak decay of $\Lambda_{c}$. 
With the inclusive process $e^{+}e^{-}\to [\Xi^{-}\to\Lambda\pi^-]+X$, where $X$ denotes anything, Belle (980~fb$^{-1}$) has reconstructed over $10^{6}$ $\Xi^{-}$ hyperons with high purity, and may reach a precision of $\mathcal{O}(10^{-3})$ level for $\alpha_{\Xi}\alpha_{\Lambda}$. With full Belle II data set (50~ab${}^{-1}$), this $\Xi^{-}$ inclusive sample will be exceed the $5\times10^7$ level. 

The STCF produces hyperons mainly from their pair production.
With 3.4 trillions of $J/\psi$ events foreseen and the branching fraction of $J/\psi\to B\bar{B}$ to be $\mathcal{O}(10^{-3})$,
a rough estimation of hyperons pair produced via $J/\psi$, as well as the maximum polarization of hyperons, are presented in Table~\ref{tab:stcfhyperon}.

\begin{table}[htbp!]
    \centering
        \caption{Expected number of hyperon pairs at STCF, where the Br is cited from Ref.~\cite{ParticleDataGroup:2022pth}. $\epsilon$ is the reconstruction efficiency of the exclusive decay from BESIII. $N_{rec}$ is the expected reconstructed number of hyperon events at STCF, with similar efficiency as BESIII, and $P_{y}^{\rm max}$ is the maximum transverse polarization of hyperon in $J/\psi$ decay.}
    \begin{tabular}{c|c|c|c|c}
    \hline\hline
       Sample  &   Br($\times10^{-3})$  & $\epsilon$ & $N_{rec}$ & $P_{y}^{\rm max}$\\
       \hline
       $J/\psi\to\Lambda\bar{\Lambda}$  &   $1.89\pm0.09$ & 40\%  &  $1\times10^{9}$  & 25\%\\
       $J/\psi\to\Sigma^{+}\bar{\Sigma}^{-}$ & $1.07\pm0.04$ & 24\% &  $2.2\times10^{8}$  & 16\%\\
       $J/\psi\to\Sigma^{0}\bar{\Sigma}^{0}$ & $1.172\pm0.032$ &  20\% & $3.3\times10^{8}$  &  \\
       $J/\psi\to\Sigma^{-}\bar{\Sigma}^{+}$ & $1.51\pm0.06$ &  6\%  & $3.0\times10^{8}$ & -\\
       $J/\psi\to\Xi^{0}\bar{\Xi}^{0}$ & $1.17\pm0.04$ & 14\% & $2.3\times10^{8}$ & 27\%\\
       $J/\psi\to\Xi^{-}\bar{\Xi}^{-}$ & $0.97\pm0.08$ &  19\% & $2.5\times10^{8}$ & 30\%\\
    \hline\hline
    \end{tabular}
    \label{tab:stcfhyperon}
\end{table}

    \item \textbf{Expected statistical sensitivity} \\
From the reconstructed number of hyperon samples at LHCb with data collected in run2, the $\CP$ observable $A_{\Xi\Lambda}=\frac{\alpha_{\Xi}\alpha_{\Lambda}+\alpha_{\bar{\Xi}}\alpha_{\bar{\Lambda}}}{\alpha_{\Xi}\alpha_{\Lambda}+\alpha_{\bar{\Xi}}\alpha_{\bar{\Lambda}}}$ 
can be constructed, where $\alpha_{\Xi}$ and $\alpha_{\Lambda}$ is the decay asymmetry parameter of $\Xi^{-}$ and $\Lambda$, respectively. The sensitivity is in a level of $\mathcal{O}(10^{-3})$. 
The expected statistical sensitivity is better than $\mathcal{O}(3\times10^{-4})$ with 50~fb$^{-1}$ integrated luminoisty reached in till run4.

At Belle and Belle II, with $\Lambda_c^+\to\Lambda\pi^+$ inclusive sample, the $\alpha$-induced $\CP$ violation observable of $\Lambda\to{}p\pi^-$ decay is constructed to be 
$A_{\Lambda}=\frac{\alpha_{\Lambda}+\alpha_{\bar{\Lambda}}}{\alpha_{\Lambda}-\alpha_{\bar{\Lambda}}}$ under the assumption of $\CP$ conservation in Cabibbo-favored decay $\Lambda_c^+\to\Lambda\pi^+$ (i.e $\alpha_{\Lambda_c^+}=-\alpha_{\bar{\Lambda}_c^-}$).
Based on 980~fb$^{-1}$ of data set at Belle, the statistical uncertainty of $A_{\Lambda}$ is determined to be $7\times10^{-3}$~\cite{Belle:2022uod}.
With 50~ab$^{-1}$ integrated luminosity collected at Belle II foreseen, the 
statistical sensitivity of $A_{\Lambda}$ is estimated to be $10^{-3}$.

At STCF, $\CP$ observables of $\Lambda$, $\Sigma^{+}$, $\Xi^{0}$ and $\Xi^{-}$ can be constructed via 
$A_{Y}=\frac{\alpha_{Y}+\alpha_{\bar{Y}}}{\alpha_{Y}-\alpha_{\bar{Y}}}$, where $\alpha_{Y}$ is the decay asymmetry parameter of the hyperon. 
The sensitivity follows the empirical formula  as in Eq.~\ref{eq11}.
%\begin{equation}
%    \sigma_{A_{Y}}\approx \sqrt{\frac{3}{2}}\frac{1}{\alpha_{Y}\sqrt{N_{\rm rec}}\sqrt{<P^{2}_{Y}>}},
%\end{equation}
%where $N_{rec}$ is the expected number of signal events after reconstruction, as expressed in Table~\ref{tab:stcfhyperon}, and 
%The second moment of the polarization  for the hyperons $<P^{2}_{Y}>$ is the second moment of the polarization for the hyperons.
With unpolarized electron-positron beam, the maximum polarization of hyperons can be found in Table~\ref{tab:stcfhyperon}. The average polarization of $\Lambda$, $\Sigma^{+}$ and $\Xi^{-}$ with unpolarized beam is 18\%, 10\%, and 22\%, respectively~\cite{Salone:2022lpt}. 
When the electron beam is 80\% polarized, the average polarization of $\Lambda$, $\Sigma^{+}$ and $\Xi^{-}$ is 70\%, 70\% and 60\%, respectively. The corresponding 
sensitivity of $A_{Y}$ is summarized in Table~\ref{tab:stcfcp}.
\begin{table}[htbp!]
    \centering
        \caption{Expected statistical sensitivity of $A_{Y}$ with different polarization at STCF.}
    \small
    \begin{tabular}{c|c|cc|cc}
    \hline\hline
      Hyperon  &   $\alpha_{Y}$   &  $<P^{2}_{Y}>_{(P_e=0)}$  & $\sigma_{A_{Y}(P_e=0)}$ &  $<P^{2}_{Y}>_{(P_e=80\%)}$ & $\sigma_{A_{Y}(P_e=80\%)}$  \\
    \hline
      $\Lambda$  &   0.75  &  18\%   &  $3\times10^{-4}$  &  70\%  & $7\times10^{-5}$ \\
      $\Sigma^{+}$  & -0.98&  10\%   &   $8\times10^{-4}$ &  70\%  & $1\times10^{-4}$ \\ 
      $\Xi^{-}$  &  -0.376 &  22\%   &   $9\times10^{-4}$ &  60\%  & $3\times10^{-4}$\\
    \hline\hline
    \end{tabular}
    \label{tab:stcfcp}
\end{table}

    \item \textbf{Systematic uncertainty}
The systematic uncertainty from LHC hyperon $\CP$ test mainly comes from
the asymmetry of detector, and the uncertainty from the simulation model. 
Several assumptions should be made under the law that Cabibbo allowed decay produces unpolarized 
hyperons, while Cabibbo suppressed decay produces polarized hyperons. The systematic uncertainty will be a dominant limiting factor of the results. 

The systematic uncertainty from Belle $\Lambda$ hyperon $\CP$ test via $\Lambda_{c}^+$ Cabibbo-favored decay is determined to be $1.1\times10^{-2}$, that is larger than statistical
uncertainty, and dominated by the source of the fit bias. Other systematic sources contribute a systematic uncertainty of $3\times10^{-3}$. Naively, the uncertainty from fit bias can be likely to improve with a better fit procedure, therefore, the final sensitivity of such $A_{\Lambda}$ measurement at Belle II is promisingly estimated to be at the $10^{-3}$ level.

At STCF, the uncertainty sources of hyperon $\CP$ test is
the selection criteria related, those including tracking, vertex fitting and
kinematic fit; the background contamination and the fit method. 
These are statistically related, and are estimated to be in a level 
of $10^{-4}\sim 10^{-5}$. 
Apart from that, the effects from regeneration, decoherence, or non-uniformity
of magnetic field will play an important role in the
systematic uncertainty, thus demands detailed detector design and simulation in future.     
\end{itemize}

Overall, the STCF offers great potential to search of new $CP$ violation source via through direct measurements of $CP$ violation and EDM measurements in hyperon sector. 
It is expected to produce billions of quantum-correlated hyperon-antihyperon with almost background free. Numerous $CP$ observables can be constructed. From the feasibility  study, the $CP$ violation in hyperon non-leptonic decays
will reach $\mathcal{O}(10^{-4})$ and the EDM of hyperons will reach
$\mathcal{O}(10^{-20})\sim\mathcal{O}(10^{-21})$ $e$ cm. 

\newpage

\section{$\CP$ violation in $\tau$ sector}
\label{sec:tau}
%\section{$CP$-violation in $\tau$ decays}

As the only lepton that is massive enough to open the hadronic decay channels, the SL $\tau$ decay provides an excellent platform  to explore the nonperturbative QCD dynamics and is one of the leading research subjects at the electron-positron colliders. In the meanwhile, the rich patterns of the $\tau$ lepton decaying into different hadrons offer valuable opportunities to study the $\CP$ violation phenomena. In this section, we first give a brief review about the current status of the various hadron form factors and structure functions entering the SL $\tau$ decays. Then we elaborate on different types of $\CP$ violation observables in the hadronic final states in $\tau$ decays, and pay special attention to the $\CP$ violation study in the $\tau\to K_S^0 \pi\nu_\tau$ channel. The $\tau$ EDM is shortly discussed as well. At the end of this section, we give some prospects for the $\CP$ violation measurements of the $\tau$ lepton at STCF. 

\subsection{Hadronic form factors in semileptonic $\tau$ decays}
\label{tau:ff}
In the SM, hadronization in two-meson %\cite{Gasser:1984ux} 
tau decays is described by two form factors, $f_{0,+}(Q^2)$, carrying angular momenta $J=0,1$, respectively ($D=d,s$ below)
\begin{equation}
\small
 \label{eq:two-mesonmatrixelement}
\begin{aligned}
& \left\langle P_1(q_1)P_2(q_2)|\bar{u}\gamma^\mu D+h.c.|0\right\rangle  \\
& =C^{12}\left\lbrace\frac{\Delta_{12}}{Q^2}f_0 ^{12}(Q^2)(q_1+q_2)^\mu  +f_+^{12}(Q^2)\left[(q_1-q_2)^\mu-\frac{\Delta_{12}}{Q^2}(q_1+q_2)^\mu\right]\right\rbrace ,
\end{aligned}
\end{equation}
where $C^{\pi^-\pi^0}=\sqrt{2}$, $Q^2=(q_1+q_2)^2$, $\Delta_{12}=m_1^2-m_2^2$, 
and the finiteness of the matrix element at the origin imposes $f_0(0)=f_+(0)$, which depart from unity only by flavor symmetry breaking corrections.

Using the angle $\alpha$, defined by the momenta of $P_1$ and $\tau$ in the hadronic rest frame, the double differential decay rate reads \cite{Kuhn:1992nz}
\begin{eqnarray}
\label{eq:two-mesondoubledifferentialdecaywidth}
\frac{\mathrm{d}\Gamma(\tau^-\to (P_1 P_2)^-\nu_\tau)}{\mathrm{d\,cos}\alpha\,\mathrm{d}\sqrt{Q^2}}=\frac{(C^{12}G_F|V_{uD}|)^2}{128\pi^3}\left(\frac{M_\tau^2}{Q^2}-1\right)^2\frac{\lambda^{1/2}(Q^2,M_1^2,M_2^2)}{2\sqrt{Q^2}M_\tau}\nonumber\\
\Bigg\lbrace|f_0^{12}(Q^2)|^2\Delta_{12}^2+|f_+^{12}(Q^2)|^2\lambda(Q^2,M_1^2,M_2^2)\left(\frac{Q^2}{M_\tau^2}+\left(1-\frac{Q^2}{M_\tau^2}\right)\mathrm{cos}^2\alpha\right)\nonumber\\
-2\mathrm{Re}[f_0^{12}(Q^2)f_+^{12}(Q^2)^*]\Delta_{12}\lambda^{1/2}(Q^2,M_1^2,M_2^2)\mathrm{cos}\alpha\Bigg\rbrace\,,
\end{eqnarray}
where $\lambda(a,b,c)=a^2+b^2+c^2-2(ab+ac+bc)$.

In the three-meson case, the most general decomposition of the matrix element encoding the  hadronization of the left-handed $W$-mediated current in presence of strong interactions is \cite{Kuhn:1992nz}
\begin{equation}
\label{eq:threemesonmatrixelement}
\left\langle P_1(q_1) P_2(q_2) P_3(q_3)|\bar{u}\gamma^\mu(1-\gamma_5)D+h.c.|0\right\rangle=V_1^\mu F_1+V_2^\mu F_2+V_3^\mu F_3+Q^\mu F_4\,,
\end{equation}
where $Q^\mu=(q_1+q_2+q_3)^\mu$, $V_{1\,\mu}=\left(g_{\mu\nu}-\frac{Q_\mu Q_\nu}{Q^2}\right)(q_2-q_1)^\nu$, $V_{2\,\mu}=\left(g_{\mu\nu}-\frac{Q_\mu Q_\nu}{Q^2}\right)(q_3-q_1)^\nu$, $V_{3\,\mu}=i\epsilon_{\mu\nu\rho\sigma}q_1^\nu q_2^\rho q_3^\sigma$. The $F_{1,2,4}$ form factors carry axial-vector quantum numbers and $F_3$ stems from the vector current (anomalous sector).

The differential decay width contains information on four angles and is most conveniently written in terms of structure functions \cite{Kuhn:1992nz}, which are bilinears of form factors depending also on kinematical invariant variables. More information on this formalism is given in the next subsection.

Beyond the Standard Model, the two-meson matrix elements can also correspond to a scalar or a tensor current \cite{Cirigliano:2009wk}. The form factor associated to the hadronization of the scalar current is related to $f_0(Q^2)$ via the divergence of the vector current hadron matrix element \cite{Garces:2017jpz}, while the tensor current form factor can be related to $f_+(s)$ in the elastic region \cite{Cirigliano:2017tqn}. For three mesons, new form factors from hadronizing possible pseudoscalar and tensor currents appear  \cite{Arteaga:2022xxy}. The former is proportional to $F_4$, while for the latter simplifications to the involved three-body unitary treatment  arise thanks to (approximate and useful) discrete symmetries of QCD.

Dispersive descriptions of the $f_{0,+}(Q^2)$ form factors in Eq.~(\ref{eq:two-mesonmatrixelement}) exploit their analytic properties and are best suited to comply with unitarity, according to
\begin{equation}
 f_{0,+}^{12}(Q^2)\,=\,P_N(Q^2)\Omega_{0,+}^{12}(Q^2)\,,  
\end{equation}
where the Omn\`es function %\cite{Omnes:1958hv}
is 
\begin{equation}
    \Omega_{0,+}^{12}(Q^2)\,=\,\mathrm{exp}\left[\frac{Q^2}{\pi}\int_{s'_0}^\infty\mathrm{d}s'\frac{\phi^{12}_{0,+}(s')}{s'(s'-Q^2)}\right]\,.
\end{equation}
In this way, knowledge of the form factor phase ($\tan\phi=$Im$\phi/$Re$\phi$) everywhere encodes this whole complex function. With a perfect description of the phase, the degree $N$ of the polynomial $P_N(Q^2)$ is immaterial. Increasing $N$ (which brings in additional subtraction constants to $f_{0,+}(0)$) reduces the weight of the high-energy region of the integrand, where data are worse. $N$ is chosen to be as small as possible, while keeping a close description of the measurements. In addition to experimental input, theory also restricts $f_{0,+}(Q^2)$. Asymptotically, they must fall off as $1/Q^2$~\cite{%Brodsky:1973kr,
Lepage:1980fj}, with their phase being an integer multiple of $\pi$ at infinite $Q^2$%~\cite{Leutwyler:2002hm}
. Watson's Theorem %\cite{Watson:1952ji}
ensures the equality of the form factors phase with the corresponding phase shift of di-meson scattering in the elastic region. Close enough to threshold, the chiral symmetry of QCD \cite{%Weinberg:1978kz,
Gasser:1983yg,Gasser:1984gg} predicts the low-$Q^2$ behaviours of the form factors. The intermediate energy region phase depends on the properties  of the various resonance states (particularly their pole positions and relative phases) that can be exchanged in the process.

The best understood form factor is  $f_{+}^{\pi\pi}(Q^2)$. Roy  equations %\cite{Roy:1971tc} 
constrain stringently the relevant (iso)vector phaseshift in the elastic region \cite{Ananthanarayan:2000ht,Garcia-Martin:2011iqs} for $Q^2\lesssim 1$ GeV$^2$, allowing for dispersive representations of $f_+(Q^2)$%\cite{Pich:2001pj}
, which are phenomenologically successful up to $M_\tau^2$ \cite{GomezDumm:2013sib,Gonzalez-Solis:2019iod}. %Inelastic effects have been the focus of refs.~\cite{Colangelo:2018mtw,Colangelo:2020lcg,Colangelo:2022prz} in the context of the leading contribution to the hadronic vacuum polarization piece of the muon $g-2$ \cite{Aoyama:2020ynm}. 
See also Refs.~\cite{Zhou:2004ms,Dai:2011bs,Guo:2011pa,Guo:2012yt,Pelaez:2019eqa,Yao:2020bxx} for discussing additional scattering channels.
A similar approach is pursued for $f_+^{K\pi}$. In this instance, %the corresponding scattering was analyzed in refs. \cite{Buettiker:2003pp,Pelaez:2016tgi,Pelaez:2020gnd} based on Roy-Steiner eqs. and 
the dispersive form factor has been obtained in  Refs. \cite{Moussallam:2007qc,Boito:2008fq, Boito:2010me,Bernard:2013jxa,%Antonelli:2013usa,
Escribano:2014joa}.
Dispersive vector form factors for partner decay modes $f_+^{K\bar{K}(K\eta^{(\prime)})}$ have been determined based on $f_+^{\pi\pi(K\pi)}$ in Refs.~\cite{Gonzalez-Solis:2019iod}(\cite{Escribano:2014joa,Escribano:2013bca}). A thorough analysis of the electromagnetic kaon form factor was recently performed in Ref. \cite{Stamen:2022uqh}. The $\tau^-\to\pi^-\eta^{(\prime)}\nu_\tau$ decays are more subtle, due to their strong $G-$parity suppression \cite{Leroy:1977pq,Pich:1987qq} and Refs.~\cite{Descotes-Genon:2014tla,Escribano:2016ntp} differ substantially in their predictions for $f_+^{\pi^-\eta^{(\prime)}}$.

Scalar and tensor two-meson form factors were computed in Ref.~\cite{Shi:2020rkz}. Specifically, $f_0^{\pi\pi}$ (which determines $f_0^{\pi^-\eta^{(\prime)}}$~\cite{ Garces:2017jpz, Descotes-Genon:2014tla}) was addressed in Refs.~\cite{Guo:2012yt,Descotes-Genon:2014tla,Doring:2013wka,Ropertz:2018stk}, although the $K\pi$ sector is better known thanks to its relevance in $K\pi$ scattering~\cite{Bernard:1990kw,Jamin:2000wn,Zheng:2003rw}, which enables a precise determination of the coupled $f_0^{K^-(\pi^0/\eta/\eta^\prime)}$ form factors~\cite{Jamin:2001zq}. 
Inelastic effects on $f_0^{K\pi}$ were the focus of Ref. \cite{VonDetten:2021rax}. Finally, $f_0^{K\bar{K}}$ was derived in Ref.~\cite{Doring:2013wka}. Tensor form factors in Refs.~\cite{Shi:2020rkz,Miranda:2018cpf,Rendon:2019awg,Gonzalez-Solis:2019lze} are also subject to chiral dynamics%, according to ref. \cite{Cata:2007ns}
.

For the three-meson final states, fully unitary dispersive determinations of the form factors are not yet available. There are, however, sound approximations implementing two-body dominance \cite{Moussallam:2007qc} (for $K\pi\pi$), and chiral constraints beyond leading order: \cite{Colangelo:1996hs,GomezDumm:2003ku,
Dumm:2009va,Shekhovtsova:2012ra,Nugent:2013hxa} ($\pi\pi\pi$),  \cite{%Gomez-Cadenas:1990vkb,
Dumm:2009kj} ($KK\pi$), \cite{GomezDumm:2012dpx} ($\eta\pi\pi$), \cite{Guo:2008sh} (vector meson and a light  pseudoscalar),  \cite{Ecker:2002cw} ($4\pi$).

One- and two-meson $\tau$ decays can bring in additional useful information (particularly regarding $\CP$ or $T$ violation) when including a real photon, or a virtual pair-converting one. The corresponding form factors and radiative corrections have been determined, according to the known chiral behaviour, in Refs.~\cite{%Cirigliano:2001er,
Cirigliano:2002pv,
Guo:2010dv,%Guevara:2013wwa,Guevara:2016trs,
Miranda:2020wdg,%GutierrezSantiago:2020bhy,
Guevara:2021tpy,
Arroyo-Urena:2021nil,Arroyo-Urena:2021dfe,Chen:2022nxm,Escribano:2023seb}.

%Form factors understanding is indebted to the diverse spectra measurements in various semileptonic $\tau$ decays of refs.~\cite{ALEPH:2005qgp,ALEPH:1996kok,ALEPH:1997fek,ALEPH:1997jrw,ALEPH:1998rgl,ALEPH:1999uux,ALEPH:2005qgp,BaBar:2005ass,BaBar:2008wlm,BaBar:2010bul,BaBar:2018qry,Belle:2006enh,Belle:2007goc,Belle:2008jjb,Belle:2008xpe,CLEO:1992nqz,CLEO:1995obd,CLEO:1994ljm,CLEO:1996rit,CLEO:1996rmd,CLEO:1997fpa,CLEO:1997kjq,CLEO:1998hhw,CLEO:1999dln,CLEO:1999heg,CLEO:1999maq,CLEO:1999rzk,CLEO:2000nrp,CLEO:2000trg,CLEO:2004hrb,CLEO:2005qyl,DELPHI:1998bhv,L3:1995cos,OPAL:1998rrm,OPAL:1999brn,OPAL:2004icu} (extending to multi-meson channels beyond what sketched here).

%\textcolor{teal}{Add something about lattice QCD evaluations of form factors? (Zhi-Hui Guo)}
The timelike form factors of the light pseudoscalar mesons have also been determined recently by the lattice QCD simulations. Most of the existing lattice calculations focus on the $\pi\pi$ electromagnetic form factors~\cite{Meyer:2011um,Feng:2014gba,Andersen:2018mau,Baroni:2018iau}, which could be related to the charged vector $\pi\pi$ form factor entering in the $\tau$ decay via the isospin transformation. However, the lattice simulations are still done with unphysically large pion masses, and hence chiral extrapolations are generally needed to provide direct inputs for the study of $\tau$ decays. Future developments in the lattice simulations to probe the electroweak timelike form factors involving kaons and multi-pions can provide alternative inputs, which will be definitely helpful for the precision studies of hadronic $\tau$ decays.  

\subsection{Structure functions in $\tau$ decays with hadronic final states}\label{sec:SFs}
This formalism was introduced in Ref.~\cite{Kuhn:1992nz} and, barring interactions of tensor type in the lepton and hadron vertices \cite{%Kuhn:1996dv,
Arteaga:2022xxy}, it gives the most general description of these decays in a model-independent way (including $\tau$ polarization).

It is based on the introduction of two reference frames ($S$ and $S'$) and the use of the Euler rotation (with angles $\alpha, \beta$ and $\gamma$) connecting them. $S$ simplifies the description of the hadron tensor and $S'$ of the $\tau$ spin and momentum. $S$ is defined by a plane containing hadron momenta and its normal. $S'$ is again specified by a plane and its normal, with the former containing the hadrons' direction in the lab system and the $\tau$ flight direction seen from the hadronic rest frame (see Fig.~1 in Ref.~\cite{Kuhn:1992nz}). The angle $\alpha$ is unobservable unless the $\tau$ momentum can be measured, which is the case for flavor factories. However, in $\tau$ pair production near threshold at STCF, the lab frame coincides with the COM one, which facilitates the description of $S'$ (see appendix B of Ref.~\cite{Kuhn:1992nz} for explicit formulae in this case).

The matrix element factorizes in its lepton and hadron parts, with the most general description of its modulus squared (factoring the lepton part out) giving rise to the structure functions, $W_i$, which are bilinears of form factors, depending also on kinematical invariant variables. $W_{E,F,H,SC,SE,SF}$ are odd under time reversal and others are sensitive to $\CP$-violation through interferences among form factors (in presence of non-vanishing weak phase differences). These observations make structure functions in SL $\tau$ decays particularly interesting for this section.

Ref.~\cite{Kuhn:1992nz} quotes the relations between the $W_i$ and the $f_i$ ($F_i$) for the two (three) meson cases. In principle, all 16 structure functions (and thus all hadronization information) can be determined in a data-driven way if decays of a polarized $\tau$ are analyzed in its rest frame \cite{Kuhn:1992nz}, which should be possible at a STCF running close to threshold production.

\subsection{$\CP$ violation observables in hadronic $\tau$ decays} 

Hadronic $\tau$ decays provide rich sources to probe possible $\CP$ violation phenomena, since the multitude of kinematical variables from multi-meson channels can be helpful to construct proper quantities to isolate the $\CP$ violation signals from the $\CP$-even ones. Generally speaking, three different types of observables are studied in literature~\cite{Kuhn:1996dv,Choi:1994ch,Choi:1998yx,Bigi:2005ts,Bigi:2012kz,Delepine:2005tw,Delepine:2018amd,Datta:2006kd,Kiers:2008mv,Cirigliano:2017tqn,Huang:1998wh,Huang:1996jr,Kilian:1994ub,LopezCastro:2009da,Tsai:1994rc,Tsai:1996bu,Wang:2014hba}, which allow to discern different mechanisms for the generation of $\CP$ violation: the integrated/full rate asymmetry, local rate asymmetry (also referred as differential angular distribution asymmetry) and the triple-product asymmetry.

In order to have true $\CP$ violation signals, at least two different amplitudes that can interfere with each other with nonvanishing relative phases are required for a given $\tau$ decay reaction, with the exception of the $K^0-\bar{K}^0$ mixing induced $\CP$ violation in the $\tau^-\to K_S^0 X^- \nu_\tau$ decays, such $X=\pi,K,\pi\pi$, etc., as discussed later. 
The SM amplitude is given by the $W$ exchange, while the other one can be mediated by a hypothetical charged Higgs, a new $W_L'$ coupled to left-handed current~\footnote{The interference of a right-handed $W_R$ and $W$ from SM is proportional to the neutrino mass. Therefore the possible effect from $W_R$ in the $\CP$ violation discussion of $\tau$ decays can be ignored in a first approximation.} and/or a leptoquark  field, among others. 

The hadronic decay amplitudes of the $\tau$ lepton can be conveniently described by numerous structure functions~\cite{Kuhn:1992nz}, which can receive contributions both from SM and BSM. On the one hand, the weak $\CP$-violating phases in the hadronic $\tau$ decays are expected to be much suppressed in the SM, since they can be only generated via higher-order weak interactions, $\eg$ the SM $\CP$ asymmetry in the $\tau^{
\pm}\to K^{\pm}\pi^0 \nu_\tau$ processes turns out to be at the level of $10^{-12}$, that is far below the current experimental sensitivity~\cite{Delepine:2005tw}. On the other hand, this also indicates that any experimental observation of the $\CP$ violation in $\tau$ decays, except the ones caused by $K^0-\bar{K}^0$ mixing, will be a smoking gun for BSM NP. As a result, it is convenient to only introduce the weak $\CP$-violating phase to the BSM part in the $\tau$ decay amplitude. Therefore, for a given $\tau$ decay reaction, one can write its amplitude sketchily as 
\begin{eqnarray}\label{eq.cpvtaut}
 T= \sum_{i} A_i e^{i\phi^{A}_i}  + \sum_{j} B_j e^{i(\delta_j+\phi^{B}_j)}\,,
\end{eqnarray}
where $A_i$ and $B_j$ stand for the magnitudes of amplitudes from SM and BSM contributions respectively, and the phases $\phi^{A}_j$ and $\phi^{B}_j$ denote the $\CP$-even strong phases and the weak $\CP$-odd phases are represented by $\delta_j$ (defined relative to the SM weak phase). In general, it is reasonable to assume $|B_j|<< |A_j|$ in Eq.~(\ref{eq.cpvtaut}), although -due to the symmetry constraints- sometimes specific form factors $A_j$ could be vanishing or suppressed in the SM. 

The nonzero $\CP$ violation observable can be calculated via the decay amplitude (Eq.~(\ref{eq.cpvtaut})) squared, $\ie$, 
\begin{eqnarray}
 |T|^2 =&&  \left|\sum_{i} A_i e^{i\phi^{A}_i} \right|^2  + 2{\rm Re}\bigg[\big(\sum_{i} A_i e^{i\phi^{A}_i}\big)  \big( \sum_{j} B_j e^{i(\delta_j+\phi^{B}_j)}\big)^*\bigg]  \nonumber \\ &&
 +  \left| \sum_{j} B_j e^{i(\delta_j+\phi^{B}_j)} \right|^2\,.
\end{eqnarray}
For the $\CP$ conjugate process, the sign of the $\CP$-odd phases $\delta_j$ should be reversed, $\ie$, by taking $\delta_j \to -\delta_j$, while the rest remain the same. The second term with $A_i B_j$ interference and the third one with $B_i B_j$ interference are responsible for the $\CP$ violation strengths. Furthermore, the final $\CP$ violation observables that can be measured in experiments should be calculated by performing the kinematical phase space integrals for the squared decay amplitude, Eq.~(\ref{eq.cpvtaut}).

To enhance the $\CP$ violation features in the multi-meson $\tau$ decay processes, it is crucial to investigate the various differential distributions of the sub-hadron systems with respect to specific kinematical variables. Many theoretical efforts have been completed to pursue such research goals. 

For the integrated rate asymmetry, one simply performs the full integration of the phase spaces for the $\tau^{\pm}$ processes and take the difference afterwards. A nonzero difference between the partial widths of the $\tau^{\pm}$ would signalize the $\CP$ violation interaction. In many cases, especially when the charged Higgs is responsible for the BSM interaction, it is usually found that only the interference parts between the BSM and (generally suppressed) scalar SM form factors in the $\tau$ decay, contribute to the integrated rate asymmetries, $\eg$, for  the $\tau\to\rho\pi\nu_\tau$, $\tau\to\omega\pi\nu_\tau$, $\tau\to a_1\pi\nu_\tau$ and $\tau^-\to K^-\pi\pi\nu_\tau$ processes~\cite{Datta:2006kd,Kiers:2008mv}, which makes the integrated rate asymmetries less appealing in such cases.

Interestingly, by taking properly weighted kinematical variables, typically the cosine/sine of some specific angles, into the integration of phase-space, it is possible to isolate the substantial interference parts between the dominant form factors from the SM and the term with nonvanishing $\CP$-violating phases, so as to enhance the $\CP$ violation signals as much as possible. Those weighted rate asymmetries are usually proportional to $\sin\delta_i \sin\phi_j$, namely both nonzero relative strong and weak phases are needed for such $\CP$ violation observables~\cite{Kuhn:1996dv,Datta:2006kd,Kiers:2008mv}. ${\it E.g.}$, several weighted asymmetries are proposed for the $\tau^{\pm}\to K^{\pm}\pi\pi\nu_\tau$~\cite{Mileo:2014pda,Kiers:2008mv} and $\tau^{\pm}\to PP'\nu_\tau$~\cite{Kuhn:1996dv} (being $P$ a light-flavor pseudoscalar meson).

Another type of $\CP$ violation observable can be constructed with the $T$-odd kinematical variable, also often referred as triple product $\xi=\vec{v}_1\cdot(\vec{v}_2\times \vec{v}_3)$, being $\vec{v}_i$ the momentum or spin of the involved particles. To take spin to construct the triple product $\xi$ means that the spin of the involved particle has to be measured and it is usually quite a challenge for the experiments~\cite{Tsai:1994rc,Tsai:1996bu}. Meanwhile, when only taking momenta in the construction of $\xi$, one needs to consider at least a four-body decay process. It should be noted that the triple product asymmetries from a single $\tau$ lepton could be also mimicked by the final-state interactions from the SM~\cite{Chen:2022nxm}. Conversely, the differences of the triple product asymmetries from $\tau^{\pm}$ will definitely signalize the genuine $\CP$ violation interactions. A significant difference of the $\CP$ violation signals between the weighted rate asymmetries and triple-product asymmetries is that the former is proportional to $\sin\delta_i \sin\phi_j$, implying nonzero phases from the strong and weak parts, while the latter is proportional to $\sin\delta_i \cos\phi_j$, indicating that the strong phase is not required in such $\CP$ violation signals. Early studies about the triple product asymmetries in the $\tau\to K\pi\pi\nu$ have been given in Ref.~\cite{Kilian:1994ub} and later refined in Refs.~\cite{Kiers:2008mv,Mileo:2014pda}. 

We also point out that in the four- and five-body $\tau$ decays, one can also access $\CP$ violation by $T$-odd observables \cite{Bigi:2012km,Shi:2019vus} without the information of the $\tau$ polarization. The global as well as the local $\CP$ violation could be probed. Another direct $\CP$ violation observable constructed by comparing the forward-backward asymmetry of $\tau^-$ and $\tau^+$, namely, $A_{FB}(\tau^-\to K^-\pi^0\nu_\tau)-\bar A_{FB}(\tau^+\to K^+\pi^0\bar\nu_\tau)$, could be also examined, although it takes a very small value in a two Higgs doublet model \cite{Kimura:2012bwp}. Note that $A_{FB}$ itself is a manifestation of parity violation \cite{Zhang:2020dla,Faustov:2019mqr,Faustov:2022ybm}. These topics are certainly worth to analyze in  future experiments with higher-statistics data samples.
  
\subsection{$\CP$ violation in $\tau\to K_S^0\pi\nu_\tau$ decays: the BaBar anomaly and the Belle measurement}
%BaBar measured the following $CP$ violating asymmetry \cite{BaBar:2011pij}
%\begin{equation}\label{eq.BaBarmeasurement}
%A_{CP}^\tau\Big|_{\mathrm{BaBar}}=\frac{\Gamma(\tau^+\to\pi^+K_S\bar{\nu}_\tau)-\Gamma(\tau^-\to\pi^-K_S\nu_\tau)}{\Gamma(\tau^+\to\pi^+K_S\bar{\nu}_\tau)+\Gamma(\tau^-\to\pi^-K_S\nu_\tau)}=-3.6(2.3)(1.1)\cdot10^{-3}\,,
%\end{equation}
%which, in the SM, is determined by neutral Kaon mixing. Specifically, through \cite{ParticleDataGroup:2022pth}
%\begin{equation}\label{eq.KLasymmetry}
%A_{CP}^{K_L}\Big|_{PDG}=\frac{\Gamma(K_L\to\pi^-\ell^+\nu_\ell)-\Gamma(K_L\to\pi^+\ell^-\bar{\nu}_\ell)}{\Gamma(K_L\to\pi^-\ell^+\nu_\ell)+\Gamma(K_L\to\pi^+\ell^-\bar{\nu}_\ell)}=3.32(6)\cdot10^{-3},
%\end{equation}
%with $A_{CP}^\tau=A_{CP}^{K_L}$, up to small corrections related to the $K_L$ reconstruction at B-factories \cite{Grossman:2011zk}. These move the SM prediction for $A_{CP}^\tau$ to $3.6(1)\cdot10^{-3}$, with a puzzling sign difference with respect to the BaBar measurement, eq.~(\ref{eq.BaBarmeasurement}).
%
%Ref.~\cite{Cirigliano:2017tqn} showed that it is extremely difficult to explain this anomaly with heavy new physics, a conclusion that has been confirmed by subsequent analyses \cite{Rendon:2019awg, Chen:2019vbr,Chen:2020uxi,Chen:2021udz}.

In the SM, there is no genuine observable $\CP$ violation signal in the hadronic $\tau$ decays. Any existence of such signal would manifest some novel kind of NP beyond SM. In fact, only the known $\CP$ violation in the neutral kaon system will lead to a nonzero $\CP$-violating signal in the $\tau$ decay channels containing the $K_S^0$ or $K_L^0$ state. As Bigi and Sanda first pointed out \cite{Bigi:2005ts}, the asymmetry
\begin{eqnarray}\label{eq:Acp:def}
&&A_{CP}^\tau=A_{CP}(\tau^{-}\to K_{S}^0\pi^{-}\nu_{\tau})\nonumber\\&&=\frac{\Gamma(\tau^{+}\to K_{S}^0\pi^{+}\bar\nu_{\tau})-\Gamma(\tau^{-}\to K_{S}^0\pi^{-}\nu_{\tau})}{\Gamma(\tau^{+}\to K_{S}^0\pi^{+}\bar\nu_{\tau})+\Gamma(\tau^{-}\to K_{S}^0\pi^{-}\nu_{\tau})} 
\end{eqnarray}
should be about $2\text{Re}\epsilon_K\approx3.3\times10^{-3}$ due to the decay rate difference in $\Gamma(\tau^{+}\to K_{L}^0\pi^{+}\bar{\nu}_{\tau})$
vs. $\Gamma(\tau^{-}\to K_{L}^0\pi^{-}{\nu_{\tau}})$. In fact, one can include any $\pi^0$ into Eq.~(\ref{eq:Acp:def}) without changing the value. It is interesting to note that the relation between the $\CP$ asymmetry of $\tau^{\pm}\to K_S^0\pi^{\pm}\nu_\tau$ and the one of $\tau^{\pm}\to K_L^0\pi^{\pm}\nu_\tau$ required by $\CP T$ symmetry is also discussed in Ref.~\cite{Bigi:2005ts}, and this issue is elegantly explained in Ref.~\cite{Calderon:2007rg}. Motivated by this prediction, BaBar collaboration measured $A_{CP}(\tau^{-}\to K_{S}^0\pi^{-}(\geq 0\pi^0)\nu_{\tau})$:  $(-0.36\pm0.23\pm0.11)\%$, with statistical and systematic error bars in order \cite{BaBar:2011pij}. Noticeably, its sign is opposite to the prediction stemming from neutral kaon mixing. In fact, the prediction of $3.3\times 10^{-3}$ should be more carefully examined, as pointed out by Grossman and Nir in Ref.~\cite{Grossman:2011zk}. That is, the above mentioned asymmetry means the time-integrated decay rate asymmetry, where the interference of the $K_S^0-K_L^0$ term is as important as the pure $K_S^0$ amplitude; the result crucially depends on the time interval over which the integration is done, and also the above theoretical value ought to be obtained by the convolution with the efficiency as a function of time (which should be part of the experimental analysis). The number $3.3\times 10^{-3}$
does correspond to the time interval $[t_1, t_2]$ where $t_1\ll\tau_{S}$ and $\tau_S\ll t_2\ll\tau_L$ with $\tau_S$ and $\tau_L$ being the lifetime of $K_S^0$ and $K_L^0$ respectively, and also to the ideal efficiency $\epsilon=1$ within this range. BaBar considered its efficiency and find that there should be a multiplicative factor of $1.08\pm0.01$ applied as a correction to the neutral kaon mixing result. Consequently, a realistic prediction is $A_{CP}(\tau^{-}\to K_{S}^0\pi^{-}\nu_{\tau})=(0.36\pm0.01)\%$, which deviates from the experimental result by 2.8 standard deviations. 
A more elaborate analysis may be to probe $A_{CP}(\tau\to\nu_\tau K_S^0\pi),\,A_{CP}(\tau\to\nu_\tau K_S^0 2\pi),\,A_{CP}(\tau\to\nu_\tau K_S^0 3\pi)$ separately to explore the existence of NP and impact of intermediate resonances~\cite{Bigi:2012km}, which would require more data.

This discrepancy aroused great interest, and various NP models were proposed. One typical example is given in Ref.~\cite{Devi:2013gya}, which found that a charged scalar interaction fails to yield the above rate asymmetry, although the $\CP$ violation could appear in angular observables; and instead, the non-standard tensor interaction could account for the measured rate asymmetry. More precisely speaking, the NP term was parametrized by its coupling strength and weak phase, and  these parameters could be accommodated by the PDG branching fraction value and the BaBar's rate asymmetry measurement. However, as pointed out by Cirigliano, Crivellin and Hoferichter \cite{Cirigliano:2017tqn}, the authors of Ref.~\cite{Devi:2013gya} overlooked the Watson final-state interaction theorem in their treatment of the form factors, and thus the conclusion therein was challenged. This theorem states that the phase of the form factor should coincide with the corresponding scattering phase shift in the elastic scattering region \cite{Watson:1954uc}. Once considering this constraint, the effect of the tensor operator estimated in Ref.~\cite{Devi:2013gya} would be strongly suppressed unless the coupling strength was very large, which -however- would conflict with the bounds from neutron EDM and $D-\bar D$ mixing \cite{Cirigliano:2017tqn}. Later, Ref.~\cite{Rendon:2019awg} confirmed the result in Ref.~\cite{Cirigliano:2017tqn} by allowing for more realistic variations of the tensor form factor (its modulus and phase) in the inelastic region. That is, the $A_{CP}$ could reach the level of $10^{-6}$ at most. Such a methodology was adopted by Refs.~\cite{Chen:2019vbr,Chen:2021udz} but with a more detailed analysis (extending to $\CP$-violating asymmetries with angular information)~\footnote{This effective treatment has just been extended to the other two-meson tau decays in Ref.~\cite{Aguilar:2024ybr}, emphasizing the interest of similar measurements for the $K_S K^\pm$ decay channels that can, moreover, provide an independent data-driven check of the BaBar anomaly.}. Again, the $\CP$ violation measured by BaBar is hard to understand, given the various constraints from branching fraction, decay spectrum, (semi-)leptonic kaon decays, neutron EDM as well as $D-\bar D$ mixing; one exception being a fine tuning (extreme cancellation between the NP contributions) or the NP occurring below the electroweak scale (see, however,  Ref.~\cite{Dighe:2019odu}).
 
{\it In one word, the anomaly in the $\CP$-violation value measured by BaBar gave rise to a bunch of dedicated studies, with extremely interesting outcomes.} Therefore more input from the experimental side should further revitalize this field, a new measurement from either Belle II or the proposed STCF considered here is absolutely essential, with results that will undoubtedly guide our future efforts to find NP signals in these searches.

Correspondingly, Belle II is studying this observable and the STCF will measure it after some years of data-taking \cite{Sang:2020ksa} and will contribute to solving this conundrum, if not settled before.

Before BABAR's result on $A_{CP}^\tau$, Belle \cite{Belle:2011sna} measured, in four $Q^2$-bins, a $\CP$-violating asymmetry as the mean value of the $\cos\beta\cos\Psi$  distributions~\footnote{Here $\beta$ is the second angle in the Euler rotation introduced in section \ref{sec:SFs} and $\Psi$ is the angle between the direction of the COM frame and the direction of the $\tau$, as seen from the
hadronic rest frame~\cite{Kuhn:1992nz}.} for the considered decays. The measured asymmetry was small ($\mathcal{O}(10^{-3,-2})$) and, except for the lowest mass bin, within one standard deviation from zero. Given the achieved precision, this result cannot shed light on the BaBar anomaly~\footnote{Neither can the earlier CLEO analysis \cite{CLEO:2001lhp}.}, which motivates further near-future measurements of $\CP$-violating asymmetries in $\tau\to K_S^0\pi\nu_\tau$ and related channels.

%\appendix

%\section{Sample Appendix Section}


\subsection{$\CP$ violation proposed via EDM}
The EDM of the $\tau$ can be introduced to the SM $\bar{\tau}\tau\gamma$ vertex with the following Lorentz-invariant ansatz:
\begin{eqnarray}
    e \Gamma^\mu = -d_\tau(q^{2})\sigma^{\mu\nu}\gamma^5 q_\nu,
\label{eq:edm}
\end{eqnarray}
where $q$ stands for the photon momentum. 
%The form factor $F_3$ is related to the $\tau$ EDM, $d_\tau$, through:
%\begin{eqnarray}
%    d_\tau = e F_3(0).
%\end{eqnarray}

In the SM, a non-zero value of $d_\tau$ is forbidden by both $P$ invariance and $T$ invariance. If $\CP T$ symmetry is assumed, observation of a non-zero value of $d_\tau$ would imply new $\CP$-violation sources beyond the KM mechanism and only upper bounds for fermion EDMs exist \cite{Workman:2022ynf}.
%Sizable $CP$ violating EDM of $\tau$ indicates the presence of new sources of $CP$-violation, which has significant implications for the existence of physics beyond the Standard Model. 
Due to its short lifetime, direct measurement of the $\tau$ lepton's static dipole moment is extremely challenging, and currently, indirect information comes from measurements of processes involving $\tau\bar{\tau}\gamma$ vertex in experiments.

For processes such as those considered by the LEP experiment ($e^+e^-\to e^+e^-\tau^+\tau^-$, $|d_
\tau| \lesssim 10^{-16}$~$e$ cm) ~\cite{DELPHI:2003nah, Cornet:1995pw, L3:2004ful, Lohmann:2005im} and by the Belle collaboration ($e^+e^-\to \tau^+\tau^-$, $|d_
\tau| \lesssim 10^{-17}$~$e$ cm) \cite{Belle:2002nla}, the focus is on measuring the total cross-section to determine the EDM of the $\tau$ lepton. As for the process $e^+e^-\to \gamma\tau^+\tau^-$, current research mostly involves measuring the total cross-section \cite{Biebel:1996ur, Gau:1997cn}. The L3 collaboration considered data from running at Z-pole, including simple photon energy distribution and angular distribution information, 
%\yy{angular dependence on $d_\tau$}, 
which can determine the $\tau$ lepton's EDM to be $d_\tau = (0\pm 1.5\pm 1.3) \times 10^{-16}$~$e$ cm \cite{L3:1998gov}.

Kinematics of the final particles can be fully utilized to improve the constraints on $d_\tau$ from experimental measurements. 
Theoretical studies based on the formalism of optimal observables applied to $\CP$-odd quantities in the process $e^+e^-\to \tau^+\tau^-$ \cite{Bernreuther:1993nd, Chen:2018cxt, Bernreuther:2021elu} predict that Belle II experiment could measure the $\tau$ lepton's EDM to the level of $|d_
\tau| < 2.04 \times 10^{-19}$~$e$ cm~\cite{Chen:2018cxt}. While for the the process $e^+e^-\to \gamma\tau^+\tau^-$, the main challenge lies in extracting useful information to the maximum extent from the rich angular distribution of final-state particles, calling for future studies.

The inherent $\CP$ properties of the $\tau$ lepton also include the weak EDM, which is introduced to the $\bar{\tau}\tau Z$ vertex in a similar way as EDM, with $d^W_\tau = e F_3~(q^2 = m^2_Z)$ being the weak EDM. By utilizing data collected near the Z-pole energy and considering the angular distribution information of the $\tau$ lepton decay products in the $e^+e^-\to\tau^+\tau^-$ process, the LEP experiments were able to constrain the weak EDM to the order of $d^W_\tau < 10^{-17}$~$e$ cm~\cite{L3:1998lhr, ALEPH:2002kbp}, leaving significant room for achieving the precision required to test the SM prediction (which is $d^W_\tau \lesssim 8\cdot 10^{-34}$~$e$ cm~\cite{Booth:1993af}). Future studies, $\eg$ 
with dedicated constructions of $\CP$-sensitive observables for $e^+e^-\to \gamma\tau^+\tau^-$ process at Z-pole, are expected to improve the constraints on the weak EDM.

Experiments like Belle II and STCF will gather more data in the future. Plans are in place to build new generations of electron-positron colliders across different regions, $\eg$ China, Japan and Europe, with their primary missions including running at the Z-pole energy for high-precision measurements. This presents a great opportunity for measuring the $\tau$ lepton's EDMs. Research on $\tau$ lepton EDM at future electron-positron colliders is still quite limited, mainly focusing on information related to scattering cross-sections \cite{Koksal:2018env, Koksal:2021gyd, Gutierrez-Rodriguez:2022mtt}. Exploring $\CP$ sensitive observables and machine learning methods for these measurements would push the boundary of future experiments in constraining $d_\tau$. 
A preliminary feasibility study shows that STCF can test the tau EDM form factor with a sensitivity of $10^{-18}$~$e$ cm in a wide energy range~\cite{xulei}.

\subsection{Prospect of $\tau$ $\CP$ violation study at STCF}

The peak of $\tau$ pair production cross section $e^{+} + e^{-} \to\tau^{+} + \tau^{-}$ is around 4.26~GeV, where it is 3.5~nb, which would yield $3.5\times10^{9}$ $\tau$ pairs per year at STCF.  The STCF has several advantages in the $\tau$ physics study, such as the excellent ratio of signal to background, the perfect detection efficiency, the well-controlled systematic uncertainty and the capability of full event reconstruction, etc. 
In the following, the feasibility study of $\CP$ violation in $\tau^{-}\rightarrow K_{S}^0\pi^{-} \nu_{\tau}$ decay from $e^{+}e^{-}\to\tau^{+}\tau^{-}$ at $\sqrt{s}=4.26$~GeV will be introduced with the fast simulation software package~\cite{Shi:2020nrf}.

%\subsubsection{MC simulation of $\tau^{-}\rightarrow K_{S}^0\pi^{-} \nu_{\tau}$}
MC events corresponding to 1~ab$^{-1}$ of integrated luminosity at $\sqrt{s}=4.26$~GeV are simulated.
The $e^{+}e^{-}\to \tau^{+}\tau^{-}$ process is generated with {\sc KKMC}~\cite{Jadach:1999vf}, which implement {\sc TAUOLA}~\cite{Was:2000st} to describe the production of $\tau$ pairs and does not include asymmetry effect.
Passage of the particles through the detector is simulated by the fast simulation software.

The signal process $\tau^{-}\to K_{S}^0\pi^{-}\nu_{\tau}$
is generated with vector and scalar configurations of $K_{S}^0\pi$ and the parameterized spectrum %decay amplitude
is described by:
\begin{eqnarray}
\label{equation3}
%\frac{d\Gamma}{d\sqrt{s}} \propto && \frac{1}{s}\left(1-\frac{s}{m^{2}_{\tau}}\right)^{2}\left(1+\frac{2s}{m^{2}_{\tau}}\right)P(s)\\ \nonumber
%				     &&\times \left\{P^{2}(s)\vert F_{V}\vert^{2}+\frac{3(m^{2}_{K_{S}}-m^{2}_{\pi})^{2}\vert F_{S}\vert^{2}}{4s(1+\frac{2s}{m^{2}_{\tau}})}\right\},
\frac{d\Gamma}{d\sqrt{s}} \propto \frac{1}{s}\left(1-\frac{s}{m^{2}_{\tau}}\right)^{2}\left(1+\frac{2s}{m^{2}_{\tau}}\right)P(s) \times \left\{P^{2}(s)\vert f_{+}\vert^{2}+\frac{3(m^{2}_{K_{S}^0}-m^{2}_{\pi})^{2}\vert f_{0}\vert^{2}}{4s(1+\frac{2s}{m^{2}_{\tau}})}\right\},
\end{eqnarray}
which is the form inherited from Eq.~\eqref{eq:two-mesondoubledifferentialdecaywidth} by integrating over the angle $\alpha$, where $s$ is the squared invariant mass of $K_{S}^0\pi^{-}$. The masses of $\tau,\, K_{S}^0$ and charged $\pi^{-}$ are $m_{\tau},\, m_{K_{S}^0}$ and $m_{\pi}$, respectively. $P(s)$ is the momentum of the $K_{S}^0$ in the ($K_{S}^0$-$\pi^{-}$) centre of mass frame, given by:
\begin{equation}
P(s) = \frac{\sqrt{(s-(m_{K_{S}^0}+m_{\pi})^{2})(s-(m_{K_{S}^0}-m_{\pi})^{2})}}{2\sqrt{s}}.
\end{equation}

The $f_{0}$ and $f_{+}$ are the scalar and vector form factors to parameterize the amplitudes of
$K^{*}_{0}(800)$, $K^{*}(892)$ and $K^{*}(1410)$:
\begin{equation}
f_{0} = a_{K^{*}_{0}(800)}\cdot BW_{K^{*}_{0}(800)},
\end{equation}
\begin{equation}
f_{+} = \frac{BW_{K^{*}(892)}+a_{K^{*}(1410)}\cdot BW_{K^{*}(1410)}}{1+a_{K^{*}(1410)}},
\end{equation}
where BW denotes Breit-Wigner function, and $a_{K^{*}_{0}(800)}$ and $a_{K^{*}(1410)}$ are complex coefficients for the fractions of the $K^{*}_{0}(800)$
and $K^{*}(1410)$ resonances as presented in Ref.~\cite{Belle:2007goc}.
It should be noted that here simplified form factors
are given, as compared to Sec.~\ref{tau:ff}, for the sake of yielding faster evaluation with sensible results. 

%\subsubsection{Optimization of event selection}

We consider signal events with $\tau^{+}$ decaying to leptons, $\tau^{+}\to l^{+}\nu_{l}\bar{\nu}_{\tau}$ ($l=e$ for $e$-tag, $l=\mu$ for $\mu$-tag), and $\tau^{-}\to K_{S}^0\pi^{-}\nu_{\tau}$ with $K_{S}^0\to\pi^{+}\pi^{-}$. 
%Each charged track is required to satisfy the vertex requisite that
%originates from the interaction point and detector acceptance of STCF.
The existence of a $K_{S}^0$ candidate is selected from pairs of oppositely charged tracks with invariant mass $0.485<M_{\pi^{+}\pi^{-}}<0.512$~GeV/$c^{2}$.
Moreover, to suppress the background from $\tau^{-}\to\pi^{+}\pi^{-}\pi^{-}\nu_{\tau}$, the flight significance of $K_{S}^0$ candidates are required to be larger than 2 and the decay length of $K_{S}^0$ larger than $0.5$~cm. $\mu^+$ is selected with an efficiency over 90\% at momentum larger than 0.9~GeV/c and 40\% at momentum 0.5~GeV/c for $\mu$-tag.
The $\pi/\mu$ mis-identification is 3\% based on current design of the STCF detector. 
To suppress the background from $\tau^{-}\to K_{S}^0\pi^{-}\pi^{0}\nu_{\tau}$ process, we veto a $\pi^{0}$ candidate with two photons with invariant mass in the region between 0.115 and 0.15~GeV/$c^{2}$. A machine learning technique is further implemented to suppress the hadronic finals tates. 
After the above selection, 
there are still backgrounds from $\tau$ decays and the hadronic final states containing multiple $\pi$ final states remain. 
%A likelihood ratio $y_{L}$ is therefore used to suppress these backgrounds. The likelihood ratio $y_{L}(\vec{x})$ is defined by
%\begin{equation}
%y_{L}(\vec{x}) = \frac{L_{S}(\vec{x})}{L_{S}(\vec{x})+L_{B}(\vec{x})},
%\end{equation}
%where $L_{S}$ and $L_{B}$ are the likelihood function for signal and background event, respectively, and $\vec{x}$ is a set of variables used for likelihood. Each likelihood function $L_{S/B}$ is the product of the probability density function (PDF) of the input variables defined by
%\begin{equation}
%L_{S/B} = \prod^{n_{\rm var}}_{k=1}P_{S/B,k}(x_{k}),
%\end{equation}
%where $P_{S/B,k}$ is the signal/background PDF of the $k$-$th$ input variable $x_{k}$. For this analysis, the set of variables $\vec{x}$ includes the number of neutral clusters, the momentum of decay products of $K_{S}^0$,
%the decay length of $K_{S}^0$,
%the length from the IP to the vertex of the $K_{S}$ decay,
%the mass of  $K_{S}^0$, the $\chi^{2}$ of  $K_{S}^0$ in vertex fit, the $E/p$ ratio of electron, the momentum of $\mu$, the cosine of the polar angle of $\mu$, the momentum of $\pi$ and the momentum of the $K_{S}^0\pi$ system. The polar angle mentioned above is defined as the angle between the track and the $e^{+}e^{-}$ beam axis.
The mass spectrum of $K_{S}^0\pi^{-}$ after the above selections is
shown in Fig.~\ref{mass1}.
%The efficiency of the likelihood requirement is 32.8\%, and the background level is less than 6\%.

\begin{figure}[htbp!]
\begin{center}
\begin{overpic}[width=7.5cm,angle=0]{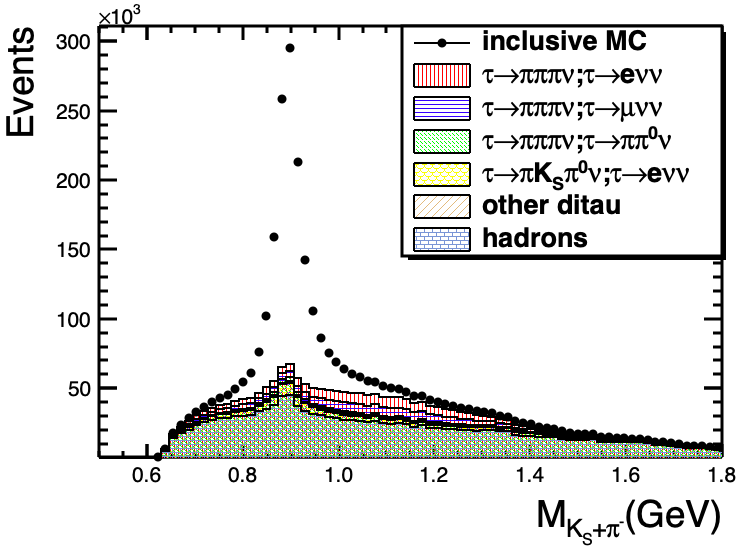}
\end{overpic}
\end{center}
\vspace{-0.5cm}
\caption{Invariant masses of $K_{S}^0\pi^{-}$ with combined $e$-tag and $\mu$-tag from $\tau^{+}$ decay after selection. Normalized according to the luminosity of 1 ab$^{-1}$ at $\sqrt{s}=4.26$ GeV.}
\label{mass1}
\end{figure}

%\subsubsection{Sensitivity of $\CP$ violation in $\tau^{-}\rightarrow K_{S}^0\pi^{-} \nu_{\tau}$ at STCF}

With the above selection criteria, the number of signal events $\tau^{-}$ and $\tau^{+}$ decay mode in 1~ab$^{-1}$ inclusive MC at $\sqrt{s}=4.26$~GeV are obtained by fitting the $K_{S}^0\pi$ invariant mass, where the signal can be parameterized by the function as shown in Eq.~(\ref{equation3}) and background is described with simulated background.
The efficiency corrected numbers for $\tau^{-}\to K_{S}^0\pi^{-}\nu_{\tau}$ and $\tau^{+}\to K_{S}^0\pi^{+}\bar{\nu}_{\tau}$ are
$3681017\pm5034$ and $3681127\pm5091$, respectively. 
It shows a good consistency with input values.
The statistical sensitivity of $\CP$ asymmetry with decay rate according to Eq.~(\ref{eq:Acp:def}) can be calculated accordingly, to be $9.7\times10^{-4}$.

 Though the analysis is performed with purely MC simulation, a discussion of possible systematic uncertainties that may bias the decay-rate asymmetry measurement between $\tau^{-}\rightarrow K_{S}^0\pi^{-} \nu_{\tau}$ and
 $\tau^{+}\rightarrow K_{S}^0\pi^{+} \bar{\nu}_{\tau}$ can be given.
 The uncertainty sources will be introduced and estimated as follows:
 \begin{itemize}
     \item  The detection asymmetry for charged particles, which can be studied from the control sample of $\tau\to\pi^{+}\pi^{-}\pi^{-}\nu_{\tau}$ decays, which is the sample with largest branching fractions. 
 The effects can be corrected by comparing the asymmetry between MC simulation and data, and the remaining uncertainty is related to the statistics uncertainty of the control sample. Since the branching fraction of the control sample is over one order of magnitude larger than the signal process, the uncertainty will be significantly smaller than the statistical uncertainty. 
     \item The uncertainty related with the selection criteria, which can be studied by varying the criteria. By performing the Barlow test~\cite{Barlow:2002yb}, and any bias from selection will be studied and corrected until the systematic accuracy matches the statistical precision.
     \item The uncertainty from the MC generator. In this analysis, no $\CP$ violation is modeled and the output asymmetry in MC is about $3\times 10^{-5}$, which can be neglected. The uncertainty associated with the background contamination. In this analysis, no decay-rate asymmetry is observed in background, and the uncertainty can be tested by applying different criteria on the background between data and MC simulation. 
     \item The different nuclear-interaction cross sections of the $K^{0}$ and $\bar{K}^{0}$ mesons. In Ref.~\cite{Ko:2010mk}, a correction to the asymmetry accounting from $K^{0}$ and $\bar{K}^{0}$ mesons interacting with the material in the detector is applied by taking the $K^{\pm}$ nucleon cross sections as an analogy, under the assumption of isospin invariance.  Similar attempts can be made at STCF according to its material setup around the beam pipe. 
 \end{itemize}

As the selection efficiency for $\tau\to K_{S}\pi\nu$ with c.m.~energy from $\sqrt{s}=4.0$~GeV to 5.0~GeV, where the cross section for $e^{+}e^{-}\to\tau^{+}\tau^{-}$ does not vary a lot around the peak value, is also quite stable, within 3\% relative difference. 
The sensitivity of the $\CP$ asymmetry is found to be proportional to $1/\sqrt{L}$. With 10~ab$^{-1}$ luminosity collected at STCF in the future, the sensitivity can be at a level of $3.1\times10^{-4}$, which is comparable to 
the uncertainty of the SM theoretical prediction. 
The clean topology and threshold production characteristics of the tau lepton pair further ensure high-precision measurements with excellent systematic uncertainty control.
Combing with the tau EDM studies at STCF, with a precision of $\mathcal{O}(10^{-18})$ $e$ cm, STCF is expected to test the $CP$ violation in the tau sector with very high precision.

\newpage

\section{$\CP$ violation in charm sector}
\label{sec:charm}
%\section{General remarks on $CP$-violation in the Standard Model}
%\subsection{The $\CKM$ matrix and its unitarity}
\subsection{$\CKM$ unitarity triangle in the charm sector}

In the SM, $\CP$ violations are generated by the weak phase in the quark flavor mixing matrix, $\ie$ the $\CKM$ matrix. The unitarity relation of the $\CKM$ matrix allows six unitarity triangles. The so-called {\it charmed} $\CKM$ unitarity triangle
illustrated in Fig.~\ref{Charmed-UT} is closely associated with the strength
of $\CP$ violation in the charm sector~\cite{Xing:2007zz},
because its three sides are just the $\CKM$
matrix elements appearing in both the box diagrams of $D^0$-$\bar{D}^0$
mixing and the penguin diagrams of the $D^0 \to \pi^+\pi^-$ decay shown in
Fig.~\ref{Box-penguin}. Unfortunately, the bottom quark plays a negligibly
small role in both the box and penguin diagrams because its contributions
are suppressed both by $m^2_b/m^2_W \sim {\cal O}(10^{-3})$ and by
\begin{eqnarray}
\left|\frac{V^{*}_{ub} V^{}_{cb}}{V^{*}_{ud} V^{}_{cd}}\right|
\simeq \left|\frac{V^{*}_{ub} V^{}_{cb}}{V^{*}_{us} V^{}_{cs}}\right|
\simeq A^2 \lambda^3 \sqrt{\rho^2 + \eta^2} \sim {\cal O}(10^{-3}) \; ,
\label{eq:box-penguin}
%     (8)
\end{eqnarray}
where $A$, $\lambda$, $\rho$ and $\eta$ are the Wolfenstein parameters %whose values can be found in PDG \cite{ParticleDataGroup:2022pth}
introduced in Eq.~(\ref{eq_Wolfenstein}). 
That is why $\CP$ violation in the charm sector is mainly associated with the first
and second families of quarks, and the magnitudes of $\CP$-violating asymmetries
between various charm quark decays and their $\CP$-conjugated processes are
expected to be of order
\begin{eqnarray}
{\cal A}^{}_{\CP\rm V} \sim
\frac{{\rm Im} \left(V^{}_{ud} V^{}_{cs} V^*_{us} V^*_{cd}\right)}{{\rm Re} \left(V^{}_{ud} V^{}_{cs} V^*_{us} V^*_{cd}\right)}
 \simeq A^2 \lambda^4 \eta \simeq 6 \times 10^{-4} \; ,
\label{eq:CPV}
%     (9)
\end{eqnarray}
up to an uncertainty factor of two or three. In other words, it is the
smallest inner angle of the charmed unitarity triangle that essentially
determines the size of charmed $\CP$ violation in the SM.
%%%%%%%%%%%%%%%%%%%%%%%%%%%% Figure 1 %%%%%%%%%%%%%%%%%%%%%%%%%%%%%%%%%%%%%
\begin{figure}[htbp!]
\begin{center}
\includegraphics[width=10cm]{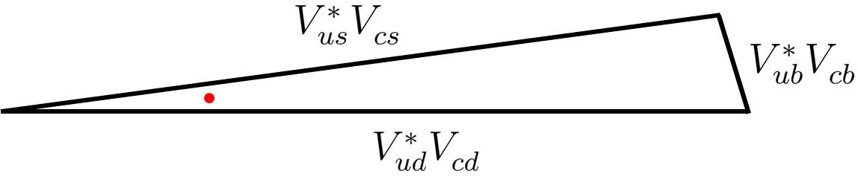}
\vspace{-0.2cm}
\caption{The {\it charmed} $\CKM$ unitarity triangle associated with
$V^*_{ud} V^{}_{cd} + V^*_{us} V^{}_{cs} + V^*_{ub} V^{}_{cb} = 0$
in the complex plane. The relevant angle determining charmed $\CP$ violation in the SM is highlighted with a red dot.}
\label{Charmed-UT}
\end{center}
\end{figure}
%%%%%%%%%%%%%%%%%%%%%%%%%%%%%%%%%%%%%%%%%%%%%%%%%%%%%%%%%%%%%%%%%%%%%%%%%%%
%%%%%%%%%%%%%%%%%%%%%%%%%%%% Figure 2 %%%%%%%%%%%%%%%%%%%%%%%%%%%%%%%%%%%%%
\begin{figure}[!htbp]
\begin{center}
\includegraphics[width=6cm]{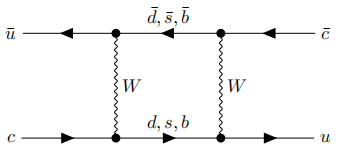}
\includegraphics[width=6cm]{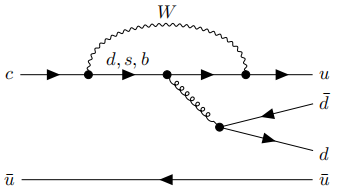}
\vspace{-0.2cm}
\caption{An illustration of the box diagrams for $D^0$-$\bar{D}^0$
mixing and the penguin diagrams for the $D^0 \to \pi^+\pi^-$ decay.}
\label{Box-penguin}
\end{center}
\end{figure}
%%%%%%%%%%%%%%%%%%%%%%%%%%%%%%%%%%%%%%%%%%%%%%%%%%%%%%%%%%%%%%%%%%%%%%%%%%%

%\subsection{Six types of $\CP$ violation}

\subsection{$\CP$ violation in the direct decays}

%Given a decay mode $D^0 \to f$ and its $\CP$-conjugated process $\bar{D}^0 \to \bar{f}$, where $f$ and $\bar{f}$ denote the respective hadronic final states, one may schematically parameterize their decay amplitudes as
%\begin{eqnarray}
%A(D^0 \to f) \hspace{-0.2cm} & = & \hspace{-0.2cm}
%A^{}_1 e^{{\rm i} \left(\delta^{}_1 + \phi^{}_1\right)} +
%A^{}_2 e^{{\rm i} \left(\delta^{}_2 + \phi^{}_2\right)} \; ,
%\nonumber \\
%A(\bar{D}^0 \to \bar{f}) \hspace{-0.2cm} & = & \hspace{-0.2cm}
%A^{}_1 e^{{\rm i} \left(\delta^{}_1 - \phi^{}_1\right)} +
%A^{}_2 e^{{\rm i} \left(\delta^{}_2 - \phi^{}_2\right)} \; ,
%\hspace{0.5cm}
%\label{eq:Direct-CPV}
%     (13)
%\end{eqnarray}
%where $\delta^{}_{1,2}$ and $\phi^{}_{1,2}$ stand respectively for
%the strong ($\CP$-conserving) and weak ($\CP$-violating) phases associated
%with the two decay modes, and $A^{}_{1,2}$ denote the
%moduli of the two amplitude components. A difference
%between the rates of $D^0 \to f$ and $\bar{D}^0 \to \bar{f}$ decays
%to their sum turns out to be
%\begin{eqnarray}
%{\cal A}^{}_f &\equiv& \frac{\Gamma(D^0 \to f) - \Gamma(\bar{D}^0 \to \bar{f})}
%{\Gamma(D^0 \to f) + \Gamma(\bar{D}^0 \to \bar{f})}
%\nonumber\\
%&=& \frac{-2 A^{}_1 A^{}_2 \sin\left(\delta^{}_2 - \delta^{}_1\right)
%\sin\left(\phi^{}_2 - \phi^{}_1\right)}
%{|A^{}_1|^2 + |A^{}_2|^2 + 2 A^{}_1 A^{}_2
%\cos\left(\delta^{}_2 - \delta^{}_1\right)
%\cos\left(\phi^{}_2 - \phi^{}_1\right)} \; . \hspace{0.2cm}
%\label{eq:Direct-CPV2}
%     (14)
%\end{eqnarray}
%So $\delta^{}_1 \neq \delta^{}_2$ and $\phi^{}_1 \neq \phi^{}_2$
As given in Eq. (\ref{eq:Acp-direct}), at least two amplitude components with different weak phases and different strong phases are the necessary conditions to
arrive at a nonzero $\CP$-violating asymmetry.
These two requirements mean that direct $\CP$ violation will
not take place unless the decay mode under consideration involves
at least two different tree-level amplitudes, or one tree-level
amplitude and the penguin amplitudes, or only the penguin amplitudes
mediated by the three down-type quarks in the loop.

For the time being, the only convincingly established signal of $\CP$
violation in the charm sector is just direct $\CP$ violation in a
combination of $D^0 \to K^+ K^-$ and $D^0 \to \pi^+\pi^-$
decays~\cite{LHCb:2019hro}:
${\cal A}^{}_{K^+ K^-} - {\cal A}^{}_{\pi^+\pi^-} =
(-1.54 \pm 0.29) \times 10^{-3}$, whose magnitude is essentially
compatible with the naive estimate made in Eq.~(\ref{eq:CPV}). %( Need to update the experimental numbers, the U-spin relation mentioned does not work! sign for new physics? With SU(3) breaking effects, SM can still accommodate data.  Maybe Fu-Sheng Yu can update this with some theory discussions?)} 
%The detailed discussions on the direct CP asymmetries will be given in Sec.~\ref{sec:dirCP}.
The time-integrated asymmetry can be further decomposed into a direct $\CP$ asymmetry $a_{CP}^{\rm dir}$
and a mixing-induced indirect $\CP$ asymmetry characterized by the parameter $\Delta Y_f$, which measures the asymmetry between $D^0\to f$ and $\overline{D}^0\to f$ effective decay widths; that is,
\be
A_{CP}(f)\approx a^{\rm dir}_f +{\langle t\rangle_f\over \tau_{D^0}} \Delta Y_f,
\en
where $\langle t\rangle_f$ is the mean decay time of the $D^0$ mesons.
Based on the LHCb average of $\Delta Y$, it follows that the direct $\CP$ asymmetry difference is given by \cite{LHCb:2019hro}
\be
\Delta a_{CP}^{\rm dir}=a_{K^+K^-}^{\rm dir}-a_{\pi^+\pi^-}^{\rm dir}=(-1.57\pm0.29)\times 10^{-3}.
\en
Notice that the $D^0$ meson is the only neutral meson system where direct $\CP$ violation can overcome indirect $\CP$ violation. 

In 2022, the time-integrated $\CP$ asymmetry in $D^0\to K^+K^-$ was measured by LHCb \cite{LHCb:2022lry} to be
\be
A_{CP}(K^+K^-)=(6.8\pm6.4\pm1.6)\times 10^{-4}. 
\en
The direct $\CP$ asymmetries in  $D^0\to K^+K^-$ and $D^0\to \pi^+\pi^-$ decays are derived by combining $A_{CP}(K^+K^-)$ with $\Delta A_{CP}$, giving
\be \label{eq:apipi,aKK}
a^{\rm dir}_{K^+K^-}=(7.7\pm5.7)\times 10^{-4}, \qquad a^{\rm dir}_{\pi^+\pi^-}=(23.2\pm6.1)\times 10^{-4}.
\en
Hence, the direct $\CP$ asymmetries deviate from zero by 1.4 and 3.8 standard deviations for $D^0\to K^+K^-$ and $D^0\to \pi^+\pi^-$, respectively. This implies a breaking of $U$-spin symmetry in charm decays. In the limit of $U$-spin symmetry, there exists a sum rule implying that $a^{\rm dir}_{K^+K^-}+a^{\rm dir}_{\pi^+\pi^-}=0$, so that the $\CP$ asymmetries of these two modes have different signs. The above measurement breaks the $U$-spin symmetry by $2.7\sigma$. If the future measurements hold the above result, it may indicate a signal of new physics \cite{Iguro:2024uuw}.  Therefore, it requires more precise measurements on the CP asymmetries of individual decay channels.

%%%%%%%%%%%%%%%%%%%%%%%%%%%%%%%%%%%%%
\begin{figure}[t]
\begin{center}
\includegraphics[width=11cm]{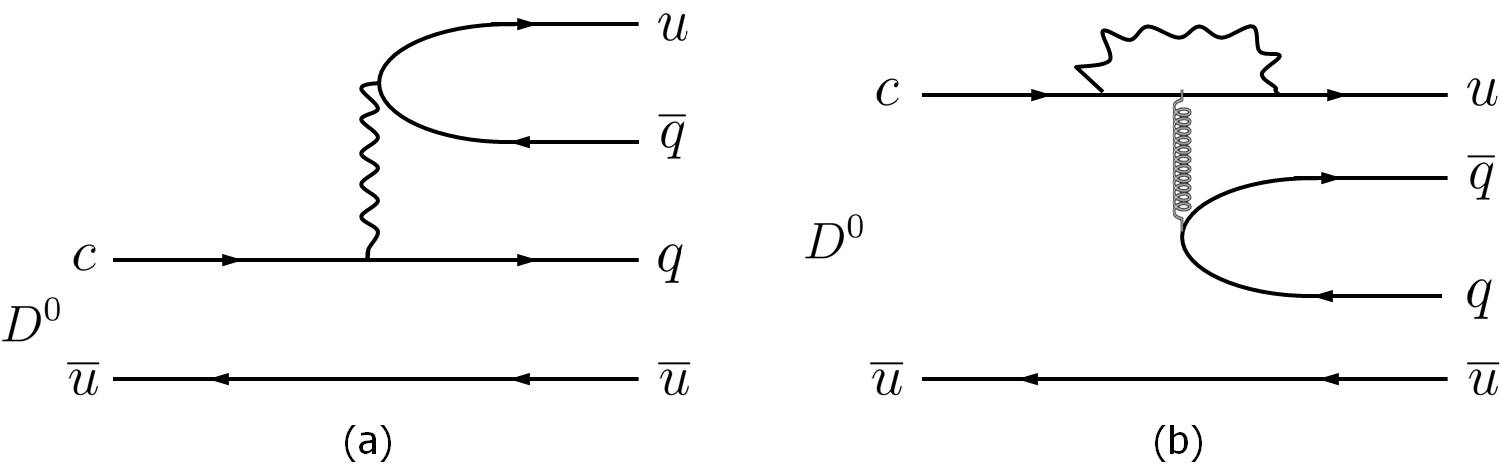}
\vspace{-0.25cm}
\caption{The tree and penguin Feynman diagrams for $D^0 \to
K^+ K^-$ and $\pi^+ \pi^-$ decays with $q = d$ and $s$~\cite{Xing:2019uzz}.}
\label{U-spin}
\end{center}
\end{figure}
%%%%%%%%%%%%%%%%%%%%%%%%%%%%%%%%%%%%%%%%%%%%%%%%%%%%%%%%%%%%%%%%%%%%%%%%%%%

In view of Fig.~\ref{U-spin}, let us write out the amplitudes of
$D^0 \to \pi^+\pi^-$ and $D^0 \to K^+ K^-$ decay modes in a universal way
as follows:
\begin{eqnarray}
A(D^0 \to f^{}_q) \hspace{-0.2cm} & = & \hspace{-0.2cm}
T^{}_q \left(V^{}_{cq} V^*_{uq}\right)
e^{{\rm i} \delta^{}_q} + P^{}_q \left(V^{}_{cb} V^*_{ub}\right)
e^{{\rm i} \delta^{\prime}_q} \; ,
\nonumber \\
A(\bar{D}^0 \to f^{}_q) \hspace{-0.2cm} & = & \hspace{-0.2cm}
T^{}_q \left(V^{*}_{cq} V^{}_{uq}\right)
e^{{\rm i} \delta^{}_q} + P^{}_q \left(V^{*}_{cb} V^{}_{ub}\right)
e^{{\rm i} \delta^{\prime}_q} \; , \hspace{0.5cm}
\label{eq:CPE-amplitude}
%     (43)
\end{eqnarray}
where $T^{}_q$ and $P^{}_q$ (for $q = d, s$) are real, $\delta^{}_q$ and
$\delta^{\prime}_q$ stand respectively for the strong phases of the tree and
penguin amplitudes, and only the dominant bottom-quark contribution to the
penguin loop has been taken into account as a reasonable
approximation~\cite{Xing:2019uzz}. If the penguin contribution is neglected
and the standard phase convention for the $\CKM$ matrix is taken, one will simply
arrive at $A(D^0 \to K^+ K^-) \simeq - A(D^0 \to \pi^+ \pi^-)$ under U-spin symmetry
because this $d \leftrightarrow s$ interchange symmetry assures $T^{}_s = T^{}_d$,
$\delta^{}_s = \delta^{}_d$ and $\delta^{\prime}_s = \delta^{\prime}_d$ to hold.
But the penguin contribution proves to be important and hence cannot be
neglected. To see this point, let us figure out the direct $\CP$-violating asymmetry 
in the following approximation: 
\begin{eqnarray}
{\cal A}^{}_{f^{}_q} \simeq {\cal U}^{}_{f^{}_q}
\simeq 2 \eta^{}_q \left(A^2 \lambda^4 \eta\right)
\frac{P^{}_q}{T^{}_q} \sin\left(\delta^{}_q - \delta^\prime_q\right) \; ,
\label{eq:CPE-CPV}
%     (44)
\end{eqnarray}
where the Wolfenstein parametrization of the $\CKM$ matrix $V$ has been used, 
and the coefficient $\eta^{}_d = -1$ for the final state 
$f^{}_d = \pi^+\pi^-$ or $\eta^{}_s = +1$ for $f^{}_s = K^+K^-$.
As a result, the combined $\CP$-violating asymmetry
\begin{eqnarray}
{\cal A}^{}_{K^+K^-} - {\cal A}^{}_{\pi^+\pi^-}
\simeq 2 \left(A^2 \lambda^4 \eta\right) \left[
\frac{P^{}_s}{T^{}_s} \sin\left(\delta^{}_s - \delta^\prime_s\right)
+ \frac{P^{}_d}{T^{}_d} \sin\left(\delta^{}_d - \delta^\prime_d\right)
\right] \;
\label{eq:CPE-CPV2}
%     (45)
\end{eqnarray}
is expected to be $(-1.54 \pm 0.29) \times 10^{-3}$, as measured in the
LHCb experiment~\cite{LHCb:2019hro}. This result requires that the
quantities in the square brackets of Eq.~(\ref{eq:CPE-CPV2}) should be as large
as about 1.2, implying that both the penguin contribution and final-state
interactions are significant (see, $\eg$,
Refs.~\cite{Cheng:2012wr,Cheng:2012xb,Li:2012cfa}). 
%In this case, the U-spin symmetry between $D^0 \to \pi^+\pi^-$ and $D^0 \to K^+K^-$ decay modes is badly broken~\cite{Xing:2019uzz}.

In 2012 there existed two independent studies in which direct $\CP$ violation in charmed meson decays was explored, based on the Topological diagrammatic approach (Topo) for tree amplitudes and the QCD-inspired approach for penguin amplitudes~\cite{Cheng:2012wr,
Cheng:2012xb}, and the Factorization-Assisted Topological-amplitude approach (FAT) \cite{Li:2012cfa}. Interestingly, both works predicted a $\Delta A_{C\!P}$ at the per mille level seven years before the LHCb's announcement of the first observation of $\CP$ violation in the charm sector.  The comparison between the
experimental measurements and the theoretical predictions is shown in Fig.~\ref{fig:CPdiff}, taken from Ref.~\cite{Saur:2020rgd}. 
These two works are the only ones with correct predictions before the observation by LHCb. 
Under topological-diagram methods, the dynamics of charm decays at the scale of $1$GeV can be understood by the data of the branching fractions. Then the penguin contributions can be reliably obtained at the same scale.  

The correct predictions are based on the precise measurements of branching fractions of $D$ meson decays by BESIII and other experiments. The dynamics of charm decays are polluted by large non-perturbative contributions which are difficult to be calculated using the current theoretical methods. But the branching fractions of $D$ meson decays can provide very useful information on the dynamics of charm decays. Using the topological diagrammatic approach and factorization hypothesis, the non-perturbative hadronic matrix elements can be well extracted from the data of branching fractions. Therefore, the measurements of branching fractions at BESIII and STCF are very important, even though these experiments cannot directly observe the $\CP$-violation of charm decays.

%====================================================================
\begin{figure}[t]
\begin{center}
\centering
 \includegraphics[scale=0.4]{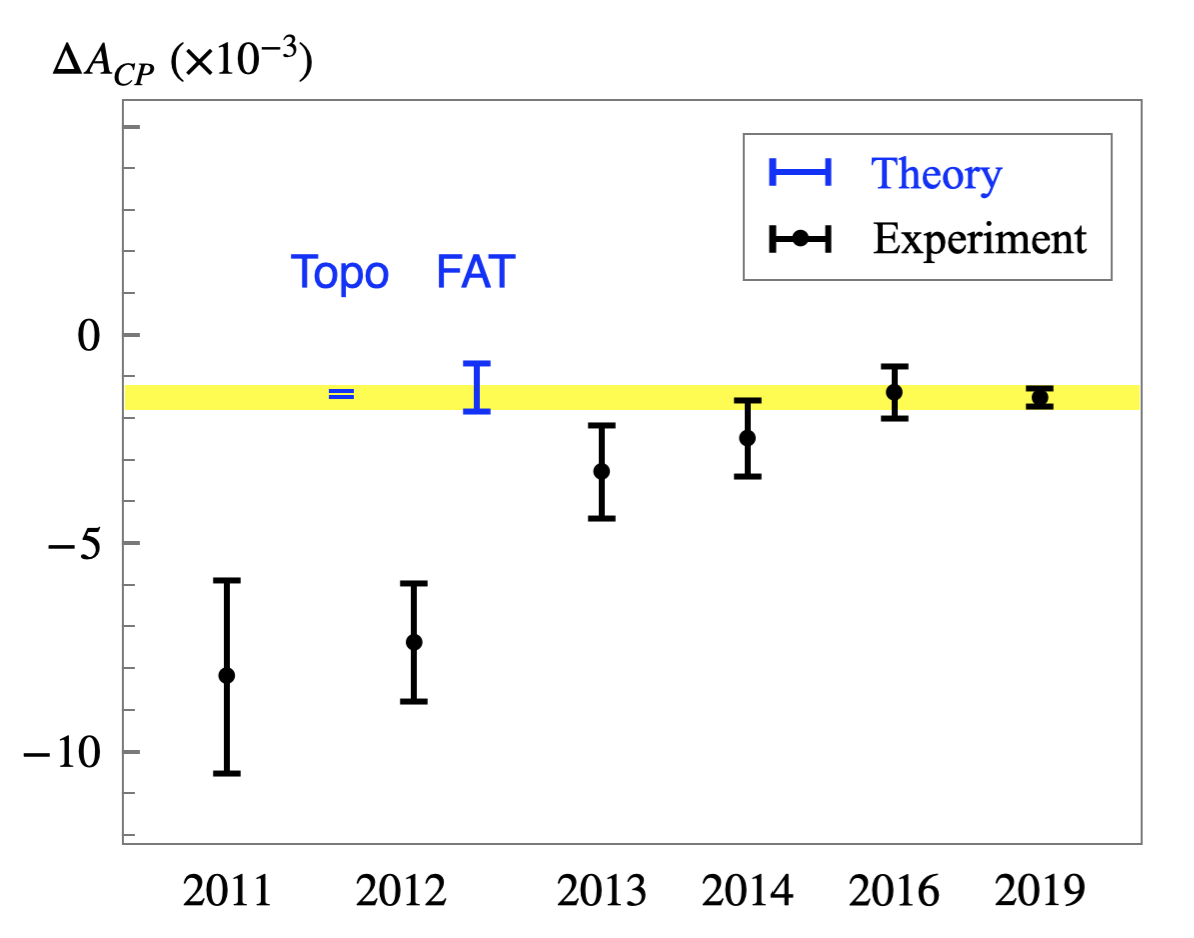}
\vspace{0.1cm}
\caption{Comparison between experimental measurements (in black) and theoretical predictions (in blue) on $\Delta A_{CP}$
as a function of time \cite{Saur:2020rgd}. Experimental results are corresponding to the world-average values for a specific year. The theoretical approaches are the Topological diagrammatic approach (Topo) under the flavor $SU(3)$ symmetry limit and the Factorization-Assisted Topological-amplitude approach (FAT), respectively. The yellow band is the 2019 experimental result, shown for comparison.}
\label{fig:CPdiff}
\end{center}
\end{figure}

%Replacing $D^0$ and $\bar{D}^0$ in Eqs.~(\ref{eq:Direct-CPV}) and(\ref{eq:Direct-CPV2}) with $D^+$ and $D^-$, 
One may similarly
discuss direct $\CP$ violation in the charged $D$-meson decays. The
magnitude of direct $\CP$ violation in such decay modes is also
expected to be of ${\cal O}(10^{-3})$ or
smaller (see, $\eg$, Refs.~\cite{Chau:1984wf,Buccella:1992sg,Buccella:1994nf,
Cheng:2012wr,Cheng:2012xb,Li:2012cfa,Qin:2013tje,Li:2019hho,Cheng:2019ggx,Cheng:2021yrn}),
but a striking and reliable signal has not been observed in the
present-day experiments.

%\subsection{Direct $\CP$ violation in the decays of charmed mesons and charmed baryons}\label{sec:dirCP}

%The direct $\CP$ violation of charm decays are expected at the order of $10^{-3}$ or even smaller, suppressed by the tiny $\CKM$ phase in the first two generations of quarks. Such small $\CP$ violation requires very large data samples of charm decays. Currently, the direct $\CP$ violation of charm decays is only observed in $D^0\to K^+K^-, \pi^+\pi^-$ by LHCb at the hadron collisions. 

The STCF, as an electron-positron collider, has only a marginal power for observation of direct $\CP$ violation in charm decays. However, its measurements on the branching fractions and any other observables would be very helpful for the theoretical prediction on the $\CP$ violation of charm-meson and charm-baryon decays, which is important in the comparison with experimental measurements to precisely test the SM and explore NP. 

%In November of 2011 LHCb announced the first evidence of $\CP$ violation in the charm sector. A nonzero value for the difference between the time-integrated $\CP$ asymmetries of the decays $D^0\to K^+K^-$ and $D^0\to\pi^+\pi^-$ \cite{LHCb:2011osy}
%\be \label{eq:LHCb:2011}
%\Delta A_{CP}\equiv A_{CP}(K^+K^-)-A_{CP}(\pi^+\pi^-)=(-0.82\pm0.21\pm0.11)\%
%\en
%was reported. The significance of the measured deviation from zero is 3.5$\sigma$. This had triggered a flurry of studies exploring whether this $\CP$ violation in the charm sector was consistent with the SM or implies NP. However, the original evidence of $\CP$ asymmetry difference was gone in 2013 and 2014 when LHCb started to use the muon tag to identify the $D^0$ and found a positive $\Delta A_{C\!P}$ \cite{LHCb:2013dkm}.
%In 2019, LHCb finally reported the first observation of
%$\CP$ asymmetry in the charm system with the result at the per mille level \cite{LHCb:2019hro}
%\be \label{eq:LHCb:2019}
%\Delta A_{CP}=(-1.54\pm0.29)\times 10^{-3}.
%\en

%=====================================================================

A lot of data have been collected for charmed anti-triplet baryons $\Lambda_c$, $\Xi_c^0$ and $\Xi^+_c$. The final-state-interaction effect can generate CP asymmetries of charm-baryon decays \cite{Jia:2024pyb}.
The leading decay amplitude for two-body charmed anti-triplet baryon $B_c$ decays into an octet low-lying baryon $B$ and an psudoscalar octet $P$ have been studied \cite{Geng:2023pkr,Zhong:2024zme,Zhong:2024qqs}. The penguin contributions in the charmed baryon decays have been addressed in \cite{He:2024pxh}, in which a possible enhancement effect to the $\CP$ violation is found similarly to what happened to $D \to \pi\pi, KK$ decays. 

The search for $\CP$ violation in charmed baryon decays has taken on new momentum
with the large samples of $\Lambda_c^+$ obtained by BESIII and LHCb. For two-body
decays of the $\Lambda_c^+$, $\CP$ violation through the measurement of $C\!P$-violating decay parameter asymmetry, ${\cal A}=(\alpha+\bar\alpha)/(\alpha-\bar\alpha)$ had been carried out with
the current result given by ${\cal A}=-0.07\pm0.19\pm0.24$ for $\Lambda_c^+\to \Lambda\pi^+$ and $\bar\Lambda_c^-\to\bar\Lambda \pi^-$ \cite{FOCUS:2005vxq} and $(1.5\pm5.2\pm1.7)\%$ for
$\Xi_c^0\to\Xi^-\pi^+$ and $\bar\Xi_c^0\to \bar\Xi^+\pi^-$ \cite{Belle:2021crz}. These measurements are still an order of magnitude away from SM expectations \cite{He:2024pxh}.

As for three-body decays,
LHCb has measured  the difference between $\CP$ asymmetries in
$\Lambda_c^+ \to p K^+ K^-$ and $\Lambda_c^+ \to p\pi^+ \pi^-$ decay channels, in analogy to $D^0\to K^+K^-$ and $D^0\to \pi^+\pi^-$ \cite{LHCb:2017hwf}.
The result  is $\Delta A_{CP}(\Lambda_c^+) =A_{CP}(p K^+ K^-)-A_{CP}(p \pi^+ \pi^-)= (0.30 \pm 0.91 \pm 0.61)\%$,
to be compared with a generic SM prediction of a fraction of 0.1\%
\cite{Bigi:2012ev}.
In order to probe the SM level, one has to multiply the available statistics
by at least a factor of 100. LHCb has also looked for local $\CP$ asymmetry in $\Xi_c^+\to pK^-\pi^+$ decays and found null results \cite{LHCb:2020zkk}.

For multi-hadrons in the final state of $\Lambda_c^+$ decays such as $\Lambda_c^+\to pK^-\pi^+\pi^0$, $\Lambda_c^+\to\Lambda\pi^+\pi^+\pi^-$ and $\Lambda_c^+\to pK_S^0\pi^+\pi^-$, $\CP$ violation can be exploited through several $T$-odd observables \cite{Wang:2024qff}. Owing to its features of high luminosity, broad c.m.~energy acceptance, abundant production and clean environment, $\STCF$ may provide a great platform for this kind of study.
A fast Monte Carlo simulation study in \cite{Shi:2019vus} by using the $e^+e^-$ annihilation data of 1 ab$^{-1}$ at $\sqrt{s}=4.64$ GeV, which are expected to be available at the future $\STCF$, indicates that a sensitivity at the level of (0.25-0.5)\% is accessible for the above-mentioned three decay modes. This will be enough to measure non-zero $C\!P$-violating asymmetries as large as 1\%.

%\subsection{$\CP$ violation in charmed baryon decays}
STCF will be able to produce also charmed baryon-antibaryon pairs. Having the c.m.~energy range up to 7 GeV, the pairs of $\Lambda_c^+\overline{\Lambda}_c^-$, $\Xi_c^0\overline{\Xi}_c^0$ and
$\Xi_c^+\overline{\Xi}_c^-$ can be created.    According to the BESIII data, there is no evident resonance that could increase production yields. For example, the cross section of $e^+e^-\to\Lambda_c^+\overline{\Lambda}_c^-$ increases quickly above the threshold and stays approximately constant with the value of about 200 pb~\cite{BESIII:2017kqg,BESIII:2023rwv}. 
Charmed baryon decays have many competing decay modes, since all branching fractions are small and do not exceed 7\%~\cite{BESIII:2015bjk}. Therefore, decay measurements require a double tagging technique that combines a set of common decay modes.
With 1~ab$^{-1}$ data sample collected at STCF, $2\times 10^8$ $\Lambda_c^+\overline{\Lambda}_c^-$ pairs can be produced, which might not be sufficient for $\CP$ symmetry studies using entangled baryon-antibaryon systems. One of the most useful decays for tagging and polarization measurement is $\Lambda_c^+\to pK^-\pi^+$ with the branching fraction as ${\cal B}=6.26(29)\%$ and the effective polarimeter constant (weighted average over the Dalitz plot) of $65$\%~\cite{LHCb:2022sck}. 

Here we comment on how the production and decay properties of charmed baryon differ from those of hyperons. 
\begin{itemize}
    \item There are many hadronic decays into three or more final-state particles. Several processes include hyperons, which allows to determine their spin via sequential weak decays.
    
    \item { The range of the $q^2$ variable in the SL decays of charmed baryons is large enough for  the form factors in Eq.~\eqref{eq:matrixelem_semil} to acquire imaginary part due to on-shell contribution of hadronic  states.
    For $\CP$ symmetry test it means that the the elements of the aligned decay matrix in Eq.~\eqref{eq:Bmatrix} terms $b_{20}$,$ b_{21}$, $b_{23}$, $b_{02}$,$ b_{12}$ and $b_{32}$ are nonzero~\cite{Batozskaya:2023rek}. This allows for $T$-odd $\CP$ tests which are complementary to the $T$-even tests such as a comparison of the partial decay widths. There are also Lattice QCD calculations for SL decays of heavy quarks~\cite{Detmold:2015aaa}.
  
    } %{\color{blue} (Fu-Sheng Yu: unclear for this paragraph.)}
    \item In the $\SM$, the direct $\CP$ violation vanishes for Cabibbo-favoured decays of charmed baryons. But for processes involving $K_{S, L}$ in the final states, CP violation can show up due to $K^0 - \bar K^0$ mixing with a size of order $10^{-3}$~\cite{Wang:2017gxe}. %Singly Cabibbo suppressed decays, which have two weak amplitudes (where one of them includes penguin graph) can generate a direct $\CP$-odd phase. It has been shown recently that the relevant SU(3) decay amplitudes can be almost completely determined \cite{Zhong:2024zme} [reference, C-Q Geng, X-G He,X-N Jin, C-W Liuand Chang Yang, PRD 109(2024)7, L071302; H-L Zhong, F-R Xu, and H-Y Cheng, arXiv: 2401.15926[hep-ph] ] using known data and re-scattering effects making predictions possible with sizeable CP asymmetry, such as $A_{CP}(\Xi^0_c \to p K^0) - A_{CP}(\Xi^0_c \to \Sigma^+ \pi^-) \sim  10^{-3}$ \cite{He:2024pxh} %[reference: X-G He and C-W Liu, arXiv: 2404.19166[hep-ph]]. But this may be still very challenging for STCF. }
    \item Exclusive reconstruction of decays with one unmeasured particle.

%    \item  since there are no prominent resonances in the considered energy range. 
    \item  Selection of the c.m.~Energy has to include effects such as the total cross section, detection efficiency and the transverse polarization of the baryon, given by the value of the relative phase $\Delta\Phi={\rm Arg}(G_E/G_M)$. For the $\Lambda_c^+\overline{\Lambda}_c^-$ channels, the cross section is largest close to threshold. The angular distribution parameter $\eta$ changes from zero at threshold to $-0.26(9)$ at $4.2619$ GeV and then raises to $+0.63(21)$ at the BESIII highest energy of $4.9559$ GeV. The $\Delta\Phi$ phase and the natural polarization is zero at threshold since $G_E=G_M$. The relative phases are not measured yet by BESIII. However, beam polarization can play an important role for the precision of the measurements. For example, the absolute value of baryon polarization at threshold is equal to the electron beam polarization and it does not depend on the production angle $\theta$~\cite{Salone:2022lpt}. 
\end{itemize}

\subsection{$\CP$ violation associated with $D^0$-$\bar{D}^0$ mixing}

The theoretical study of $\CP$ violation in $D^0$-$\bar{D}^0$ mixing can
be traced back to the seminal papers by Okun, Zakharov and
Pontecorvo~\cite{Okun:1975di} and by Pais and Treiman~\cite{Pais:1975qs}
in 1975, just one year after the experimental discoveries of the
charm quark~\cite{E598:1974sol,SLAC-SP-017:1974ind}. The key point is that
the box diagrams illustrated in Fig.~\ref{Box-penguin} allow
$D^0$ and $\bar{D}^0$ mesons to form the mass eigenstates
\begin{eqnarray}
|D^{}_1\rangle \hspace{-0.2cm} & = & \hspace{-0.2cm}
p |D^0\rangle + q |\bar{D}^0\rangle \; ,
\nonumber \\
|D^{}_2\rangle \hspace{-0.2cm} & = & \hspace{-0.2cm}
p |D^0\rangle - q |\bar{D}^0\rangle \; , \hspace{0.5cm}
\label{eq:D-mixing}
%     (10)
\end{eqnarray}
where the complex parameters $p$ and $q$ satisfy the normalization condition
$|p|^2 + |q|^2$, and $\CPT$ invariance has been assumed. Let us use the
dimensionless parameters,
\begin{eqnarray}
x \hspace{-0.2cm} & \equiv & \hspace{-0.2cm}
\frac{\Delta m}{\Gamma} \equiv
\frac{m^{}_2 - m^{}_1}{\Gamma} \; , \hspace{0.5cm}
\nonumber \\
y \hspace{-0.2cm} & \equiv & \hspace{-0.2cm}
\frac{\Delta \Gamma}{2\Gamma} \equiv
\frac{\Gamma^{}_2 - \Gamma^{}_1}{2\Gamma} \; ,
\label{eq:x-y}
%     (11)
\end{eqnarray}
to describe the effects of $D^0$-$\bar{D}^0$ mixing, where
$m^{}_{1,2}$ and $\Gamma^{}_{1,2}$ denote the masses and decay widths
of $D^{}_{1,2}$, and $\Gamma = (\Gamma^{}_1 + \Gamma^{}_2)/2$
is the mean decay width of these two particles. It is notoriously
difficult to reliably calculate $x$ and $y$, simply because the
long-distance effects on $D^0$-$\bar{D}^0$ mixing at the hadron level
are essentially nonperturbative and hence involve a lot of
theoretical uncertainties (see, $\eg$,
Refs.~\cite{Georgi:1992as,Ohl:1992sr,Bigi:2000wn,Falk:2001hx,Falk:2004wg}).
Some order-of-magnitude estimates have
given $x \sim y \sim {\cal O}(10^{-3})$, consistent with their
allowed regions extracted from current experimental data shown in
Fig.~\ref{PDG-xy}.
%%%%%%%%%%%%%%%%%%%%%%%%%%%% Figure 3 %%%%%%%%%%%%%%%%%%%%%%%%%%%%%%%%%%%%%
\begin{figure}[htbp!]
\begin{center}
\includegraphics[width=8cm]{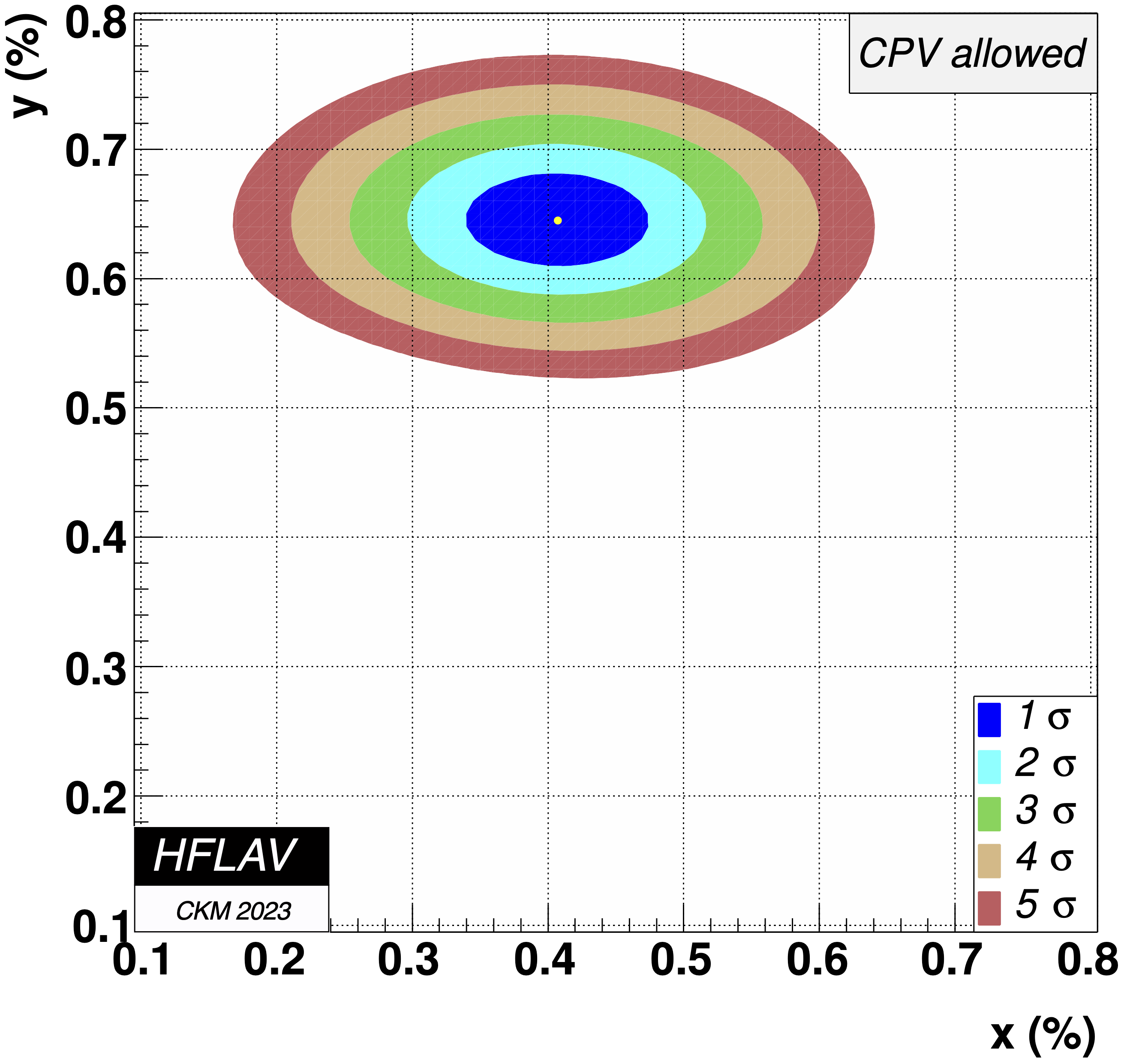}
%\vspace{-1.5cm}
\caption{Constraints on the $D^0$-$\bar{D}^0$ mixing parameters $x$ and $y$
as obtained by the HFLAV group from current data~\cite{HFLAV:2022pwe}.}
\label{PDG-xy}
\end{center}
\end{figure}
%%%%%%%%%%%%%%%%%%%%%%%%%%%%%%%%%%%%%%%%%%%%%%%%%%%%%%%%%%%%%%%%%%%%%%%%%%%

The definition of $D^{}_{1,2}$ in Eq.~(\ref{eq:D-mixing}) tells us
that these two mass eigenstates would be the $\CP$ eigenstates if
$p = \pm q$ held. So $|p| \neq |q|$ is a rephrasing-invariant measure
of $\CP$ violation in $D^0$-$\bar{D}^0$ mixing. In the SL decays
of neutral $D$ mesons, the $\CP$-violating observable
is~\cite{Okun:1975di,Pais:1975qs}
\begin{eqnarray}
\Delta^{}_{\rm mixing} \equiv \frac{|p|^4 - |q|^4}{|p|^4 + |q|^4} \; .
\label{eq:CPV-mixing}
%     (12)
\end{eqnarray}
It is expected that the magnitude of $\Delta^{}_{\rm mixing}$ is
at most of ${\cal O}(10^{-3})$ in the SM, but the calculations
suffer from large long-distance
uncertainties~\cite{Bigi:2000wn,Falk:2001hx,Falk:2004wg}. Fig.~\ref{PDG-pq}
illustrates the present experimental constraints on the argument
and modulus of $q/p$, which are consistent with the SM expectation.
NP might contribute to $D^0$-$\bar{D}^0$ mixing via the box
diagrams and enhance the size of $\Delta^{}_{\rm mixing}$ to some extent
(see, $\eg$, Refs.~\cite{Golowich:2007ka,Blum:2009sk,Pakvasa:2009zzb}),
but a significant enhancement seems quite unlikely in this regard.
%%%%%%%%%%%%%%%%%%%%%%%%%%%% Figure 4 %%%%%%%%%%%%%%%%%%%%%%%%%%%%%%%%%%%%%
\begin{figure}[htbp!]
\begin{center}
\includegraphics[width=8cm]{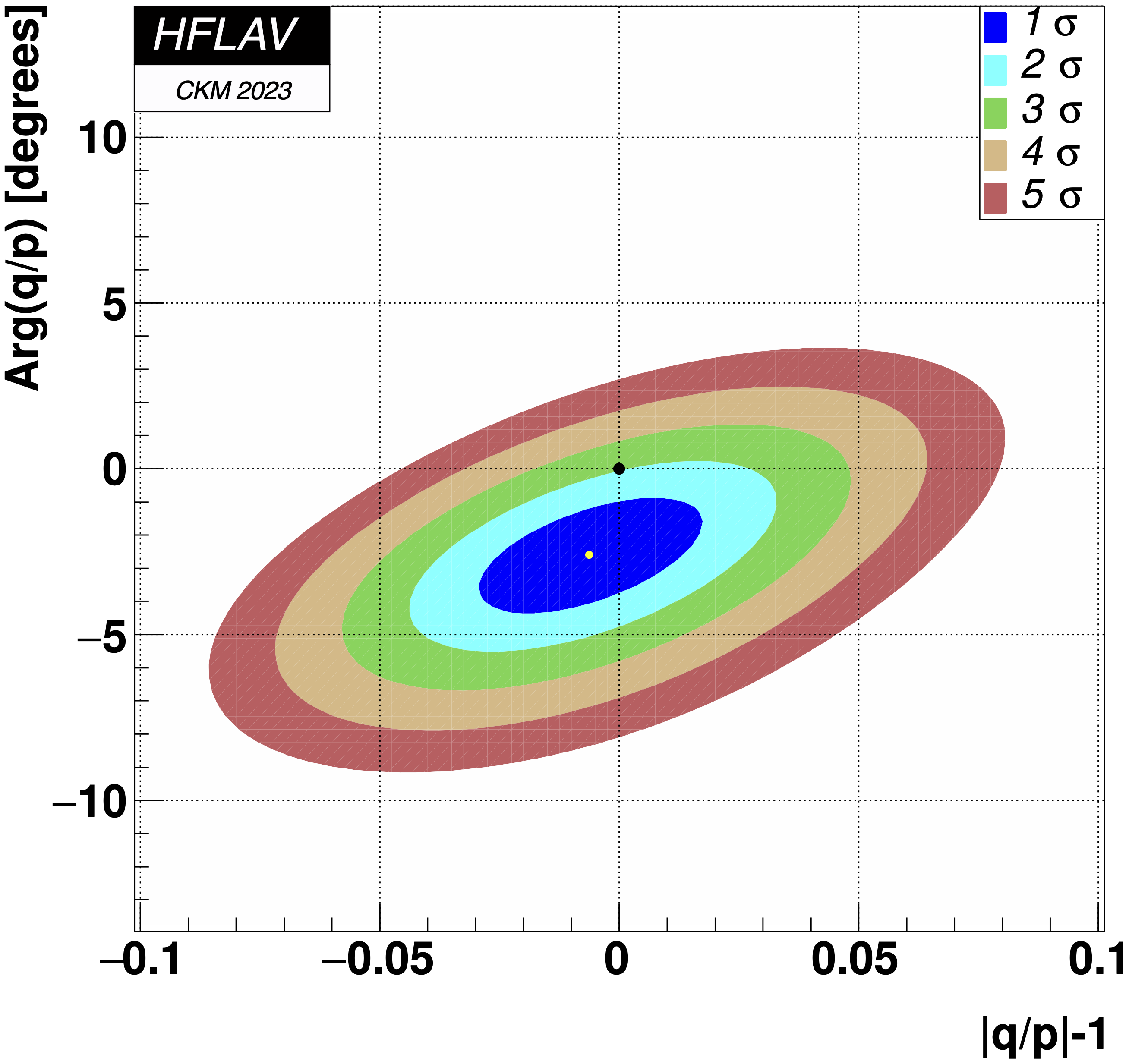}
%\vspace{-1.5cm}
\caption{Constraints on the modulus and argument of $q/p$ in
$D^0$-$\bar{D}^0$ mixing as obtained by the HFLAV group from currently
available experimental data~\cite{HFLAV:2022pwe}.}
\label{PDG-pq}
\end{center}
\end{figure}
%%%%%%%%%%%%%%%%%%%%%%%%%%%%%%%%%%%%%%%%%%%%%%%%%%%%%%%%%%%%%%%%%%%%%%%%%%%

In the $e^+e^-$ collisions at the threshold of $D\bar D$, it has a particular advantage to investigate the indirect $\CP$ violation of $D^0-\bar D^0$ mixing where $D^0$ and $\bar D^0$ are quantum correlated to each other.  

\paragraph{Formulas for incoherent neutral $D$ meson decays}

Given the parameters of $D^0$-$\bar{D}^0$ mixing defined in
Eqs.~(\ref{eq:D-mixing}) and (\ref{eq:x-y}) %, the proper-time evolution
the time-independent decay rates can be expressed as~\cite{Xing:1996pn}
\begin{eqnarray}
\Gamma(D^0 \to f) \hspace{-0.2cm} & \propto & \hspace{-0.2cm}
|A(D^0 \to f)|^2
\left[\frac{1}{1 - y^2} \cdot \frac{1 + |\lambda^{}_f|^2}{2}
+ \frac{y}{1 - y^2} {\rm Re}\lambda^{}_f \right.
\nonumber \\
\hspace{-0.2cm} & & \hspace{-0.2cm}
+ \left. \frac{1}{1 + x^2} \cdot \frac{1 - |\lambda^{}_f|^2}{2}
- \frac{x}{1 + x^2} {\rm Im}\lambda^{}_f \right] \; ,
\nonumber \\
\Gamma(\bar{D}^0 \to f) \hspace{-0.2cm} & \propto & \hspace{-0.2cm}
|A(D^0 \to f)|^2
\left[\frac{1}{1 - y^2} \cdot \frac{1 + |\lambda^{}_f|^2}{2}
+ \frac{y}{1 - y^2} {\rm Re}\lambda^{}_f \right. \hspace{0.5cm}
\nonumber \\
\hspace{-0.2cm} & & \hspace{-0.2cm}
- \left. \frac{1}{1 + x^2} \cdot \frac{1 - |\lambda^{}_f|^2}{2}
+ \frac{x}{1 + x^2} {\rm Im}\lambda^{}_f \right]\left|\frac{p}{q}\right|^2 \; .
\label{eq:D-decay-rate2}
%     (27)
\end{eqnarray}
Here, we define \begin{eqnarray}
\lambda^{}_f \hspace{-0.2cm} & \equiv & \hspace{-0.2cm}
\frac{q}{p} \cdot \frac{A(\bar{D}^0 \to f)}{A(D^0 \to f)} \; ,
\nonumber \\
\bar{\lambda}^{}_{\bar f} \hspace{-0.2cm} & \equiv & \hspace{-0.2cm}
\frac{p}{q} \cdot \frac{A(D^0 \to \bar{f})}{A(\bar{D}^0 \to \bar{f})} \; ,
\hspace{0.5cm}
\label{eq:Indirect-CPV}
%     (15)
\end{eqnarray}
and $f$ and $\bar{f}$ are required to be common to the decay of
$D^0$ (or $\bar{D}^0$).
Therefore, one may calculate the time-independent decay rates
of neutral $D$ mesons into $\bar{f}$, the $\CP$-conjugate state of $f$,
in the same way.

As shown in Fig.~\ref{PDG-xy}, the $D^0$-$\bar{D}^0$ mixing parameters $x$
and $y$ are both positive and smaller than $1\%$. It is therefore safe
to make analytical approximations for the time-independent decay rates of
$D^0 \to f$, $D^0 \to \bar{f}$, $\bar{D}^0 \to f$ and $\bar{D}^0 \to \bar{f}$
by neglecting the terms of ${\cal O}(x^2)$ and ${\cal O}(y^2)$. With the help of
Eq.~(\ref{eq:D-decay-rate2}), we easily arrive at
\begin{eqnarray}
\Gamma(D^0 \to f) \hspace{-0.2cm} & \propto & \hspace{-0.2cm}
|A(D^0 \to f)|^2
\left(1 + y {\rm Re}\lambda^{}_f - x {\rm Im}\lambda^{}_f \right) \; ,
\nonumber \\
\Gamma(\bar{D}^0 \to \bar{f}) \hspace{-0.2cm} & \propto & \hspace{-0.2cm}
|A(\bar{D}^0 \to \bar{f})|^2
\left(1 + y {\rm Re}\bar{\lambda}^{}_{\bar{f}} - x {\rm Im}
\bar{\lambda}^{}_{\bar{f}} \right) \; ; \hspace{0.5cm}
\label{eq:D-decay-rate3}
%     (28)
\end{eqnarray}
and
\begin{eqnarray}
\Gamma(\bar{D}^0 \to f) \hspace{-0.2cm} & \propto & \hspace{-0.2cm}
|A(D^0 \to f)|^2
\left(|\lambda^{}_f|^2 + y {\rm Re}\lambda^{}_f + x {\rm Im}\lambda^{}_f \right)
\left( 1 + \Delta^{}_{\rm mixing}\right) \; ,
\nonumber \\
\Gamma(D^0 \to \bar{f}) \hspace{-0.2cm} & \propto & \hspace{-0.2cm}
|A(\bar{D}^0 \to \bar{f})|^2
\left(|\bar{\lambda}^{}_{\bar{f}}|^2 + y {\rm Re}\bar{\lambda}^{}_{\bar{f}}
+ x {\rm Im}\bar{\lambda}^{}_{\bar{f}} \right)
\left( 1 - \Delta^{}_{\rm mixing}\right) \; , \hspace{0.5cm}
\label{eq:D-decay-rate4}
%     (29)
\end{eqnarray}
where $\bar{\lambda}^{}_{\bar f}$ has been defined in Eq.~(\ref{eq:Indirect-CPV}),
and $|p/q|^4 = (1 + \Delta^{}_{\rm mixing})/
(1 + \Delta^{}_{\rm mixing}) \simeq 1 + 2\Delta^{}_{\rm mixing}$
derived from Eq.~(\ref{eq:CPV-mixing}) has been used. As the magnitude of
$\Delta^{}_{\rm mixing}$ is at most of ${\cal O}(10^{-3})$ in the SM, it
can also be omitted from Eq.~(\ref{eq:D-decay-rate4}) in most cases.

\paragraph{Formulas for coherent $(D^0 \bar{D}^0)^{}_{ C = \pm 1}$ decays}

On the $\psi (3770)$ or $\psi (4140)$ resonance, which possess the quantum
number $J^{\rm PC} = 1^{- -}$, a $C$-odd or $C$-even $D^0 \bar{D}^0$ pair can
be coherently produced as follows~\cite{Bigi:1989ah,Fry:1993hms}:
\begin{eqnarray}
\psi (3770) \hspace{-0.2cm} & \longrightarrow & \hspace{-0.2cm}
\left(D^0\bar{D}^0\right)_{\rm C = -1} \; ,
\nonumber \\
\psi (4140) \hspace{-0.2cm} & \longrightarrow & \hspace{-0.2cm}
D^0 \bar{D}^{* 0} \hspace{0.2cm} {\rm or} \hspace{0.2cm} D^{* 0} \bar{D}^0
\longrightarrow \left(D^0\bar{D}^0\right)_{\rm C = -1} + \pi^0 \; ,
\nonumber \\
\psi (4140) \hspace{-0.2cm} & \longrightarrow & \hspace{-0.2cm}
D^0 \bar{D}^{* 0} \hspace{0.2cm} {\rm or} \hspace{0.2cm} D^{* 0} \bar{D}^0
\longrightarrow \left(D^0\bar{D}^0\right)_{\rm C = +1} + \gamma \; .
\hspace{0.5cm}
\label{eq:CP-coherent}
%     (20)
\end{eqnarray}
For such a coherent pair at rest, its time-dependent wave function can be
written as
\begin{eqnarray}
|D^0\bar{D}^0(t)\rangle^{}_{C} 
=
\frac{1}{\sqrt 2} \left[|D^0 ({\bf K}, t)\rangle
%\otimes 
|\bar{D}^0 (-{\bf K}, t)\rangle +
 {C} |D^0 (-{\bf K}, t)\rangle 
%\otimes 
 |\bar{D}^0 ({\bf K}, t)\rangle
\right],
\label{eq:DDpair-wave-function}
%     (21)
\end{eqnarray}
where $\pm \bf K$ stand for the three-momentum vectors of the two neutral
$D$-mesons, and ${ C} = \pm 1$ denotes the even or odd
charge-conjugation parity of this coherent system. The joint decay mode
\begin{eqnarray}
\left(D^0\bar{D}^0\right)_{\CP \pm }
\longrightarrow \left(f_1 f_2\right)_{\CP\mp} \; ,
\label{eq:CP-forbidden}
%     (22)
\end{eqnarray}
where $f^{}_{1,2}$ stand for the proper hadronic $\CP$ eigenstates
($\eg$, $\pi^+\pi^-$, $K^+K^-$ and $K^{}_{\rm S} \pi^0$), must
be a $\CP$-forbidden process. In other words, such a decay cannot take place
unless there exists $\CP$ violation in the $D^0 \to f^{}_{1,2}$ and
$\bar{D}^0 \to f^{}_{1,2}$ decays. As $\CP$-odd and $\CP$-even $D^0 \bar{D}^0$ pairs are expected to be
produced in a huge amount at $\STCF$, it will
be important to measure the $\CP$-forbidden $D^0\bar{D}^0$ decays
on the $\psi (3770)$ and $\psi (4140)$ resonances, although such
a measurement is very challenging in practice~\cite{Xing:2019uzz}.

One may also start from
Eq.~(\ref{eq:DDpair-wave-function}) and consider that one of the two neutral
$D$ mesons (with momentum $\bf K$) decays into a final state $f^{}_1$ at
%proper time %$t^{}_1$ 
and the other (with $-{\bf K}$) to $f^{}_2$, and therefore derive the time-independent joint decay rate as
\begin{eqnarray}
\Gamma(f^{}_1; f^{}_2)^{}_{C}
\hspace{-0.2cm} & \propto & \hspace{-0.2cm}
|A(D^0 \to f^{}_1)|^2 \hspace{0.05cm} |A(D^0 \to f^{}_2)|^2
\nonumber\\
\hspace{-0.2cm} & & \hspace{-0.2cm}
\times\left[\frac{1 + {C} y^2}{\left(1 - y^2\right)^2}
\left(|\xi^{}_{ C}|^2 + |\zeta^{}_{ C}|^2\right)
+ \frac{2\left(1 + { C}\right) y}{\left(1 - y^2\right)^2}
{\rm Re}\left(\xi^*_{ C} \zeta^{}_{ C}\right) \right.
\nonumber \\
\hspace{-0.2cm} & & \hspace{0.2cm}
- \left. \frac{1 - {C} x^2}{\left(1 + x^2\right)^2}
\left(|\xi^{}_{C}|^2 - |\zeta^{}_{ C}|^2\right)
+ \frac{2 \left(1 + {C}\right) x}{\left(1 + x^2\right)^2}
{\rm Im}\left(\xi^*_{ C} \zeta^{}_{ C}\right) \right] \; , \hspace{0.5cm}
\label{eq:D-joint-decay-rate2}
%     (33)
\end{eqnarray}
where
\begin{eqnarray}
\xi^{}_{ C} \hspace{-0.2cm} & \equiv & \hspace{-0.2cm}
\frac{p}{q} \left(1 + { C} \lambda^{}_{f^{}_1}
\lambda^{}_{f^{}_2}\right) \; ,
\nonumber \\
\zeta^{}_{ C} \hspace{-0.2cm} & \equiv & \hspace{-0.2cm}
\frac{p}{q} \left(\lambda^{}_{f^{}_2} + { C}
\lambda^{}_{f^{}_1} \right) \; , \hspace{0.5cm}
\label{eq:D-joint-decay-amplitude2}
%     (31)
\end{eqnarray}

%and the definition of $\lambda^{}_{f^{}_{1,2}}$ is the same as that
%made in Eq.~(\ref{eq:Indirect-CPV}).

It is obvious that the two interference terms,
${\rm Re}\left(\xi^*_{ C} \zeta^{}_{ C}\right)$ and
${\rm Im}\left(\xi^*_{ C} \zeta^{}_{ C}\right)$, disappear in the
${ C} = -1$ case, no matter what the final states $f^{}_1$ and
$f^{}_2$ are. Of course, the joint decay rates of
$\Gamma(f^{}_1; \bar{f}^{}_2)^{}_{ C}$,
$\Gamma(\bar{f}^{}_1; f^{}_2)^{}_{ C}$ and
$\Gamma(\bar{f}^{}_1; \bar{f}^{}_2)^{}_{ C}$ can similarly be
calculated, where $\bar{f}^{}_{1,2}$ denote the $\CP$-conjugate states
of $f^{}_{1,2}$.

Taking into account the approximation $|p/q|^2 \simeq 1 + \Delta^{}_{\rm mixing}$
and neglecting the terms of ${\cal O}(x^2)$, ${\cal O}(y^2)$ and smaller
in Eq.~(\ref{eq:D-joint-decay-rate2}), we find
\begin{eqnarray}
\Gamma(f^{}_1; f^{}_2)^{}_{C = -1}
\hspace{-0.2cm} & \propto & \hspace{-0.2cm}
2 |A(D^0 \to f^{}_1)|^2 \hspace{0.05cm} |A(D^0 \to f^{}_2)|^2
\nonumber\\
\hspace{-0.2cm} & & \hspace{-0.2cm}
\times \left[ |\lambda^{}_{f^{}_1}|^2 + |\lambda^{}_{f^{}_2}|^2
- 2 \left({\rm Re}\lambda^{}_{f^{}_1} {\rm Re}\lambda^{}_{f^{}_2}
+ {\rm Im}\lambda^{}_{f^{}_1} {\rm Im}\lambda^{}_{f^{}_2}\right)\right]
\nonumber \\
\hspace{-0.2cm} & & \hspace{-0.2cm}
\times
\left( 1 + \Delta^{}_{\rm mixing}\right) \; ,
\nonumber \\
\Gamma(f^{}_1; f^{}_2)^{}_{ C = +1}
\hspace{-0.2cm} & \propto & \hspace{-0.2cm}
2 |A(D^0 \to f^{}_1)|^2 \hspace{0.05cm} |A(D^0 \to f^{}_2)|^2
\nonumber \\
\hspace{-0.2cm} & & \hspace{-0.2cm}
\times
\bigg\{ |\lambda^{}_{f^{}_1}|^2 + |\lambda^{}_{f^{}_2}|^2
+ 2 \left({\rm Re}\lambda^{}_{f^{}_1} {\rm Re}\lambda^{}_{f^{}_2}
+ {\rm Im}\lambda^{}_{f^{}_1} {\rm Im}\lambda^{}_{f^{}_2}\right)
\nonumber \\
 & & 
 + \left. 2 y \left[ \left(1 + |\lambda^{}_{f^{}_1}|^2\right)
{\rm Re}\lambda^{}_{f^{}_2} + \left(1 + |\lambda^{}_{f^{}_2}|^2\right)
{\rm Re}\lambda^{}_{f^{}_1}\right] \right.
\nonumber \\
& & 
+ \left. 2 x \left[ \left(1 - |\lambda^{}_{f^{}_1}|^2\right)
{\rm Im}\lambda^{}_{f^{}_2} + \left(1 - |\lambda^{}_{f^{}_2}|^2\right)
{\rm Im}\lambda^{}_{f^{}_1}\right] \right. \bigg\}
\nonumber \\
 & & 
\times
\left( 1 + \Delta^{}_{\rm mixing}\right) \; . \hspace{0.5cm}
\label{eq:D-joint-decay-rate3}
\end{eqnarray}
The above formulas can be further simplified if $f^{}_1 = f^{}_2$ is
taken. Here a key observation is that a combination of the measurements of
$\Gamma(f^{}_1; f^{}_2)^{}_{ C = -1}$ and
$\Gamma(f^{}_1; f^{}_2)^{}_{ C = +1}$ at $\STCF$ will
allow us to extract much more information about $D^0$-$\bar{D}^0$ mixing and
$\CP$ violation in neutral $D$-meson decays, although the events of coherent
$D^0$ and $\bar{D}^0$ mesons on the $\psi(4140)$ resonance are much fewer
than those on the $\psi(3770)$ resonance.

The decays of $D^0 \to K^+\pi^-$ and $K^-\pi^+$ are of particular
interest for the study of both $D^0$-$\bar{D}^0$ mixing and $\CP$ violation.
The reason is simply that both $D^0$ and $\bar{D}^0$ can decay into
the final states $K^\pm \pi^\mp$ via just the tree-level quark diagrams,
assuring that their amplitudes are hardly contaminated by any kind of new
physics~\cite{Blaylock:1995ay}. Let us factorize the transition
amplitudes of $D^0 \to K^\pm \pi^\mp$ and $\bar{D}^0 \to K^\pm \pi^\mp$
as~\cite{Xing:1996pn}:
\begin{eqnarray}
A(D^0 \to K^-\pi^+) \hspace{-0.2cm} & = & \hspace{-0.2cm}
\left(V^{}_{cs} V^*_{ud}\right) T^{}_a e^{{\rm i}\delta^{}_a} \; ,
\nonumber \\
A(D^0 \to K^+\pi^-) \hspace{-0.2cm} & = & \hspace{-0.2cm}
\left(V^{}_{cd} V^*_{us}\right) T^{}_b e^{{\rm i}\delta^{}_b} \; ,
\nonumber \\
A(\bar{D}^0 \to K^+\pi^-) \hspace{-0.2cm} & = & \hspace{-0.2cm}
\left(V^{*}_{cs} V^{}_{ud}\right) T^{}_a e^{{\rm i}\delta^{}_a} \; ,
\nonumber \\
A(\bar{D}^0 \to K^-\pi^+) \hspace{-0.2cm} & = & \hspace{-0.2cm}
\left(V^{*}_{cd} V^{}_{us}\right) T^{}_b e^{{\rm i}\delta^{}_b} \; , \hspace{0.5cm}
\label{eq:Kpi-amplitude}
%     (35)
\end{eqnarray}
where $T^{}_{a,b}$ denote the real hadronic matrix elements, and
$\delta^{}_{a,b}$ are the corresponding strong phases. Defining
$h^{}_{K\pi} \equiv T^{}_b/T^{}_a$ and $\delta^{}_{K\pi} \equiv
\delta^{}_b - \delta^{}_a$ for the decay modes under consideration
and $q/p \equiv |q/p| e^{{\rm i}\phi}$ for the weak phase of
$D^0$-$\bar{D}^0$ mixing, we obtain
\begin{eqnarray}
\lambda^{}_{K^-\pi^+} \hspace{-0.2cm} & = & \hspace{-0.2cm}
\frac{q}{p} \cdot \frac{A(\bar{D}^0 \to K^-\pi^+)}
{A(D^0 \to K^-\pi^+)} \simeq -\lambda^2 h^{}_{K\pi} \left|\frac{q}{p}\right|
e^{{\rm i}\left(\delta^{}_{K\pi} + \phi\right)} \; ,
\nonumber \\
\bar{\lambda}^{}_{K^+\pi^-} \hspace{-0.2cm} & = & \hspace{-0.2cm}
\frac{p}{q} \cdot \frac{A(D^0 \to K^+\pi^-)}
{A(\bar{D}^0 \to K^+\pi^-)} \simeq -\lambda^2 h^{}_{K\pi} \left|\frac{p}{q}\right|
e^{{\rm i}\left(\delta^{}_{K\pi} - \phi\right)} \; , \hspace{0.5cm}
\label{eq:Kpi-amplitude2}
%     (36)
\end{eqnarray}
for $f = K^-\pi^+$ and $\bar{f} = K^+\pi^-$, where the standard phase convention
for the $\CKM$ matrix elements taken in %Eqs.~(\ref{eq:V})
%and (\ref{eq:Wolf})
Eq.~(\ref{eq_Wolfenstein}) (namely, $V^{}_{ud} \simeq V^{}_{cs} \simeq 1$ and
$V^{}_{cd} \simeq -V^{}_{us} \simeq -\lambda$) has been used. The magnitude of $h^{}_{K\pi}$ is expected to be of ${\cal O}(1)$, but 
an explicit calculation of this quantity involves some uncertainties 
arising from the relevant hadronic matrix elements. 

On the $\psi(3770)$ resonance, one may easily obtain the time-independent joint
decay rates
\begin{eqnarray}
\Gamma(K^-\pi^+; K^-\pi^+)^{}_{C = -1} \hspace{-0.2cm} & \propto & \hspace{-0.2cm}
2 |A(D^0 \to K^-\pi^+)|^4 r \left|\frac{p}{q}\right|^2 \; ,
\nonumber \\
\Gamma(K^+\pi^-; K^+\pi^-)^{}_{C = -1} \hspace{-0.2cm} & \propto & \hspace{-0.2cm}
2 |A(D^0 \to K^-\pi^+)|^4 r \left|\frac{q}{p}\right|^2 \; , \hspace{0.5cm}
\label{eq:Kpi-coherent}
%     (39)
\end{eqnarray}
which are highly suppressed due to the smallness of $r \equiv (x^2 + y^2)/2$.
On the $\psi(4140)$ resonance, the corresponding joint decay rates turn
out to be~\cite{Xing:1996pn}
\begin{eqnarray}
\Gamma(K^-\pi^+; K^-\pi^+)^{}_{C = +1} \hspace{-0.2cm} & \propto &\hspace{-0.2cm}
2 |A(D^0 \to K^-\pi^+)|^4 \left\{ 3 r \left|\frac{p}{q}\right|^2 + 4 \lambda^4
h^2_{K\pi} \right.
\nonumber \\
\hspace{-0.2cm} & & \hspace{-0.2cm}
- \left. 4 \lambda^2 h^{}_{K\pi} \left|\frac{p}{q}\right| \left[
y \cos(\delta^{}_{K\pi} + \phi) + x \sin(\delta^{}_{K\pi} + \phi)
\right] \right\} \; ,
\nonumber \\
\Gamma(K^+\pi^-; K^+\pi^-)^{}_{C = +1} \hspace{-0.2cm} & \propto & \hspace{-0.2cm}
2 |A(\bar{D}^0 \to K^+\pi^-)|^4 \left\{ 3 r \left|\frac{q}{p}\right|^2 + 4 \lambda^4
h^2_{K\pi} \right.
\nonumber \\
\hspace{-0.2cm} & & \hspace{-0.2cm}
- \left. 4 \lambda^2 h^{}_{K\pi} \left|\frac{q}{p}\right| \left[
y \cos(\delta^{}_{K\pi} - \phi) + x \sin(\delta^{}_{K\pi} - \phi)
\right] \right\}  ,\;\;\;\; \hspace{0.5cm}
\label{eq:Kpi-coherent2}
%     (40)
\end{eqnarray}
as a good approximation. The difference between these two joint decay rates
is also a direct measure of $D^0$-$\bar{D}^0$ mixing and $\CP$ violation.

It is well known that the above decay modes have played an important role
in the experimental constraints of $D^0$-$\bar{D}^0$ mixing, but so
far no $\CP$ violation has been observed in them~\cite{ParticleDataGroup:2022pth}.

\paragraph{$\CP$ violation in $D^0 \to \pi^+\pi^-$ and $K^+K^-$ decays}

Both $f^{}_d \equiv \pi^+\pi^-$ and $f^{}_s \equiv K^+K^-$ are the $\CP$-even
eigenstates, as $\bar{f}^{}_q = f^{}_q$ holds in either case (for $q = d$ or
$s$). In particular, the decay modes $D^0 \to \pi^+\pi^-$
and $D^0 \to K^+K^-$ satisfy the so-called U-spin symmetry --- a kind of
invariance under the $d \leftrightarrow s$ interchange, as shown in Fig.~\ref{U-spin}.
For such a decay mode, one may simply define
%%%%%%%%%%%%%%%%%%%%%%%%%%%%%%%%%%%%%%%%%%%%%%%%%%%%%%%%%%%%%%%%%%%%%%%
%\footnote{Note that ${\cal U}^{}_f$ is equivalent to the direct $\CP$-violating asymmetry ${\cal A}^{}_f$ defined in Eq.~(\ref{eq:Direct-CPV2}) in the approximation of $|q/p| \simeq 1$, and hence it is essentially a measure  of the strength of direct $\CP$ violation.}
%%%%%%%%%%%%%%%%%%%%%%%%%%%%%%%%%%%%%%%%%%%%%%%%%%%%%%%%%%%%%%%%%%%%%%%
\begin{eqnarray}
{\cal U}^{}_f \equiv \frac{1 - |\lambda^{}_f|^2}{1 + |\lambda^{}_f|^2} \; , \quad
{\cal V}^{}_f \equiv \frac{-2 {\rm Im}\lambda^{}_f}{1 + |\lambda^{}_f|^2} \; , \quad
{\cal W}^{}_f \equiv \frac{2 {\rm Re}\lambda^{}_f}{1 + |\lambda^{}_f|^2} \; ,
\label{eq:CP-eigenstate}
%     (41)
\end{eqnarray}
which satisfy the sum rule ${\cal U}^2_f + {\cal V}^2_f + {\cal W}^2_f = 1$~\cite{Xing:1996pn}.
It is obvious that ${\cal U}^{}_f$ and ${\cal V}^{}_f$ correspond respectively to
direct and indirect $\CP$-violating effects, while ${\cal W}^{}_f$ is a $\CP$-conserving
quantity. %With the help of Eq.~(\ref{eq:CP-eigenstate}), the decay rates in
%Eq.~(\ref{eq:D-decay-rate}) can be rewritten as
%\begin{eqnarray}
%\Gamma(D^0(t) \to f) \hspace{-0.2cm} & \propto & \hspace{-0.2cm}
%|\tilde{A}(D^0 \to f)|^2
%e^{-\Gamma t} \left[ 1 + {\cal U}^{}_f + \left({\cal V}^{}_f x +
%{\cal W}^{}_f y\right) \Gamma t\right] \; ,
%\nonumber \\
%\Gamma(\bar{D}^0(t) \to f) \hspace{-0.2cm} & \propto & \hspace{-0.2cm}
%|\tilde{A}(D^0 \to f)|^2
%e^{-\Gamma t} \left[ 1 - {\cal U}^{}_f - \left({\cal V}^{}_f x -
%{\cal W}^{}_f y\right) \Gamma t\right] 
%\nonumber\\
%&&\hspace{3cm}\times\left(1 + \Delta^{}_{\rm mixing}\right)
%\; , \hspace{0.5cm}
%\label{eq:D-decay-CPE}
%     (42)
%\end{eqnarray}
%to the accuracy of ${\cal O}(x)$ and ${\cal O}(y)$, where
%$|\tilde{A}(D^0 \to f)|^2 \equiv 2|A(D^0 \to f)|^2(1 + |\lambda^{}_f|^2)$
%has been defined for the sake of simplicity. Given tha $\Delta^{}_{\rm mixing}
%\simeq 0$ in the SM, the difference between these two decay rates provides
%a straightforward way to extract ${\cal U}^{}_f$ and ${\cal V}^{}_f$.
%Unfortunately, the term proportional to ${\cal V}^{}_f$ is suppressed by
%the smallness of $x$. That is why only direct $\CP$ violation has so far been
%determined from a combination of $D^0 \to \pi^+\pi^-$ and
%$D^0 \to K^+K^-$ decays~\cite{LHCb:2019hro}.
%%%%%%%%%%%%%%%%%%%%%%%%%%%% Figure 5 

It is naturally expected that the joint decays of a pair of coherent
$D^0$ and $\bar{D}^0$ mesons into $(K^+K^-)(\pi^+\pi^-)$ on the $\psi(3770)$
resonance cannot happen unless there exists $\CP$ violation in them. Taking into account
Eqs.~(\ref{eq:D-joint-decay-rate3}) and (\ref{eq:CP-eigenstate}), we find
\begin{eqnarray}
\Gamma(K^+K^-; \pi^+\pi^-)^{}_{C=-1} \hspace{-0.2cm} & \propto & \hspace{-0.2cm}
|\tilde{A}(D^0 \to K^+K^-)|^2 |\tilde{A}(D^0 \to \pi^+\pi^-)|^2
\left|\frac{p}{q}\right|^2 
\nonumber \\
\hspace{-0.3cm} & & \hspace{-0.3cm}
\times\left(\frac{1 - {\cal W}^{}_{KK} {\cal W}^{}_{\pi\pi}}{1 - y^2} \right.- \left. \frac{{\cal U}^{}_{KK} {\cal U}^{}_{\pi\pi}
+ {\cal V}^{}_{KK} {\cal V}^{}_{\pi\pi}}{1 + x^2}\right)
\; ,\;\;\;\;\;\;
\label{eq:CPE-Kpi}
%     (46)
\end{eqnarray}
as well as 
\begin{eqnarray}
\Gamma(K^+K^-; K^+K^-)^{}_{C=-1} \hspace{-0.2cm} & \propto & \hspace{-0.2cm}
|\tilde{A}(D^0 \to K^+K^-)|^4 \left|\frac{p}{q}\right|^2
\nonumber\\
&&
\times\left(\frac{1}{1 - y^2} - \frac{1}{1 + x^2}\right)
\left({\cal U}^{2}_{KK} + {\cal V}^{2}_{KK}\right)
 \; , \hspace{0.5cm}
\nonumber \\
\Gamma(\pi^+\pi^-; \pi^+\pi^-)^{}_{ C=-1} \hspace{-0.2cm} & \propto & \hspace{-0.2cm}
|\tilde{A}(D^0 \to \pi^+\pi^-)|^4 \left|\frac{p}{q}\right|^2
\nonumber\\
&&
\times\left(\frac{1}{1 - y^2} - \frac{1}{1 + x^2}\right)
\left({\cal U}^{2}_{\pi\pi} + {\cal V}^{2}_{\pi\pi}\right)
 \; .
\label{eq:CPE-KK+pipi}
%     (47)
\end{eqnarray}
Note that the $D^0$-$\bar{D}^0$ mixing parameters $x$ and $y$ play an important role
in Eq.~(\ref{eq:CPE-KK+pipi}), simply because $1/(1 - y^2) - 1/(1 + x^2) 
\simeq x^2 + y^2$ holds. 
In practice, it will be extremely challenging to observe such a suppressed
signal of $\CP$ violation on the $\psi(3770)$ resonance even in $\STCF$~\cite{Xing:2019uzz}. But we remark that this kind of $\CP$
violation is conceptually interesting and deserves our special attention and 
penetrating search, as it is simply a rate rather than a conventional $\CP$-violating 
asymmetry.

\paragraph{$\CP$ violation in $D^0 \to K^{* +}K^-$ and $K^+K^{* -}$ decays}

In the SM the singly Cabibbo-suppressed decays $D^0 \to K^{* \pm} K^\mp$
can take place through both the tree-level and penguin diagrams. The former
is essentially $\CP$-conserving because it is proportional to the $\CKM$ factor
$V^{}_{cs} V^*_{us}$, and the latter may not be very small with respect to
the former as can be seen in the $D^0 \to K^+K^-$ case discussed above. 
But for the sake of simplicity and illustration, here we assume that the 
tree-level contribution to $D^0 \to K^{* \pm} K^\mp$ decays 
is dominant. In this approximation, we have~\cite{Xing:2007sd}
\begin{eqnarray}
\lambda^{}_{K^{*+}K^-} \hspace{-0.2cm} & = & \hspace{-0.2cm}
\frac{q}{p} \cdot \frac{A(\bar{D}^0 \to K^{*+}K^-)}
{A(D^0 \to K^{*+}K^-)} \simeq h^{}_{K^*K} \left|\frac{q}{p}\right|
e^{{\rm i}\left(\delta^{}_{K^*K} + \phi\right)} \; ,
\nonumber \\
\bar{\lambda}^{}_{K^{*-}K^+} \hspace{-0.2cm} & = & \hspace{-0.2cm}
\frac{p}{q} \cdot \frac{A(D^0 \to K^{*-}K^+)}
{A(\bar{D}^0 \to K^{*-}K^+)} \simeq h^{}_{K^*K} \left|\frac{p}{q}\right|
e^{{\rm i}\left(\delta^{}_{K^*K} - \phi\right)} \; , \hspace{0.5cm}
\label{eq:K*K-amplitude}
%     (48)
\end{eqnarray}
where $h^{}_{K^*K}$ is real and of ${\cal O}(1)$, $\delta^{}_{K^*K}$ denotes
the relevant strong phase difference, and the standard phase convention
$\arg(V^{}_{cs} V^*_{us}) \simeq 0$ has been taken as an excellent approximation.
%With the help of Eq.~(\ref{eq:D-decay-rate}), we obtain the time-dependent
%decay rates of $D^0 \to K^{*\pm}K^\mp$ and $\bar{D}^0 \to K^{*\pm}K^\mp$ as
%follows:
%\begin{eqnarray}
%\Gamma (D^0(t) \to K^{*+}K^-) \hspace{-0.2cm} & \propto & \hspace{-0.2cm}
%|A(D^0 \to K^{*+}K^-)|^2 e^{-\Gamma t} 
%\nonumber\\
%&&
%\times \left [ 1 + h^{}_{K^*K} \left
%|\frac{q}{p} \right | \left (y^\prime_- \cos\phi - x^\prime_+
%\sin\phi \right ) \Gamma t \right ] \; ,
%\nonumber \\
%\Gamma (\bar{D}^0(t) \to K^{*-}K^+) \hspace{-0.2cm} & \propto & \hspace{-0.2cm}
%|A(\bar{D}^0 \to K^{*-}K^+)|^2 e^{-\Gamma t} 
%\nonumber\\
%&&
%\times \left [ 1 + h^{}_{K^*K} \left
%|\frac{p}{q} \right | \left (y^\prime_- \cos\phi + x^\prime_+
%\sin\phi \right ) \Gamma t \right ] \; ; \hspace{0.5cm}
%\label{eq:K*K-rate}
%       (49)
%\end{eqnarray}
%and
%\begin{eqnarray}
%\Gamma (D^0(t) \to K^{*-}K^+) \hspace{-0.2cm} & \propto & \hspace{-0.2cm}
%|A (\bar{D}^0 \to K^{*-}K^+)|^2 e^{-\Gamma t} 
%\nonumber\\
%&&
%\times \left [ h^2_{K^*K} + h^{}_{K^*K}
%\left |\frac{q}{p} \right | \left (y^\prime_+ \cos\phi -
%x^\prime_- \sin\phi \right ) \Gamma t \right ] \; ,
%\nonumber \\
%\Gamma (\bar{D}^0(t) \to K^{*+}K^-) \hspace{-0.2cm} & \propto & \hspace{-0.2cm}
%|A (D^0 \to K^{*+}K^-)|^2 e^{-\Gamma t} 
%\nonumber\\
%&&
%\times \left [ h^2_{K^*K} + h^{}_{K^*K} \left
%|\frac{p}{q} \right | \left (y^\prime_+ \cos\phi + x^\prime_-
%\sin\phi \right ) \Gamma t \right ] \; , \;\;\;\;\hspace{0.5cm}
%\label{eq:K*K-rate2}
%       (50)
%\end{eqnarray}
where only the terms of ${\cal O}(x)$ and ${\cal O}(y)$ are kept, 
and the so-called ``strong-phase-rotated" $D^0$-$\bar{D}^0$
mixing parameters $x^\prime_{\pm}$ and $y^\prime_{\pm}$ are defined as
\begin{eqnarray}
x^\prime_{\pm} \hspace{-0.2cm} & = & \hspace{-0.2cm}
x \cos\delta^{}_{K^*K} \pm y \sin\delta^{}_{K^*K} \; ,
\nonumber \\
y^\prime_{\pm} \hspace{-0.2cm} & = & \hspace{-0.2cm}
y \cos\delta^{}_{K^*K} \pm x \sin\delta^{}_{K^*K} \; . \hspace{0.5cm}
\label{eq:K*K-mixing}
%       (51)
\end{eqnarray}
So it is possible to determine $h^{}_{K^*K}$ and constrain the magnitudes
of both $D^0$-$\bar{D}^0$ mixing and $\CP$ violation from the measurements
of these four decay modes. In particular, a remarkable difference between
$y^\prime_+$ and $y^\prime_-$ will imply that both $x$ and
$\delta^{}_{K^*K}$ should not be very small~\cite{Grossman:2006jg}.

On the $\psi(3770)$ resonance, where a pair of $D^0$ and
$\bar{D}^0$ mesons with $C = -1$ can be coherently produced, their
joint decays into the final states $(K^{*\pm}K^\mp)(K^{*\pm}K^\mp)$ have the
following rates to the accuracy of ${\cal O}(x^2)$ and
${\cal O}(y^2)$~\cite{Xing:2007sd}:
\begin{eqnarray}
\Gamma(K^{*+}K^-; K^{*+}K^-)^{}_{C = -1} \hspace{-0.2cm} & \propto & \hspace{-0.2cm}
2|A(D^0 \to K^{*+}K^-)|^4 r 
\nonumber\\
&&\hspace{-0.3cm}
\times \left [ \left | \frac{p}{q} \right |^2 - 2 h^2_{K^*K}
\cos \left (\delta^{}_{K^*K} + \phi \right ) + h^4_{K^*K}
\left | \frac{q}{p} \right |^2 \right ] \; ,
\nonumber \\
\Gamma(K^{*-}K^+; K^{*-}K^+)^{}_{C = -1} \hspace{-0.2cm} & \propto & \hspace{-0.2cm}
2|A(\bar{D}^0 \to K^{*-}K^+)|^4 r 
\nonumber\\
&&\hspace{-0.3cm}
\times \left [ \left | \frac{q}{p} \right |^2 - 2 h^2_{K^*K}
\cos \left (\delta^{}_{K^*K} - \phi \right ) + h^4_{K^*K}
\left | \frac{p}{q} \right |^2 \right ] \; , \;\;\;\;\hspace{0.5cm}
\label{eq:K*K-coherent}
%       (52)
\end{eqnarray}
where $r \equiv (x^2 + y^2)/2$ as defined below Eq.~(\ref{eq:Kpi-coherent}). On the
$\psi(4140)$ resonance, with $ C = +1$ for the coherent $D^0\bar{D}^0$ pairs,
the rates of their joint decays into $(K^{*\pm}K^\mp)(K^{*\pm}K^\mp)$ are given
as follows to the accuracy of ${\cal O}(x)$ and
${\cal O}(y)$~\cite{Xing:2007sd}:
\begin{eqnarray}
\Gamma(K^{*+}K^-; K^{*+}K^-)^{}_{C = +1} \hspace{-0.2cm} & \propto & \hspace{-0.2cm}
8|A(D^0 \to K^{*+}K^-)|^4 h^{}_{K^*K} 
\nonumber\\
&&\hspace{-0.3cm}
\times
\left [ h^{}_{K^*K} +
\left|\frac{p}{q} \right| \left(y^\prime_+ \cos\phi + x^\prime_- \sin\phi\right)
\right.
\nonumber \\
\hspace{0.2cm} & & \hspace{0.2cm}
+ \left. h^2_{K^*K} \left|\frac{q}{p}\right|
\left(y^\prime_- \cos\phi - x^\prime_+ \sin\phi\right) \right ] \; ,
\nonumber \\
\Gamma(K^{*-}K^+; K^{*-}K^+)^{}_{C = +1} \hspace{-0.2cm} & \propto & \hspace{-0.2cm}
8|A(\bar{D}^0 \to K^{*-}K^+)|^4 h^{}_{K^*K} 
\nonumber\\
&&\hspace{-0.3cm}
\times\left [ h^{}_{K^*K} +
\left|\frac{q}{p} \right| \left(y^\prime_+ \cos\phi - x^\prime_- \sin\phi\right)
\right. \hspace{0.5cm}
\nonumber \\
\hspace{0.2cm} & & \hspace{0.2cm}
+ \left. h^2_{K^*K} \left|\frac{p}{q}\right|
\left(y^\prime_- \cos\phi + x^\prime_+ \sin\phi\right) \right ] \; ,
\label{eq:K*K-coherent2}
%       (53)
\end{eqnarray}
%where $x^\prime_\pm$ and $y^\prime_\pm$ have been defined in Eq.~(\ref{eq:K*K-mixing}).
where only the terms of ${\cal O}(x)$ and ${\cal O}(y)$ are kept, 
and the so-called ``strong-phase-rotated" $D^0$-$\bar{D}^0$
mixing parameters $x^\prime_{\pm}$ and $y^\prime_{\pm}$ are defined as
\begin{eqnarray}
x^\prime_{\pm} \hspace{-0.2cm} & = & \hspace{-0.2cm}
x \cos\delta^{}_{K^*K} \pm y \sin\delta^{}_{K^*K} \; ,
\nonumber \\
y^\prime_{\pm} \hspace{-0.2cm} & = & \hspace{-0.2cm}
y \cos\delta^{}_{K^*K} \pm x \sin\delta^{}_{K^*K} \; . \hspace{0.5cm}
\label{eq:K*K-mixing}
%       (51)
\end{eqnarray}
So it is possible to determine $h^{}_{K^*K}$ and constrain the magnitudes
of both $D^0$-$\bar{D}^0$ mixing and $\CP$ violation from the measurements
of these four decays. In particular, a remarkable difference between
$y^\prime_+$ and $y^\prime_-$ will imply that both $x$ and
$\delta^{}_{K^*K}$ should not be very small~\cite{Grossman:2006jg}.
A nonzero difference between the two joint decay rates in either Eq.~(\ref{eq:K*K-coherent})
or Eq.~(\ref{eq:K*K-coherent2}) is a clear signal of $D^0$-$\bar{D}^0$ mixing and $\CP$
violation.

It is worth emphasizing that studying the $K^{*\pm}K^\mp$ events of neutral $D$-meson
decays are important as they can be complementary to the
$K^\pm\pi^\mp$ and $K^+K^-$ (or $\pi^+\pi^-$) events for the
experimental searches for both $D^0$-$\bar{D}^0$ mixing and $\CP$
violation. A quite similar idea, which utilizes the $D^{*\pm}D^\mp$
events of neutral $B$-meson decays to probe $\CP$ violation
and test the factorization hypothesis~\cite{Xing:1998ca},
has already been applied to the Belle \cite{Belle:2002pgi} and BaBar
\cite{BaBar:2005xvn} experiments.

\subsection{$\CP$ violation from the interplay between decay and mixing}

As for the neutral $D$-meson decays, $\CP$ violation may arise from the
"double-slit" interference effect of the form $D^0 \to \bar{D}^0
\to f$ versus $D^0 \to f$ (or $\bar{D}^0 \to D^0 \to \bar{f}$ versus
$\bar{D}^0 \to \bar{f}$)~\cite{Bigi:1986dp,Du:1986ai,Xing:1996pn,
Asner:2005wf,Du:2006jc,Bigi:2007zz}, where the final hadronic state $f$
and its $\CP$-conjugated state $\bar{f}$ are unnecessarily the $\CP$ eigenstates.
In this case, the interplay between the direct decay and $D^0$-$\bar{D}^0$
mixing can be described by the following rephasing-invariant quantities $\lambda^{}_f$ and $\bar{\lambda}^{}_{\bar f}$ as defined in Eq.~\ref{eq:Indirect-CPV}.
%\begin{eqnarray}
%\lambda^{}_f \hspace{-0.2cm} & \equiv & \hspace{-0.2cm}
%\frac{q}{p} \cdot \frac{A(\bar{D}^0 \to f)}{A(D^0 \to f)} \; ,
%\nonumber \\
%\bar{\lambda}^{}_{\bar f} \hspace{-0.2cm} & \equiv & \hspace{-0.2cm}
%\frac{p}{q} \cdot \frac{A(D^0 \to \bar{f})}{A(\bar{D}^0 \to \bar{f})} \; ,
%\hspace{0.5cm}
%\label{eq:Indirect-CPV}
%     (15)
%\end{eqnarray}
%where $f$ and $\bar{f}$ are required to be common to the decay of
%$D^0$ (or $\bar{D}^0$). 
Then the difference
\begin{eqnarray}
{\rm Im} \lambda^{}_f - {\rm Im} \bar{\lambda}^{}_{\bar f} \neq 0 \;
\label{eq:Indirect-CPV2}
%     (16)
\end{eqnarray}
characterizes the effects of {\it indirect} $\CP$ violation, which
may include both the contribution of $\Delta^{}_{\rm mixing}$ in
Eq.~(\ref{eq:CPV-mixing}) and that from direct $\CP$ violation in
Eq.~(\ref{eq:Direct-CPV2}).

Note that $D^0$-$\bar{D}^0$ mixing affects ${\rm Im} \lambda^{}_f$
and ${\rm Im}\bar{\lambda}^{}_{\bar f}$ even in the special case of
$\Delta^{}_{\rm mixing} = 0$ ($\ie$, $|q/p| = 1$), if $q/p$ has a
nontrivial complex phase. In this case $\bar{\lambda}^{}_{\bar f}
= \lambda^*_f$ will hold if $D^0 \to f$ and $\bar{D}^0 \to f$
are both dominated by a single amplitude component and
$\bar{f} = f$ is a hadronic $\CP$ eigenstate ($\eg$, $K^+K^-$ and
$\pi^+\pi^-$). Such a {\it purely indirect} $\CP$-violating effect,
without the contamination from either $\CP$ violation in $D^0$-$\bar{D}^0$
mixing or direct $\CP$ violation in the direct decays of $D^0$ and
$\bar{D}^0$ mesons, is of great interest in the charm sector.
A typical example of this kind in the bottom sector is $\CP$ violation
in $B^0_d \to \bar{B}^0_d \to J/\psi K^{}_{\rm S}$ versus
$B^0_d \to J/\psi K^{}_{\rm S}$ decays, which has played an important
role in testing the KM mechanism of $\CP$ violation in
the SM.

%\subsection{$\CP$ violation in the $\CP$-forbidden coherent
%$D^0 \bar{D}^0$ decays}

\subsection{$\CP$ violation due to the final-state $K^0$-$\bar{K}^0$ mixing}

In some decay modes of neutral or charged $D$ mesons with $K^{}_{\rm S}$
or $K^{}_{\rm L}$ in the final states, $\CP$ violation of ${\cal O}(10^{-3})$
arising from $K^0$-$\bar{K}^0$ mixing is expected to show up. It is well
known that the mass and flavor eigenstates of neutral K mesons are
correlated with each other via
\begin{eqnarray}
|K^{}_{\rm S}\rangle \hspace{-0.2cm} & = & \hspace{-0.2cm}
\frac{1}{\sqrt{2 \left(1 + |\epsilon^{}_K|^2\right)}}
\left[\left(1 + \epsilon^{}_K\right) |K^0\rangle +
\left(1 - \epsilon^{}_K\right) |\bar{K}^0\rangle\right] \; ,
\nonumber \\
|K^{}_{\rm L}\rangle \hspace{-0.2cm} & = & \hspace{-0.2cm}
\frac{1}{\sqrt{2 \left(1 + |\epsilon^{}_K|^2\right)}}
\left[\left(1 + \epsilon^{}_K\right) |K^0\rangle -
\left(1 - \epsilon^{}_K\right) |\bar{K}^0\rangle\right] \; ,
\hspace{0.5cm}
\label{eq:K-mixing}
%     (17)
\end{eqnarray}
where $\epsilon^{}_K$ is the complex $K^0$-$\bar{K}^0$ mixing
parameter, and the strength of $\CP$ violation in $K^0$-$\bar{K}^0$
mixing is characterized by $|\epsilon^{}_K| =
(2.228 \pm 0.011) \times 10^{-3}$ and
$\arg(\epsilon^{}_K) = 43.52^\circ \pm
0.05^\circ$~\cite{ParticleDataGroup:2022pth}. Let us consider the
decay modes $D^\pm \to X^\pm K^{}_{\rm S}$ with $X$ being a
SL or nonleptonic state, and assume them to occur mainly
through the tree-level quark diagrams. In this case the amplitudes
of these two transitions can be parameterized as~\cite{Xing:1995jg}
\begin{eqnarray}
A(D^+ \to X^+ K^{}_{\rm S}) \hspace{-0.2cm} & = & \hspace{-0.2cm}
\kappa\left[
\left(1 + \epsilon^*_K\right) T^{}_1 e^{{\rm i} \left(\delta^{}_1
+ \phi^{}_1\right)} +
\left(1 - \epsilon^*_K\right) T^{}_2 e^{{\rm i} \left(\delta^{}_2
+ \phi^{}_2\right)} \right] \; ,
\nonumber \\
A(D^- \to X^- K^{}_{\rm S}) \hspace{-0.2cm} & = & \hspace{-0.2cm}
\kappa\left[
\left(1 - \epsilon^*_K\right) T^{}_1 e^{{\rm i} \left(\delta^{}_1
- \phi^{}_1\right)} +
\left(1 + \epsilon^*_K\right) T^{}_2 e^{{\rm i} \left(\delta^{}_2
- \phi^{}_2\right)} \right] \; ,
\hspace{0.5cm}
\label{eq:DK-amplitude}
%     (18)
\end{eqnarray}
where $\kappa=1/\sqrt{2 \left(1 + |\epsilon^{}_K|^2\right)} $, $T^{}_{1,2}$, $\delta^{}_{1,2}$ and $\phi^{}_{1,2}$ are the
moduli, strong and weak phases of the two amplitude components
associated respectively with $K^0$ and $\bar{K}^0$ in the
$D^+$ decay mode. Then the $\CP$-violating asymmetry between
$D^- \to X^- K^{}_{\rm S}$ and $D^+ \to X^+ K^{}_{\rm S}$ is given as
\begin{eqnarray}
{\cal A}^{}_{X^\pm K^{}_{\rm S}} \hspace{-0.2cm} & \equiv & \hspace{-0.2cm}
\frac{\Gamma(D^- \to X^- K^{}_{\rm S}) - \Gamma(D^+ \to X^+ K^{}_{\rm S})}
{\Gamma(D^- \to X^- K^{}_{\rm S}) + \Gamma(D^+ \to X^+ K^{}_{\rm S})}
\nonumber \\
\hspace{-0.2cm} & = & \hspace{-0.2cm}
2 \frac{{\rm Re}\epsilon^{}_K \left(T^2_2 - T^2_1\right)
+ 2 {\rm Im}\epsilon^{}_K T^{}_1 T^{}_2
\cos\left(\phi^{}_2 - \phi^{}_1\right)
\sin\left(\delta^{}_2 - \delta^{}_1\right)}
{T^2_1 + T^2_2 + 2 T^{}_1 T^{}_2 \cos\left(\phi^{}_2 - \phi^{}_1\right)
\cos\left(\delta^{}_2 - \delta^{}_1\right)}
\hspace{0.5cm}
\nonumber \\
\hspace{-0.2cm} & & \hspace{-0.2cm}
+ \hspace{0.06cm} 2 \frac{T^{}_1 T^{}_2 \sin\left(\phi^{}_2 - \phi^{}_1\right)
\sin\left(\delta^{}_2 - \delta^{}_1\right)}
{T^2_1 + T^2_2 + 2 T^{}_1 T^{}_2 \cos\left(\phi^{}_2 - \phi^{}_1\right)
\cos\left(\delta^{}_2 - \delta^{}_1\right)} \; ,
\label{eq:DK-amplitude}
%     (19)
\end{eqnarray}
in which the first term arises from $K^0$-$\bar{K}^0$ mixing and the
second term results from the direct $D^\pm$ decays. The former is very
likely to be comparable with and even dominant over the latter in magnitude
for some explicit decay modes of this
category~\cite{Xing:1995jg,Bigi:1994aw,Lipkin:1999qz,Yu:2017oky}.
Of course, one may similarly consider the $D^\pm \to X^\pm K^{}_{\rm L}$
transitions and their $\CP$-violating asymmetries. As for the decay modes
of $D^0$ and $\bar{D}^0$ mesons, $K^0$-$\bar{K}^0$ mixing in their final
states can also give rise to similar $\CP$-violating effects characterized
by ${\rm Re}\epsilon^{}_K$ and ${\rm Im}\epsilon^{}_K$.

In practice, a precision measurement of the $\epsilon^{}_K$-induced $\CP$
violation in charged or neutral $D$-meson decays will be very helpful to
calibrate the experimental systematics of $\STCF$ sensitive
to a $\CP$-violating asymmetry of ${\cal O}(10^{-3})$, in order to
search for the genuine charmed $\CP$ violation of the same order in
$D$-meson decays.

\subsection{$\CP$ violation due to $D^0-\bar{D}^0$ and $K^0-\bar{K}^0$ oscillating interference}

In Cascade decays involving both neutral $D$ mesons and neutral kaons, $\CP$ violation may arise from the interference effect of different $D$ and kaon oscillating paths, such as $D^0 \to \bar{K}^0 \to K^0$ and $D^0\to \bar{D}^0\to {K}^0$. The decay products of the primary decay associated with kaons, such as $\pi^0$, $\rho^0$ $\pi^+\pi^-$, etc., have been omitted, and the final state $f$ is not necessarily a $\CP$ eigenstate. This $\CP$-violation effect is referred to as double-mixing $\CP$ violation, as introduced in Refs.~\cite{Shen:2023nuw,Song:2024jjn}. It offers an opportunity for a two-dimensional time-dependent analysis of $\CP$-violation effects, depending on the weak decaying time durations of the primary and secondary decaying mesons. Furthermore, as $\CP$ violation arises from the interplay between decay and mixing, its presence does not necessitate a nonzero strong phase, providing a direct test of the KM mechanism of $\CP$ violation in the SM.

The cascade decay $D^0\to Kf_{\rm CP} \to (\pi^-\ell^+\nu_\ell)f_{\rm CP}$ can serve as an illustrative example. Owing to the frequent kaon oscillations, the dominant amplitude of the process is $D^0\to \bar{K}^0 f_{\rm CP} \to K^0 f_{\rm CP} \to (\pi^-\ell^+\nu_\ell)f_{\rm CP}$. The subdominant amplitudes are $D^0\to K^0 f_{\rm CP} \to (\pi^-\ell^+\nu_\ell)f_{\rm CP}$ and $D^0\to \bar{D}^0 \to K^0 f_{\rm CP} \to (\pi^-\ell^+\nu_\ell)f_{\rm CP}$, which are suppressed by $\lambda^2$ and $x_D(y_D)$, respectively. The $\CP$-violation effects can be induced by the magnitude squares and interferences of the three amplitudes, including the $\CP$ violation from neutral meson mixing, $\CP$ violation from the interplay between decay and mixing, as discussed previously, and also the double-mixing $\CP$ violation~\cite{Shen:2023nuw,Song:2024jjn}. The double-mixing $\CP$ violation arises from the interplay between $D$ mixing and kaon mixing, specifically between the $D^0\to \bar{K}^0 f_{\rm CP} \to K^0 f_{\rm CP} \to (\pi^-\ell^+\nu_\ell)f_{\rm CP}$ and $D^0\to \bar{D}^0 \to K^0 f_{\rm CP} \to (\pi^-\ell^+\nu_\ell)f_{\rm CP}$ amplitudes. It can be characterized by the difference $(q/p)_D^*(p/q)_K A(D^0\to \bar{K}^0 f_{\rm CP}) - (p/q)_D^*(q/p)_K A(\bar{D}^0\to {K}^0 f_{\rm CP}) \neq 0$. The corresponding time dependence of the double-mixing $CP$ violation is given by 
\begin{eqnarray}\label{eq:acpdm}
A^{\rm dm}_\mathrm{CP}(t_1,t_2) \propto  e^{-\Gamma_Dt_D-\Gamma_Kt_K}  \left[ \sinh{\Delta\Gamma_Dt_D\over 2} \; S_h(t_K) +   \sin{(\Delta m_Dt_D)}\; S_n(t_K) \right] ,\;
\end{eqnarray}
where 
\begin{eqnarray}
\label{eq:shn}
S_h(t_K) 
% \hspace{-0.2cm} & = & \hspace{-0.2cm}   -  \sin{(\Delta m_Kt_K)}  \sin \Phi_{DK} + {1\over 2}\sinh{\Delta\Gamma_Kt_K\over2} \left( \left|{q_D \over p_D}{p_K\over q_K}\right| - \left|{p_D\over q_D}{q_K\over p_K}\right| \right) \cos\Phi_{DK}   \nonumber \\
\hspace{-0.2cm} & \approx & \hspace{-0.2cm}   -  \sin{(\Delta m_Kt_K)}  \sin \Phi_{DK}
+ 2 \sinh{\Delta\Gamma_Kt_K\over2}  \mathrm{Re} (\epsilon_K-\epsilon_D) \cos\Phi_{DK}    ,\;  \nonumber \\
S_n(t_K) 
% \hspace{-0.2cm} & = & \hspace{-0.2cm} \sinh{\Delta\Gamma_K t_K\over2}  \sin\Phi_{DK}  +{1\over 2}\sin{(\Delta m_Kt_K)}  \left( \left|{q_D \over p_D}{p_K\over q_K}\right| - \left|{p_D\over q_D}{q_K\over p_K}\right| \right) \cos\Phi_{DK}  \nonumber \\
\hspace{-0.2cm} & \approx & \hspace{-0.2cm}   \sinh{\Delta\Gamma_Kt_K\over2}  \sin\Phi_{DK}  + 2\sin{(\Delta m_K t_K)}   \mathrm{Re} (\epsilon_K-\epsilon_D) \cos\Phi_{DK} ,  \;
\end{eqnarray}
with $\Phi_{DK}= \phi_D+\phi_K+ \omega $ and $e^{i\omega}\equiv A(D^0\to \bar{K}^0 f_{\rm CP})/A(\bar{D}^0\to {K}^0 f_{\rm CP})$. Apart from the overall exponential factor, the four terms exhibit distinct sine and hypersine dependencies on $t_D$ and $t_K$, allowing for a two-dimensional time-dependent analysis to extract their contributions individually.  
Similarly, in the decay $D^0\to Kf_{\rm CP} \to (\pi^+\ell^-\bar{\nu}_\ell)f_{\rm CP}$, the double-mixing $\CP$ violation arises from the interplay between $D^0\to \bar{K}^0 f_{\rm CP} \to (\pi^+\ell^-\bar{\nu}_\ell)f_{\rm CP}$ and $D^0\to \bar{D}^0 \to K^0 f_{\rm CP} \to \bar{K}^0 f_{\rm CP} \to (\pi^+\ell^-\bar{\nu}_\ell)f_{\rm CP}$. It is characterized by the difference $(q/p)_D^*(q/p)_K^* A(D^0\to \bar{K}^0 f_{\rm CP}) - (p/q)_D^*(p/q)_K^* A(\bar{D}^0\to {K}^0 f_{\rm CP}) \neq 0$, and the corresponding time-dependent functions in Eq.~(\ref{eq:acpdm}) are given by 
\begin{eqnarray}\label{eq:shn}
S_h(t_K)
% \hspace{-0.2cm} & = & \hspace{-0.2cm}  -   \sin{(\Delta m_Kt_K)}  \sin \Phi_{DK} + {1\over2} \sinh{\Delta\Gamma_Kt_K\over2} \left( \left|{q_D \over p_D}{q_K\over p_K}\right| - \left|{p_D\over q_D}{p_K\over q_K}\right| \right) \cos\Phi_{DK}    \nonumber \\
\hspace{-0.2cm} & \approx & \hspace{-0.2cm} -  \sin{(\Delta m_Kt_K)}  \sin \Phi_{DK} - 2 \sinh{\Delta\Gamma_Kt_K\over2} \mathrm{Re} (\epsilon_K+\epsilon_D) \cos\Phi_{DK}    ,  \nonumber \\
S_n(t_K)
%\hspace{-0.2cm} & = & \hspace{-0.2cm}  - 2 \sinh{\Delta\Gamma_Kt_K\over2}  \sin\Phi_{DK}  - {1\over2} \sin{(\Delta m_Kt_K)}  \left( \left|{q_D \over p_D}{q_K\over p_K}\right| - \left|{p_D\over q_D}{p_K\over q_K}\right| \right) \cos\Phi_{DK}  \nonumber \\
\hspace{-0.2cm} & \approx & \hspace{-0.2cm} - \sinh{\Delta\Gamma_2t_2\over2}  \sin\Phi_{DK}  + 2\sin{(\Delta m_2t_2)}   \mathrm{Re} (\epsilon_K+\epsilon_D) \cos\Phi_{DK}    . 
\end{eqnarray}

%\subsection{Indirect $\CP$ violation associated with $D^0$-$\bar{D}^0$ mixing}

\subsection{Prospects of charm $\CP$ violation studies at STCF}
%\paragraph{Mixing and time-independent $\CP$ violation}

Mixing and indirect $\CP$ violation effects in the neutral $D$ meson system can be investigated with the quantum-correlated (QC) neutral $D\bar{D}$ pairs produced at STCF. On the resonance $\psi(4010)$ where the production cross-section of neutral $D\bar{D}^{*}$ is 3320 $\rm{pb}$~\cite{CLEO:2008ojp}, the integrated luminosity of the data sample that STCF is expected to collect annually is around 1 $\rm{ab}^{-1}$, corresponding to $3.32 \times 10^{9}$ $D\bar{D}^{*}$ pairs. The $D\bar{D}$ pairs are $C$-odd and even correlated with the subsequent decays $D^{*0}\to D^0\pi^0$ and $D^{*0}\to D^0\gamma$, respectively. Taking the branching fraction of $D^{*0}\to D^0\gamma$, the number of pairs of $C$-even $D\bar{D}$ can reach approximately $1.2 \times 10^{9}$. At the c.m.~energy $\sqrt{s}=4.01$ GeV, the production cross-section of $D^{*+}\bar{D}^{-}$ is $3300$ pb~\cite{CLEO:2008ojp}. The number of $D^{*+}\bar{D}^{-}$ pairs produced at STCF annually is about $N_{D^{*+}\bar{D}^{-}}=3.30 \times 10^{9}$. Taking the branching fraction of $D^{*+}\to D^0\pi^+$, the number of $D^0$ mesons is approximately $2.23 \times 10^{9}$. Flavour-specific neutral $D$ meson decays can be accessed by reconstructing the soft pion from $D^{*}$ decay or the charged $D$ meson. 

The two $D$ mesons decay coherently with the double decay rates given by Eq.~(\ref{eq:D-joint-decay-rate3}). A "double tag" (DT) method is needed to exploit the QC effects, where the signal $D$ decay is reconstructed against the tag-side decay. Multiple-body decays are more sensitive to the charm mixing and indirect $\CP$ violation measurements due to the amplitude variations in phase space. Time-integrated measurements, referred as "single tag" (ST) method, with flavor-specific neutral $D$ meson decays can add to the sensitivity. The decay rates are given by Eq.~(\ref{eq:D-decay-rate4}). We summarize the status of the future prospects at STCF with the three and four-body decays $D\to K^-\pi^+\pi^0$, $K_S^0\pi^+\pi^-$ and $K^-\pi^+\pi^+\pi^-$. They proceed through tree-level decays, where the direct $\CP$ violation can be neglected.

\paragraph{Measurements of the $D\to K^-\pi^+\pi^+\pi^-$ decay}

With the $C$-even correlated $D\bar{D}$ pairs, the $D\to K^-\pi^+\pi^+\pi^-$ decay can be reconstructed against the flavour tags $K^{\mp}\pi^{\pm}\pi^{+}\pim$, $K^{\mp}\pi^{\pm}\pi^0$, $K^{\mp}\pi^{\pm}$, the $\CP$-even eigenstates $K^+K^-$, $\pi^{+}\pi^{-}$ and $\CP$-odd eigenstates $K_{S}^0\pi^0$, $K_{S}^0\eta$, $K_{S}^0\omega$ and $K_{S}^0\eta^{'}$ and the self-conjugate decay $K_{S}^0\pi^+\pi^-$. By exploiting the signal decays in four phase-space regions, the sensitivity is obtained as listed in Table~\ref{tab:fitting_results_k3pi} and 1, 2 and 3$\sigma$ regions of the parameters are shown in Fig.~\ref{fig:xy_k3pi}. The $\CP$-conserving strong-phase differences between $D^0$ and $\bar{D}^0$ decays in phase-space regions are required as inputs to such studies. These parameters can be accessed with the $C$-odd correlated $D\bar{D}$ pairs produced at the same energy point~\cite{BESIII:2021eud}.%, where the charm mixing and $\CP$ violation effects are minimal.

\begin{table}[htbp]
    \centering
    \caption{Statistical sensitivity of the charm mixing and indirect $\CP$ violation parameters by analyzing the $C$-even correlated and flavor specific $D\to K^-\pi^+\pi^+\pi^-$ decays at STCF, and comparison with the LHCb Upgrade I.}
    \label{tab:fitting_results_k3pi}
    \renewcommand{\arraystretch}{1.2}
    \begin{tabular}{l|cccc}
        \hline \hline
                         & STCF (DT) & STCF (DT\&ST) & LHCb $(50~\textrm{fb}^{-1})$  \\ \hline
        $x(\%)$          & $0.056$ & $0.047$  & $-$    \\
        $y(\%)$          & $0.027$ & $0.025$  & $-$    \\
        $r_{CP}$         & $0.050$ & $0.042$  & $0.005$   \\
        $\phi(^{\circ})$ & $3.64$  & $3.10$   & $0.30$  \\
        \hline \hline
    \end{tabular}
\end{table}

\begin{figure}[htbp!]
    \centering
    %\subfigure{
    \includegraphics[width=0.48\textwidth]{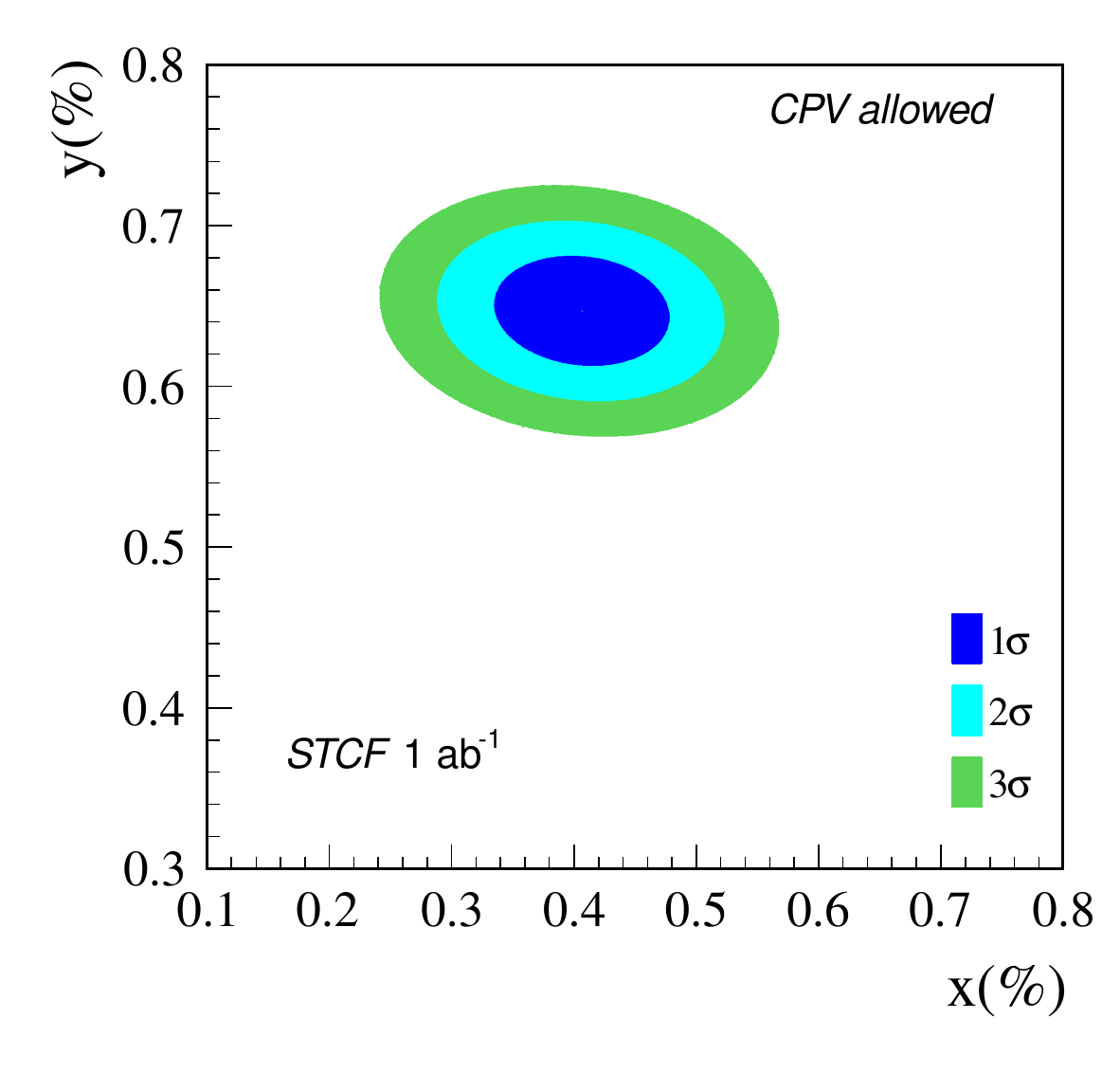}
    %}
    %\subfigure{
    \includegraphics[width=0.48\textwidth]{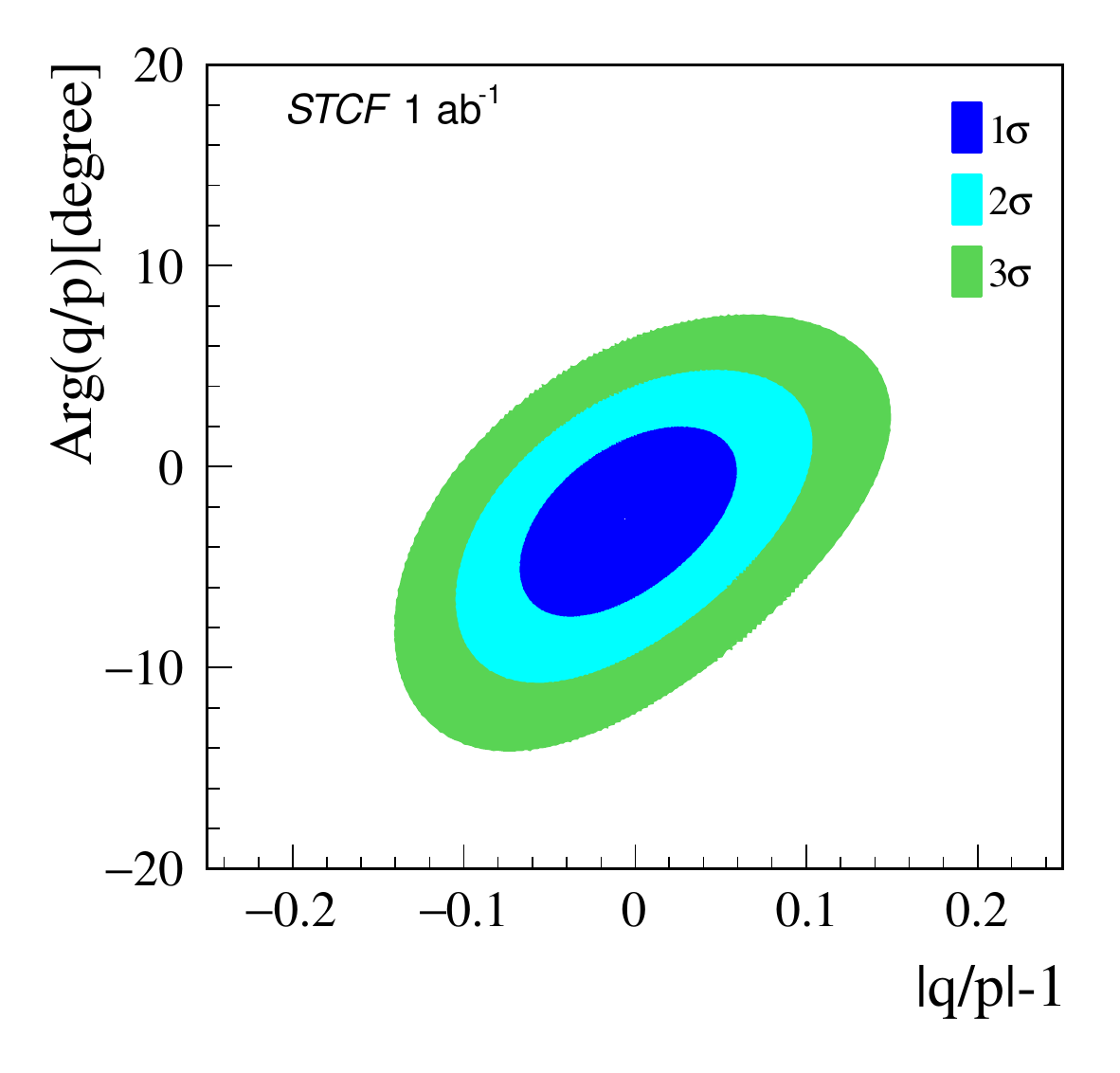}
    %}
    \caption{1, 2 and 3$\sigma$ regions of the charm mixing parameters and indirect $\CP$ violation parameters by analyzing the $D\to K^-\pi^+\pi^+\pi^-$ decay.}
    \label{fig:xy_k3pi}
\end{figure}

\paragraph{Measurements of the $D\to K_S^0\pi^+\pi^-$ decay}

Studies with $D\to K_S^0\pi^+\pi^-$ decay also include the various tags used for $D\to K^-\pi^+\pi^+\pi^-$ decay. With one-year data taking, the sensitivity that STCF is expected to achieve is shown in Table~\ref{tab:fitting_results_kspipi} and 1, 2 and 3$\sigma$ regions of the parameters are shown in Fig.~\ref{fig:xy_kspipi}. The strong-phase parameters in the exploit eight phase-space regions can be accessed with the corresponding $C$-odd correlated $D\bar{D}$ decays~\cite{BESIII:2020khq}. Utilizing the same method, a recent time-dependent measurement performed by the LHCb experiment determined that $x=\left(0.398^{+0.056}_{-0.054}\right)\%$, $y=\left(0.46^{+0.15}_{-0.14}\right)\%$ and $r_{CP}=0.996\pm 0.052$, $\phi=\left(3.2^{+2.7}_{-2.9}\right)^{\circ}$~\cite{LHCb:2021ykz}.

\begin{table}[htbp]
    \centering
    \caption{Statistical sensitivity of the charm mixing and indirect $\CP$ violation parameters by analyzing the $C$-even correlated and flavor specific $D\to K_S^0\pi^+\pi^-$ decays at STCF and comparison with the LHCb Upgrade I and Belle II.}
    \label{tab:fitting_results_kspipi}
    \renewcommand{\arraystretch}{1.2}
    \begin{tabular}{l|cccc}
        \hline \hline
                         & STCF (DT) & STCF (DT\&ST) & LHCb $(50~\textrm{fb}^{-1})$ & Belle II $(50~\textrm{ab}^{-1})$ \\ \hline
        $x(\%)$          & $0.076$ & $0.069$  & $0.012$                     & $0.030$                        \\
        $y(\%)$          & $0.054$ & $0.050$  & $0.013$                     & $0.020$                        \\
        $r_{CP}$         & $0.084$ & $0.077$  & $0.011$                     & $0.022$                        \\
        $\phi(^{\circ})$ & $5.00$  & $4.57$   & $0.48$                      & $1.50$                         \\
        \hline \hline
    \end{tabular}
\end{table}

\begin{figure}[htbp!]
    \centering
    %\subfigure{
    \includegraphics[width=0.48\textwidth]{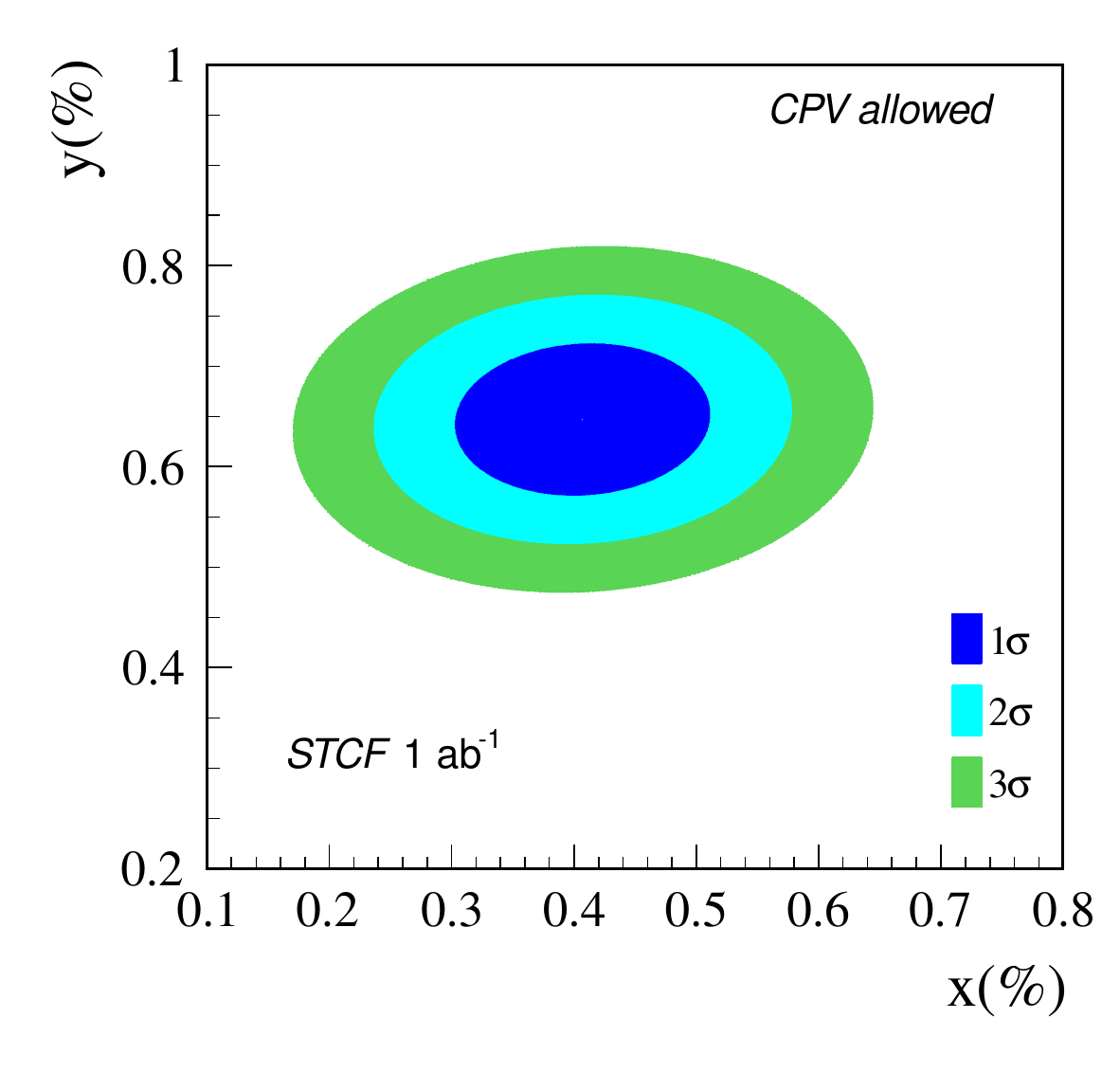}
    %}
    %\subfigure{
    \includegraphics[width=0.48\textwidth]{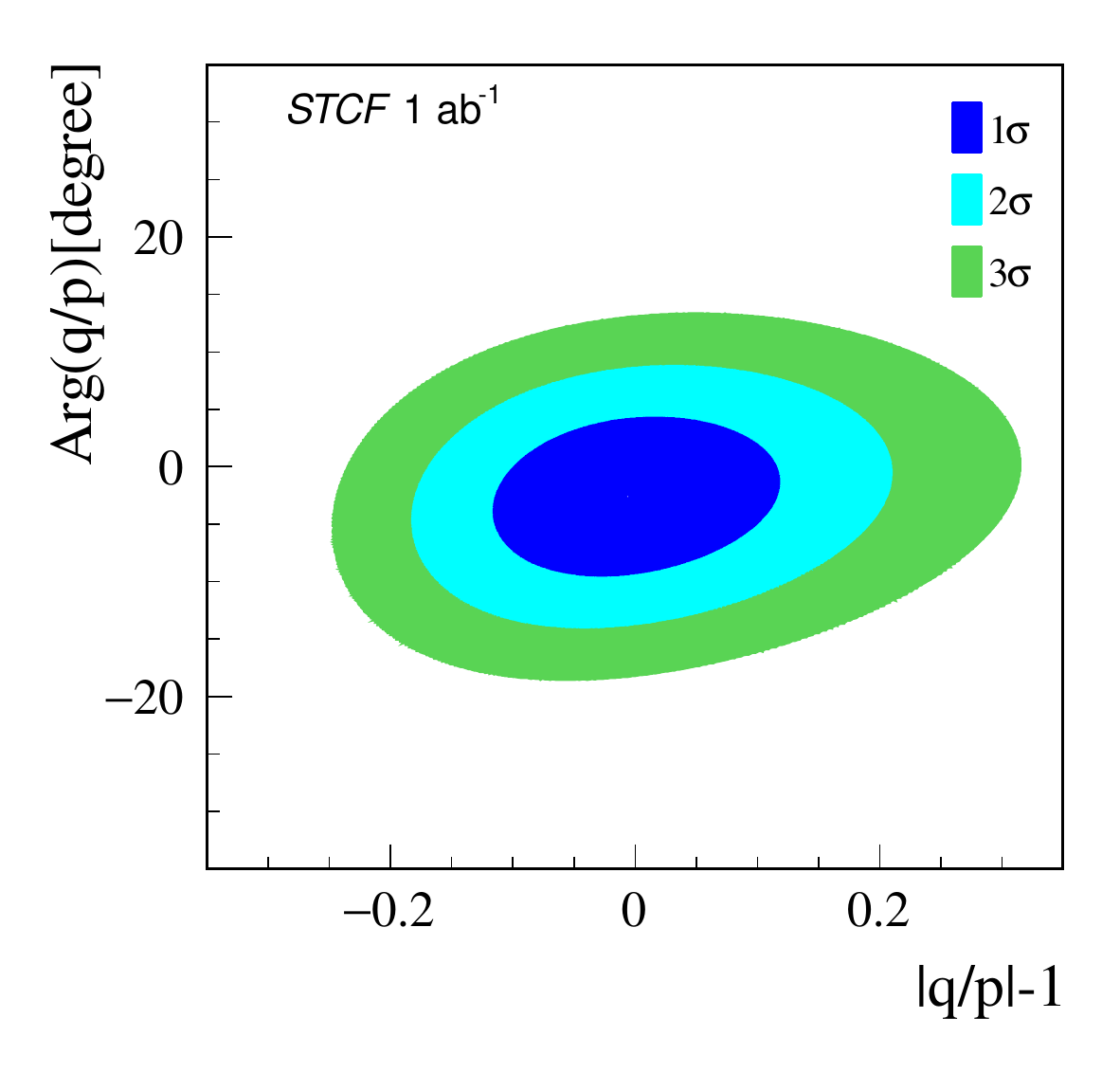}
    %}
    \caption{1, 2 and 3$\sigma$ regions of the charm mixing parameters and indirect $\CP$ violation parameters by analyzing the $D\to K_S^0\pi^+\pi^-$ decay.}
    \label{fig:xy_kspipi}
\end{figure}

\paragraph{Measurements of the $D\to K^-\pi^+\pi^0$ decay}

Similar to the $D\to K^-\pi^+\pi^+\pi^-$ decay, the $D\to K^-\pi^+\pi^0$ decay can be studied with the $C$-even correlated $D\bar{D}$ pairs and the flavor-specific $D$ mesons. Currently, the sensitivity is obtained by taking the inputs of average strong-phase and associated hadronic parameters over the whole phase-space~\cite{BESIII:2021eud}. This decay is found to be dominant for the charm mixing and indirect $\CP$ violation study at STCF, with the sensitivity shown in Table~\ref{tab:fitting_results_kspipi} and 1, 2 and 3$\sigma$ regions of the parameters shown in Fig.~\ref{fig:xy_kpipi0}. A binned study can further improve the sensitivity and again the strong-phase parameters can be determined simultaneously with the $C$-odd correlated $D\bar{D}$ decays.

\begin{table}[htbp!]
    \centering
    \caption{Statistical sensitivity of the charm mixing and indirect $\CP$ violation parameters by analysing the $C$-even correlated and flavor specific $D\to K^-\pi^+\pi^0$ decays at STCF.}
    \label{tab:fitting_results_kpipi0}
    \renewcommand{\arraystretch}{1.2}
    \begin{tabular}{l|cc}
        \hline \hline
                         & STCF~(DT) & STCF~(DT\&ST)\\ \hline
        $x(\%)$          & $0.055$  & $0.044$ \\
        $y(\%)$          & $0.019$  & $0.017$ \\
        $r_{CP}$         & $0.048$  & $0.034$ \\
        $\phi(^{\circ})$ & $3.34$   & $2.51$ \\
        \hline \hline
    \end{tabular}
\end{table}

\begin{figure}[htbp!]
    \centering
    %\subfigure{
    \includegraphics[width=0.48\textwidth]{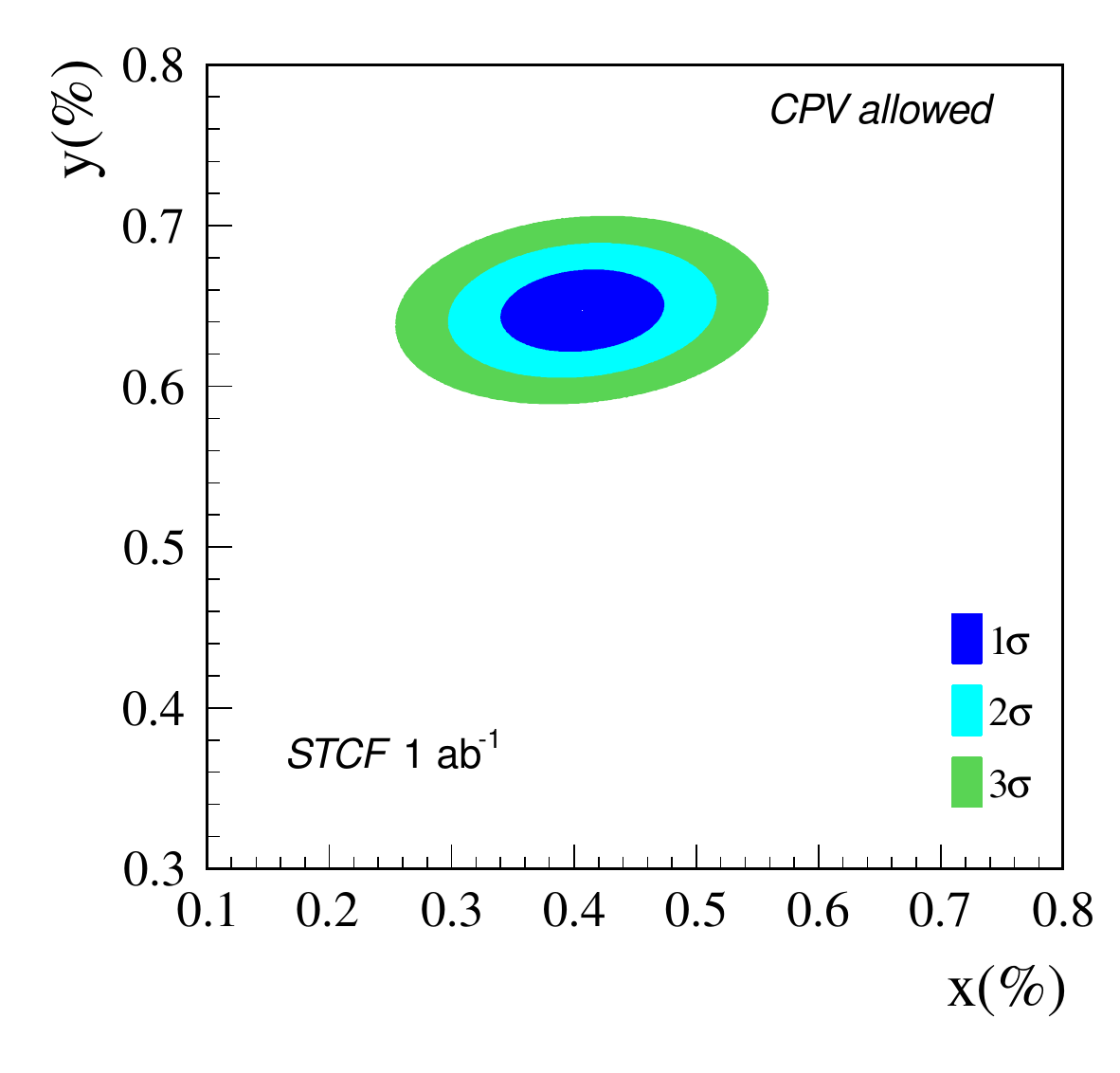}
    %}
    %\subfigure{
    \includegraphics[width=0.48\textwidth]{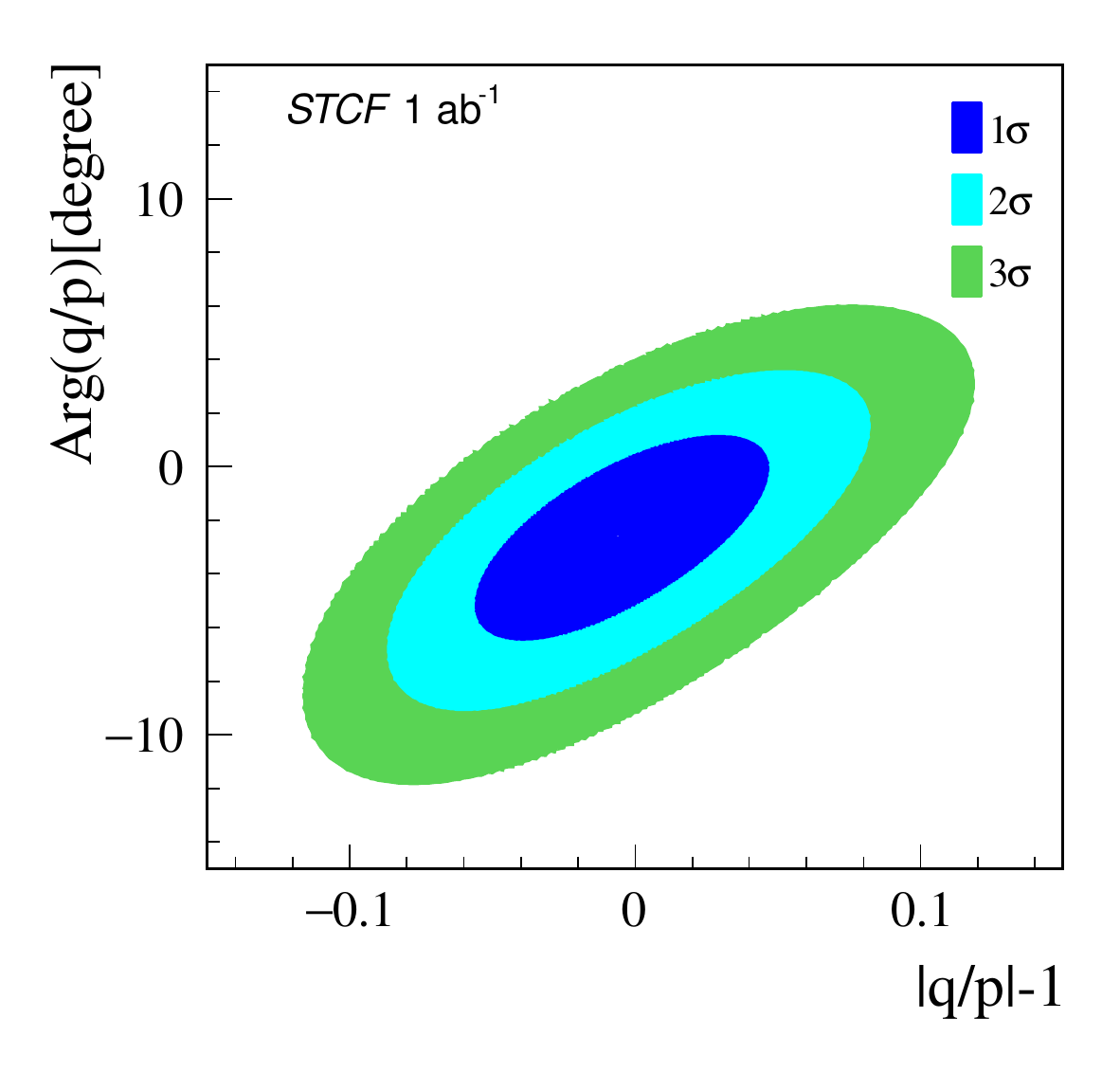}
    %}
    \caption{1, 2 and 3$\sigma$ regions of the charm mixing parameters and indirect $\CP$ violation parameters by analyzing the $D\to K^-\pi^+\pi^0$ decay.}
    \label{fig:xy_kpipi0}
\end{figure}

\paragraph{Overall prospects}

A global fit of the DT $D\to K^-\pi^+\pi^0$, $K_S^0\pi^+\pi^-$ and $K^-\pi^+\pi^+\pi^-$ signal decays and flavor-specific ST gives the sensitivities of the charm mixing and indirect $\CP$ violation parameters to be $\sigma(x)=0.036\%$, $\sigma(y)=0.015\%$, $\sigma(r_{\cp})=0.028$ and $\sigma(\phi)=2.14^{\circ}$, as shown in Fig.~\ref{fig:xy_this_work}. Such studies indicate that indirect charm $\CP$ violation can be studied at the $10^{-4}$ level in STCF and would complement the studies in upgraded LHCb and Belle II experiments. 

\begin{figure}[htbp!]
    \centering
    %\subfigure{
    \includegraphics[width=0.48\textwidth]{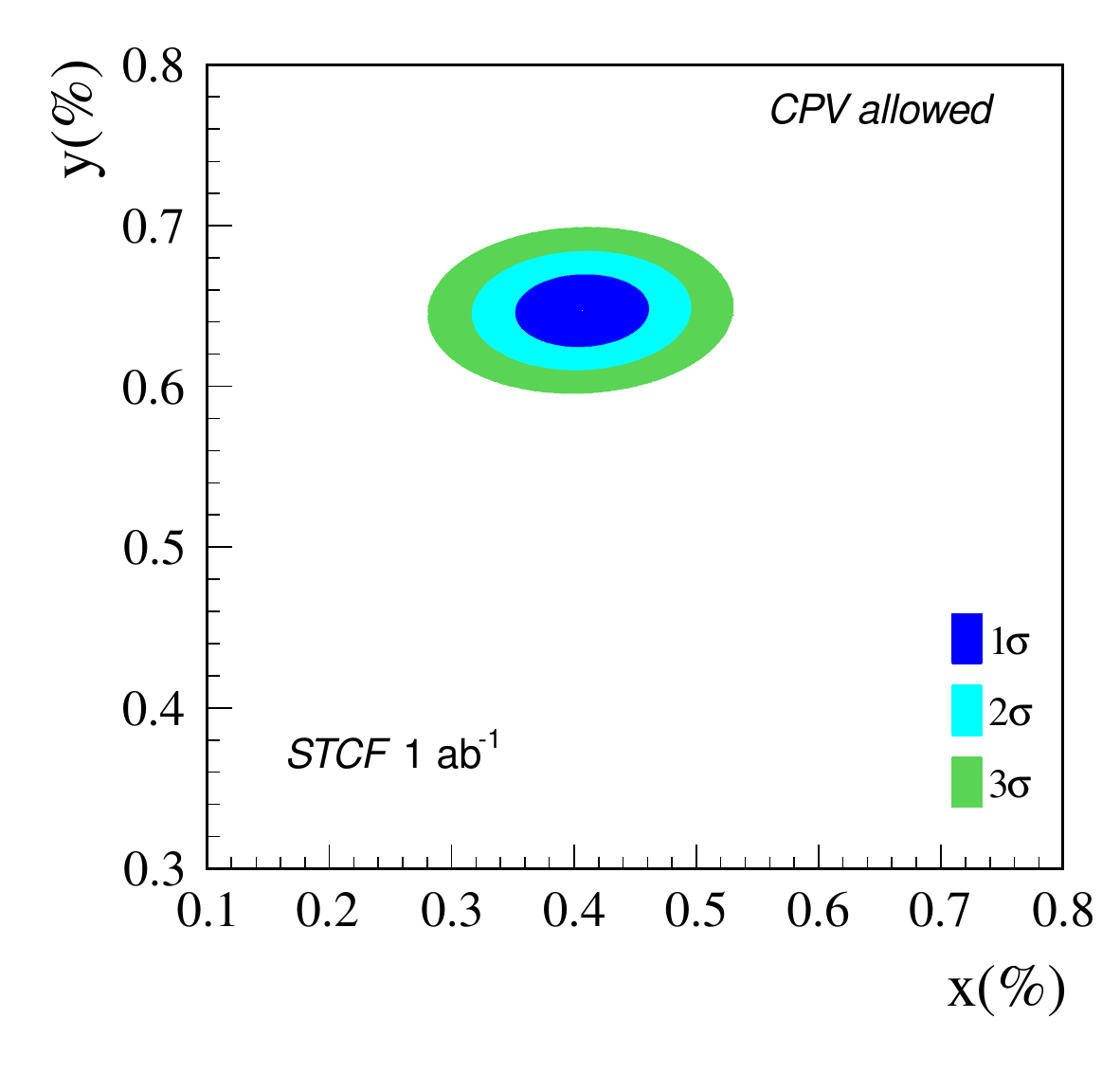}
    %}
    %\subfigure{
    \includegraphics[width=0.48\textwidth]{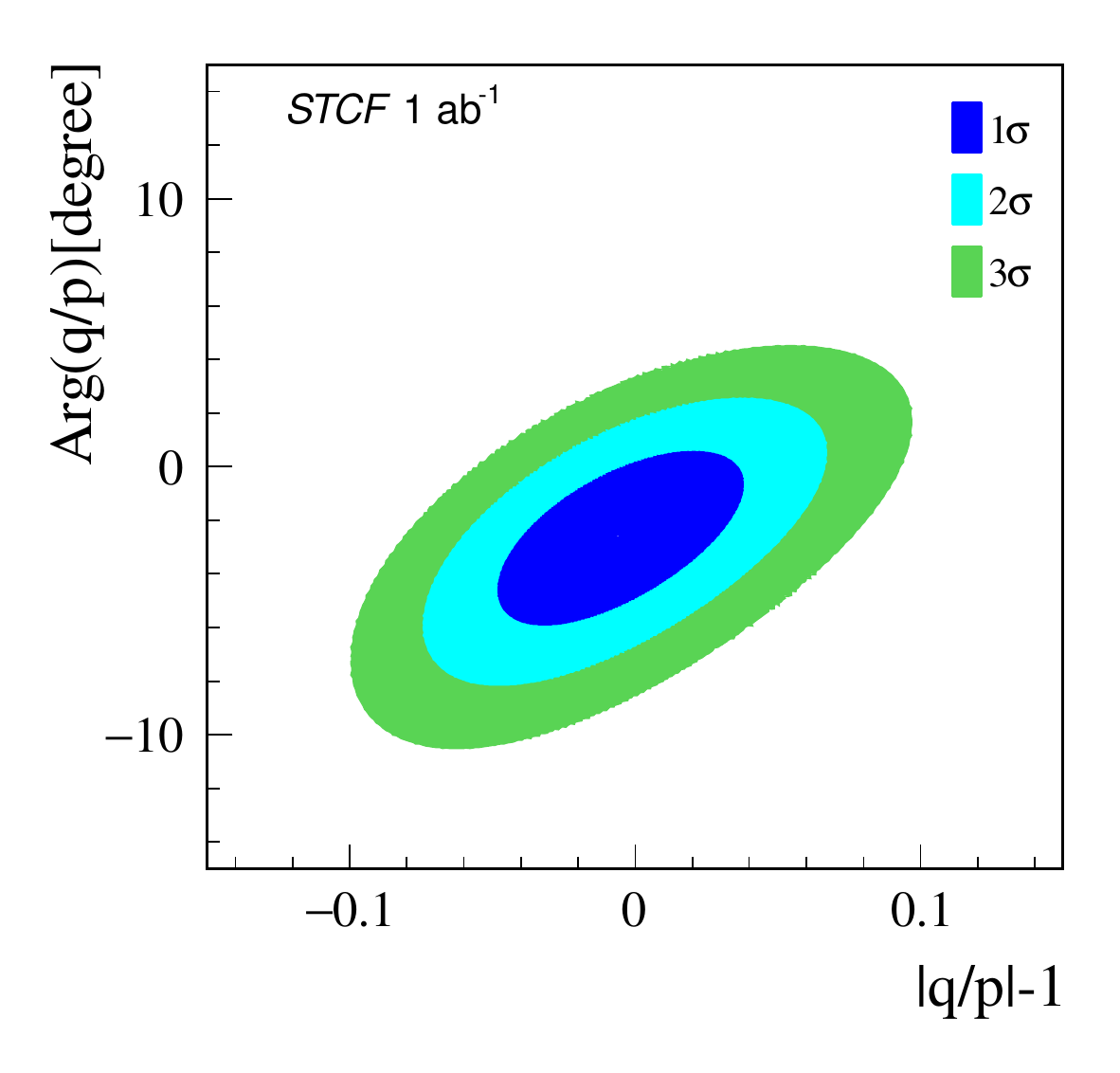}
    %}
    \caption{1, 2 and 3$\sigma$ regions of the charm mixing parameters and indirect $\CP$ violation parameters by analyzing the $D\to K^-\pi^+\pi^0$, $K_S^0\pi^+\pi^-$ and $K^-\pi^+\pi^+\pi^-$ decays.}
    \label{fig:xy_this_work}
\end{figure}

Prerequisites for the charm mixing and indirect $\CP$ violation studies at charm and $B$ factories are the strong-phase differences with ultimate uncertainties. The $C$-odd correlated $D\bar{D}$ data samples collected with the BESIII and CLEO experiments at the $\psi(3770)$ energy point have been the best laboratory to perform such measurements. Currently, the uncertainties of the dilution parameters are non-vanishing. STCF would collect more than 100 times the $C$-odd correlated $D\bar{D}$ pairs at BESIII per year, which would be another important contribution to the whole community.

\newpage

\section{Tests of the $\CP T$ invariance with $\jpsi$ decays}
\label{sec:kaon}
The $\CP T$~theorem ~\cite{Schwinger:1951xk}
states that any quantum field theory  that is {\it Lorentz invariant}, has {\it local point-like interaction vertices},
and is {\it hermitian} ({\it i.e.}, conserves probability) is invariant under the combined operations of
$C$,~$P$~and~$T$. Since the three quantum field theories that make up the SM---QED, QCD, and
Electroweak theory---all satisfy these criteria, $\CP T$~symmetry is a fundamental and inescapable prediction of the theory.
Any deviation from $\CP T$ invariance would be unambiguous evidence for new, beyond the
SM physics.  Given the central role that the $\CP T$ theorem plays in our understanding of the most basic
properties of space and time, it is essential that its validity should be tested at the highest level of sensitivity
that the experimental technology permits.

Fortunately, the $\Kz$-$\Kzbar$ mixing process, coupled with the requirements of unitarity, provides sensitive ways to search for violations of the $\CP T$ theorem. This is embodied in a remarkable equation called the
{\it Bell-Steinberger relation}, that unambiguously relates a fundamental $\CP T$ violating amplitude called $\delta$ to
well defined measurable quantities. The relation is exact and, unlike most other experimental tests of the SM, there are no loose ends such as long-distance QCD-corrections or theoretical assumptions about anything other
than unitarity and time independence of the Hamiltonian. If a non-zero value of $\delta$ is established, the only possible
interpretations are that either the $\CP T$ theorem is invalid  or unitarity is violated, or that kaon mass or lifetime
changes with time.

The measurement involves studies of the proper-time-dependence of the  decay rates of large numbers of neutral kaons that
have well established  ({\it i.e.}, tagged) strangeness quantum numbers at the time of their production $\Kz(\tau)$ for
\Str=+1 and $\Kzbar(\tau)$ for \Str=$-$1.  The most sensitive limits to date are from experiments in the 1990s that used data
samples that contained tens of millions of $\Kz$ and $\Kzbar$ decays. A $\sim$10$^{12}$-event sample of $\jpsi$  decays
will contain $\sim$2 billion $\jpsi$$\rt$$\Km\pip\Kz$ events  and an equivalent number of $\jpsi$$\rt$$\Kp\pim\Kzbar$ events where
the neutral kaon decays to $\pipi$, and $\sim$1 billion events for each mode in which the kaon decays to $\piz\piz$. In
these reactions, the initial strangeness of the neutral kaon is tagged by the sign of the accompanying charged-kaon's
electric charge: a $\Km$ tags an \Str=+1 $\Kz(\tau)$ and  a $\Kp$ tags an \Str=$-$1 $\Kzbar(\tau)$. These data samples
would support a revisit of previous measurements with $\sim$fifty-times larger data samples. 

In this report we discuss:

\vspace{1mm}
\noindent
{\it i})~why we expect $\CP T$ to be violated somewhere below the Planck mass-scale;

\vspace{1.5mm}
\noindent
{\it ii})~why the neutral kaon system is well suited for tests of the $\CP T$ theorem;

\vspace{1.5mm}
\noindent
{\it iii})~the quantum-mechanics of neutral kaons with restrictions on $\CP T$ relaxed;

\vspace{1.5mm}
\noindent
{\it iv})~estimated experimental sensitivities with 10$^{12}$ $\jpsi$ decays;

\vspace{1.5mm}
\noindent
{\it v})~applications of the Bell-Steinberger relation at a $\tau$-charm factory. 

\subsection{$\CP T$ and the Theory of Everything}

One of the requirements for a $\CP T$-invariant theory is that it is {\it local}, which means that the
couplings at each vertex occur at a single point in space-time. But theoretical physics always has
troubles with point-like quantities. For example, the classical Coulombic self-energy of the electron is
\begin{equation}
  W_e=\frac{e^2}{4\pi\eps_0 r_e},
\end{equation}
which diverges for $r_e$$\rt$0.  The {\it classical radius of the electron}, {\it i.e.}, the value of $r_e$
that makes $W_e$=\,$m_ec^2$, is $r^{\rm c.r.e.}_e$=\,2.8$\times$10$^{-13}$~cm (2.8 fermis),  which is three times the
radius of the proton, and $\sim$3000 times larger than the experimental upper limit on the electron radius, which
is of $\mathcal{O}(10^{-16}\,{\rm cm})$~\cite{ZEUS:2003eqd}. Infinities associated with point-like objects persist
in quantum field theories, where they are especially troublesome. In second- and higher-order perturbation theory, all diagrams that have virtual-particle loops involved integrals over all possible configurations of the 
loops that conserve energy and momentum.  Whenever two of the point-like vertices coincide,
the integrands become infinite and cause the integrals to diverge.

In the QED, QCD and Electroweak quantum field theories that make up the SM, these infinities are
removed by the well established methods of renormalization~\cite{Bethe:1947id,Dyson:1949bp,Wilson:1973jj}. In all these three theories, the perturbative expansions are in increasing powers of a dimensionless coupling strength, $\alpha_{\rm QED}, \alpha_s$ and, $\alpha_{\rm EW}=\sqrt{2}M^2_WG_F/\pi$.\footnote{Specifically
  not just $G_F$, which has dimensions of mass$^{-2}$.}
As a result of this, in the renormalization procedure, relations that exist between different orders of the
perturbation expansion reduce the number of observed quantities that are needed to subtract off divergences.
In QED, for example, there are only two, the electron's mass, and charge (four, if the diagram includes
muons and tau-leptons).   However, in quantum theories of gravity, where a massless spin=2 {\it graviton} plays
the role of the photon in QED, the  expansion constant is Newton's gravitational constant $G_N$\,=\,$\hbar c/M^2_{\rm P}$,
where $M_{\rm P}\equiv\sqrt{\hbar c/G_N}=1.24\times$10$^{19}$~GeV is the {\it Planck mass}. Because this
has dimension mass$^{-2}$, every order in the perturbation expansion has different dimensions and, thus, would need a
distinct observed quantity to carry out the subtraction. This means that a complete renormalization would require an
infinite number of observed quantities, with the end effect that a quantum theory of gravity with point-like vertices
would, in principle, be {\it nonrenormalizable}~\cite{Weinberg:1980kq}. A viable theory of everything would have to
include a mechanism for avoiding these divergences by somehow smearing out the vertices, for example by treating particles
as loops of strings like in Fig.~\ref{fig:String-vertex}, and this {\it non-locality} would eliminate one of the necessary
conditions for  $\CP T$ invariance. Thus, a Theory of Everything, which, by definition, would have to include
gravity~\footnote{Even cavemen were aware of gravity.}, would not be local and include $\CP T$ violations at some mass scale.

\begin{figure}[!]
\centering
\includegraphics[width=0.7\textwidth]{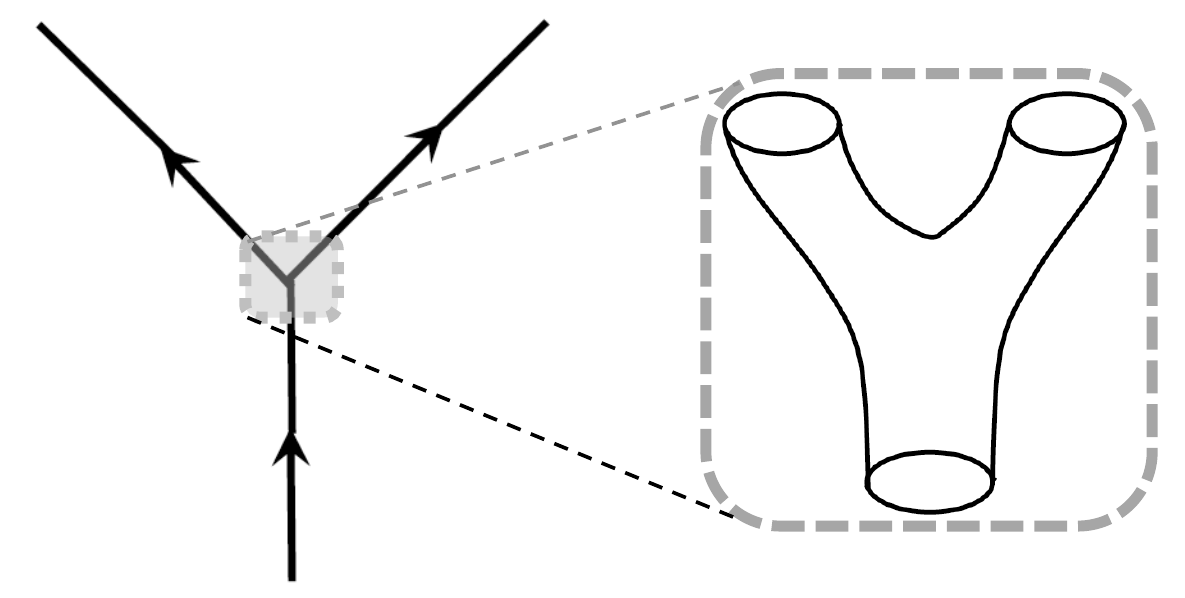}
\caption{\footnotesize In string theories, elementary particles are tiny loops
  of oscillating strings with no point-like vertices and their associated infinities.}
 \label{fig:String-vertex}
 \end{figure}

Although difficulties associated with non-renormalizability ({\it i.e.}, higher-order perturbative effects) will
never show up at scales below the Planck mass, the fact that this problem exists at all demonstrates that there
is nothing especially sacred about $\CP T$-invariance, and nothing prevents it from being violated at a lower mass scale.
Because of this close connection with the fundamental assumptions of the SM, stringent experimental tests
of $\CP T$~invariance should have high priority.

\subsection{Neutral $K$ mesons and tests of the $\CP T$~theorem}

The main consequences of $\CP T$~symmetry are that particle and antiparticle masses and lifetimes are equal, being their electric charges opposite and higher electromagnetic multipoles identical. 
Since lifetime differences can only come from on-mass-shell intermediate states and do not probe short-distance
high-mass physics, these are unlikely to exhibit any $\CP T$-violating asymmetry. Instead, the focus here is on the
possibility that particle and antiparticle masses may be different.

The particles with the best measured masses are the stable electron and proton, and, according the
PDG20~\cite{Zyla:2020zbs}:
\begin{eqnarray}
    \label{eqn:eebar-limit}
  |m_{\bar{e}}-m_{e}|&<&4\times 10^{-9}~{\rm MeV},\\
    \label{eqn:ppbar-limit}
  |m_{\bar{p}}-m_{p}|&<&7\times 10^{-7}~{\rm MeV}.
\end{eqnarray}
However, these limits do not provide the best tests of $\CP T$; the most stringent experimental restriction on
$\CP T$ violation comes from the difference between the $\Kzbar$ and $\Kz$ masses:
\begin{equation}
  \label{eqn:KKbar-limit}
  |M_{\Kzbar}-M_{\Kz}|=<5\times 10^{-16}~{\rm MeV},
\end{equation}
which is $7$-$9$~orders of magnitude stricter than those from the electron and proton mass measurements,
even though the value of $M_{\Kz}$ itself is only known to~$\pm 13$~keV. This is because of the
Fig.~\ref{fig:k-mix_c-quark-KM} diagrams, taken together with the quantum mechanics of $\Kz$-$\Kzbar$ mixing, map the
$M_{\Kzbar}$$-$$M_{\Kz}$ difference into the quantity $\Delta M$\,$\equiv$\,$M_{\KL}$$-$$M_{\KS}$$\approx$\,3.5$\times$10$^{-12}$~MeV,
which is 14 orders of magnitude lower than $M_{\Kzbar}$ or $M_{\Kz}$ and the independent quantity 
$\Delta\Gamma$$\equiv$\,$\Gamma_{\KS}$$-$$\Gamma_{\KL}$$\approx$\,7.4$\times$10$^{-12}$~MeV (which is, coincidentally,
$\approx$2$\times\Delta M$).

\begin{figure}[htbp!]
\centering
\includegraphics[width=0.45\textwidth]{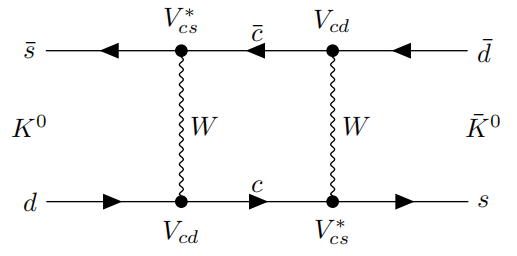}
\includegraphics[width=0.45\textwidth]{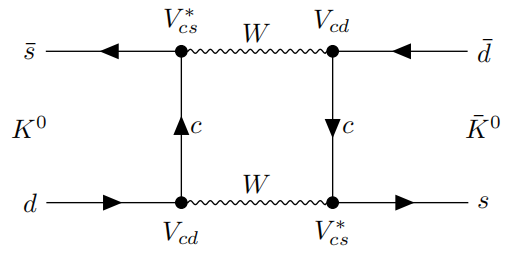}
\caption{\footnotesize The box diagrams for the short-distance contributions to  $\Kz$-$\Kzbar$
mixing.}
 \label{fig:k-mix_c-quark-KM}
  \end{figure}

\subsection{The neutral kaon mass eigenstates with no $\CP T$-invariance related restrictions} 

Any arbitrary neutral kaon state in its rest frame can be expressed as a linear combination of the strangeness
eigenstates, $\ket{\psi}$=\,$\alpha_1(\tau)\ket{\Kz}$\,+\,$\alpha_2(\tau)\ket{\Kzbar}$, where $\alpha_1(\tau)$ and
$\alpha_2(\tau)$ are complex functions that are normalized as $|\alpha_1(0)|^2$+\,$|\alpha_2(0)|^2$=\,1. In the
Wigner-Weisskopf formulation of quantum mechanics for exponentially decaying systems, the time-dependence is included
by using $\alpha_i(\tau)$\,=\,$\alpha^k_ie^{-i\lambda_k\tau}$ where $\alpha^k_i$ are complex constants and the
Schr\"{o}dinger equation in the form
\begin{equation}
  \label{eqn:time-dep-schrod}
  \boldsymbol{\Ham}\psi(\tau)=\Bigg(\boldsymbol{\Mass}-\frac{i}{2}\Gmm\Bigg)\boldsymbol{\Psi}_k e^{-i\lambda_k\tau}
  =i\frac{d\psi_K}{d\tau}=\lambda_k\boldsymbol{\Psi}_ke^{-i\lambda_K\tau},
\end{equation}
where the Hamiltonian $\boldsymbol{\Ham}$, the {\it mass matrix} $\boldsymbol{\Mass}$ and the {\it decay matrix}
$\Gmm$ are 2$\times$2 matrices and $\boldsymbol{\Psi}_k$ is a two-dimensional spinor in ($\Kz,\Kzbar$) space:
\begin{equation}
\boldsymbol{\Mass}=\begin{pmatrix} M_{11} & M_{12}\\ M^*_{12} & M_{22}\end{pmatrix},~~~~~
\Gmm=\begin{pmatrix} \Gamma_{11} & \Gamma_{12}\\ \Gamma^*_{12} & \Gamma_{22}\end{pmatrix}~~~~{\rm and}~~~~
\boldsymbol{\Psi}_k=\begin{pmatrix} \alpha^k_1 \\ \alpha^k_2\end{pmatrix}.
\end{equation}
Here $\boldsymbol{\Mass}$ and $\Gmm$ are hermitian, which is ensured by setting $M_{21}$\,=\,$M^*_{12}$ and
$\Gamma_{21}$\,=\,$\Gamma^*_{12}$, while $\boldsymbol{\Ham}$, which describes decaying particles and does not conserve
probability, is not hermitian. Symmetry under $\CP$ requires $M_{12}$ to be real; $\CP T$ invariance requires
$M_{11}$=\,$M_{22}$ and $\Gamma_{11}$=\,$\Gamma_{22}$. The mass eigenvalues and eigenstates for the most general case,
{\it i.e.}, neither $\CP$ or $\CP T$ symmetry are~\cite{Bigi:2000yz,Schubert:2014ska}
\begin{eqnarray}
  \label{eqn:mass-eigenstates-noCPT}
 \lambda_S=M_S-\rootionehalf\Gamma_S,~~~\ket{\KS}={\textstyle \frac{1}{\sqrt{2(1+|\eps_S|^2)}}}
  \big[\big(1+{\eps_S}\big)\ket{\Kz}
      +\big(1-{\eps_S}\big)\ket{\Kzbar}\big]\,,\;\;\;\;&&\nonumber\\
 \lambda_L=M_L-\rootionehalf\Gamma_L,~~~\ket{\KL}={\textstyle \frac{1}{\sqrt{2(1+|\eps_L|^2)}}}
    \big[\big(1+{\eps_L}\big)\ket{\Kz}
               -\big(1-{\eps_L}\big)\ket{\Kzbar}\big],\;\;\;\;&&
\end{eqnarray}  
\noindent
where $\eps_S$\,=\,$\eps$+$\delta$, $\eps_L$=\,$\eps$$-$$\delta$, and 
\begin{equation}
  \label{eqn:def-eps}
  \eps=-\frac{i Im(M_{12})+{\textstyle \frac{1}{2}} Im(\Gamma_{12})}{{\textstyle \frac{i}{2}}\Delta\Gamma+\Delta M},
\end{equation}
is the familiar neutral kaon mass-matrix $\CP$ violation parameter, while
\begin{equation}
  \label{eqn:deltaCPT}
  \delta=\frac{(M_{\Kzbar}-M_{\Kz})-i(\Gamma_{\Kzbar}-\Gamma_{\Kz})/2}{2\Delta M -i\Delta\Gamma},
\end{equation}
is the equivalent mass-matrix $\CP T$ violation parameter.  A unique and important feature is that the $\KL$ and $\KS$
eigenstates are not orthogonal; according to Eq.~(\ref{eqn:mass-eigenstates-noCPT}),
\begin{equation}
  \label{eqn:KLKS-CPT}
  \langle\KS |\KL\rangle=2  Re(\eps) -2i Im(\delta),
\end{equation}
where here, and in (most of) the following, we drop second-order terms in $|\eps|$.

%\subsubsection{Properties of $\eps$ and $\delta$}

In a commonly used phase convention, $Im\Gamma_{12}$ is defined to be zero, in which case the
phase of $\eps$ is directly related to the well measured quantities\footnote{We use\,\cite{Workman:2022ynf}
  $\Delta M$\,=\,$(0.5289\pm 0.0010)$$\times$10$^{10}$s$^{-1}$\,\&\,$\Delta\Gamma$=\,$(1.1149\pm 0.0005)$$\times$10$^{-10}$s$^{-1}$.}
$\Delta M$ and $\Delta\Gamma$ via
\begin{equation}
  \phi_{\rm SW} = \tan^{-1}\bigg(\frac{2\Delta M}{\Delta\Gamma}\bigg)=(43.5\pm 0.1)^\circ,
  ~~~\leftarrow~{\rm the~{\it Superweak~phase}}
\end{equation}
that, since $\Delta M$$\approx$$\Delta\Gamma/2$, is very nearly 45$^\circ$.
Using this approximate relation for $\phi_{\rm SW}$, we can rewrite  the Eq.~(\ref{eqn:deltaCPT}) expression for $\delta$ as
\begin{equation}
  \label{eqn:deltaCPT-modified}
  \delta \approx \frac{i(M_{\Kzbar}-M_{\Kz})+(\Gamma_{\Kzbar}-\Gamma_{\Kz})/2}{2\sqrt{2}\Delta M}e^{i\phi_{\rm SW}}
     = (i\delta_{\perp} + \delta_{\parallel})e^{i\phi_{\rm SW}},
  \end{equation}
where the real quantities $\delta_{\perp}$
and $\delta_{\parallel}$ are defined as
\begin{equation}
\label{eqn:delta-perp-par-def}
  \delta_{\perp} \approx \frac{M_{\Kzbar}-M_{\Kz}}{2\sqrt{2}\Delta M}~~~~~{\rm and}~~~~
   \delta_{\parallel} = \frac{(\Gamma_{\Kzbar}-\Gamma_{\Kz})}{4\sqrt{2}\Delta M}.
\end{equation}
In the complex plane, $\delta_{\perp}$, which is the short-distance component of $\delta$, is perpendicular
to $\eps$, and $\delta_{\parallel}$, the long-distance component, is  parallel to $\eps$. Thus, a signature for a non-zero
short-distance $\CP T$ violation would be a difference between the phase of $\eps_L$ and $\phi_{\rm SW}$ by an amount
\begin{equation}
\Delta\phi^{CPT}=\phi_{\eps_L}-\phi_{\rm SW}=\frac{\delta_{\perp}}{|\eps|},
\end{equation}
which means that
\begin{eqnarray}
  \label{eqn:DeltaM_veltaPhi}
  &~&M_{\Kzbar}-M_{\Kz}=2\sqrt{2}|\eps|\Delta M\Delta\phi^{CPT}\\
  \nonumber
  &~&~~~~~~~~~~~~~~~~~~~~~~=3.8\times 10^{-16}{\rm (MeV/deg)}\times\Delta\phi^{CPT}.
\end{eqnarray}
The tiny proportionality constant between the mass difference  and $\Delta\phi^{CPT}$ is the reason why the
Eq.~(\ref{eqn:KKbar-limit}) limit on $M_{\Kzbar}$$-$$M_{\Kz}$ is many orders of magnitude more stringent those for
the $m_{\bar{p}}$$-$$m_{p}$ (Eq.~(\ref{eqn:ppbar-limit}))    and $m_{\bar{e}}$$-$$m_{e}$ (Eq.~(\ref{eqn:eebar-limit})).
This is a feature that is unique to the neutral kaon system and does not apply to any other particle. 

\subsection{Interference measurements of the $\phi_{+-}$ and $\phi_{00}$ phases}

\begin{figure}[t!]
\centering
\includegraphics[width=0.8\textwidth]{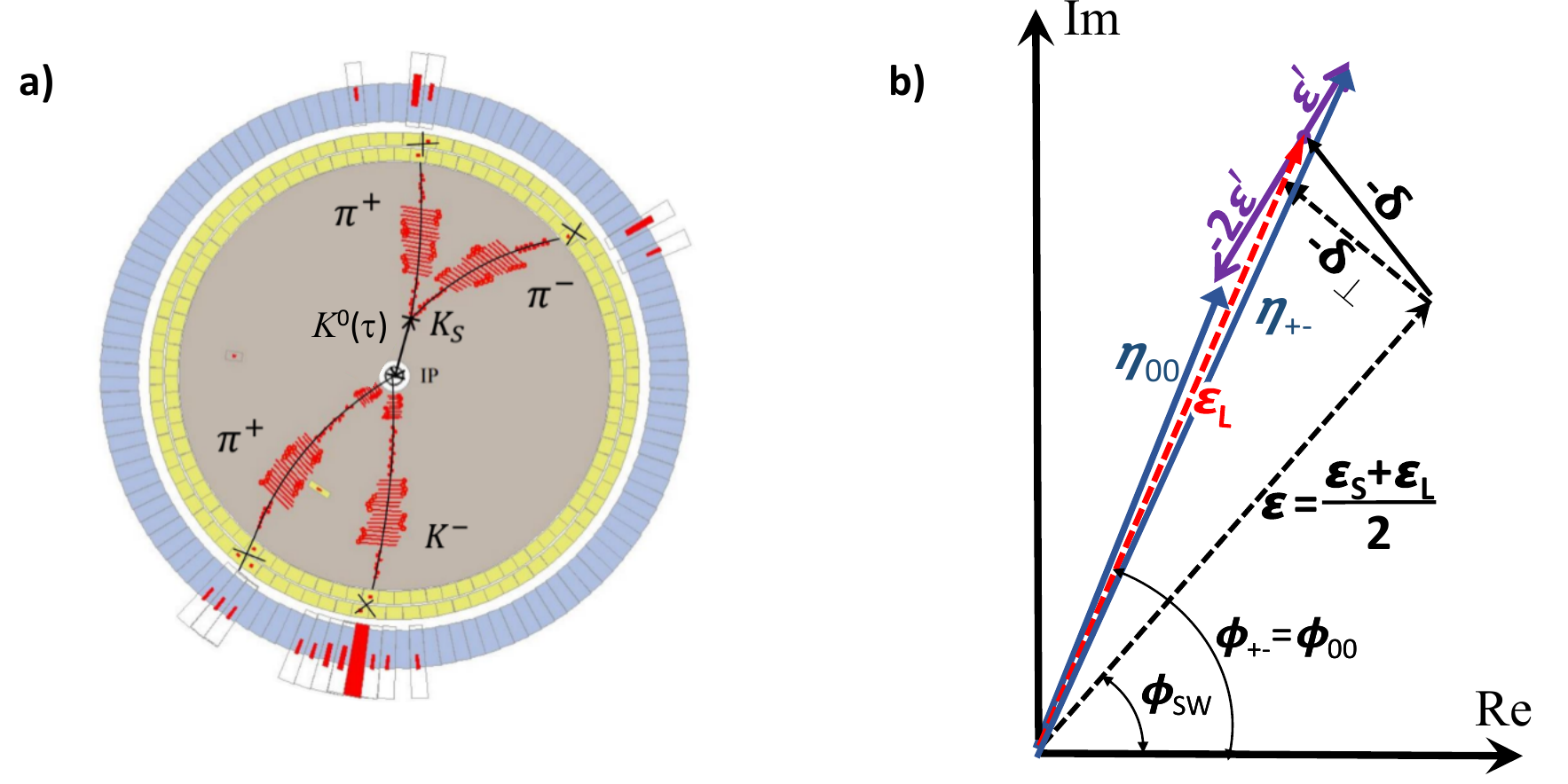}
\caption{\footnotesize {\bf a)}  A simulated $\jpsi\rt K^-\pip\Kz(\tau)$; $\Kz(\tau)\rt\pipi$ event in
  the BESIII detector.
  {\bf b)}  The relative arrangements in the complex plane of the complex quantities discussed in the text.
  Here, for display purposes, the magnitudes of $\delta$ and $\epsp$ relative to $\eps$ are exaggerated.}
 \label{fig:k0kpi-Wu-Yang}
 \end{figure}

A simulated $\jpsi$$\rt$$\Km\pip\Kz(\tau)$; $\Kz(\tau)$$\rt$$\pipi$ event in the BESIII detector is shown in
Fig.~\ref{fig:k0kpi-Wu-Yang}(a). Although it looks superficially like a $\Km\pip\KS$ event, the neutral kaon is not
a $\KS$ mass eigenstate and does not have a simple exponential decay curve. Instead, the neutral kaons produced in these
reactions are coherent mixtures of interfering $\ket{\KS}$ and $\ket{\KL}$ eigenstates:

\begin{eqnarray}
  \ket{\Kz (\tau)}&=&{\textstyle \frac{1}{\sqrt{2}}}\big[(1+\eps_L)\ket{\KS}e^{-i\lambda_s\tau}
                 +(1+\eps_S)\ket{\KL}e^{-i\lambda_L\tau}\big]\\
  \nonumber
  \ket{\Kzbar (\tau)}&=&{\textstyle \frac{1}{\sqrt{2}}}\big[(1-\eps_L)\ket{\KS}e^{-i\lambda_s\tau}
                 -(1-\eps_S)\ket{\KL}e^{-i\lambda_L\tau}\big],
\end{eqnarray}
and have decay curves with opposite-sign $\KS$-$\KL$ interference terms:
\begin{equation}
\label{eq:k0_decayrate}
  \begin{bmatrix} R(\tau)_{\Kz\rt\pi\pi}\\ \bar{R}(\tau)_{\Kzbar\rt\pi\pi}\end{bmatrix}
  \propto(1\mp 2 Re(\eps_L))\big[e^{-\Gamma_S\tau}+|\eta_{j}|^2e^{-\Gamma_L\tau}
        \nonumber
        \pm 2|\eta_{j}|e^{-\bar{\Gamma}\tau}\cos\big(\Delta M\tau - \phi_{j}\big)\big],
\end{equation}
where $\bar{\Gamma}$=\,$(\Gamma_S$+$\Gamma_L)/2$\,$\approx$\,$\Gamma_S/2$  
and $|\eta_{j}|$ and $\phi_j$
are the magnitudes and phases of $\eta_{+-}$ and $\eta_{00}$:
\begin{equation}
  \eta_{+-} = \frac{\braket{\pipi}{\KL}}{\braket{\pipi}{\KS}}=\eps-\delta+\epsp~~~~{\rm and}~~~~
  \eta_{00} = \frac{\braket{\piz\piz}{\KL}}{\braket{\piz\piz}{\KS}}=\eps-\delta-2\epsp.
  \end{equation}
Here $\epsp$=$\braket{\pi\pi}{\Ktwo}/\braket{\pi\pi}{\Kone}$ is the ratio of the direct-$\CP$-violating
amplitude of the $\CP$-odd $\ket{\Ktwo}$\,=$\smallrootonehalf\big(\ket{\Kz}$$-$$\ket{\Kzbar}\big)$ eigenstate
to $\CP$-even $\pi\pi$ final states, to that for $\CP$-allowed $\pi\pi$ decays of its $\CP$-even counterpart:
$\ket{\Kone}\,$=$\smallrootonehalf\big(\ket{\Kz}$+$\ket{\Kzbar}\big)$. Measurements~\cite{NA48:2002tmj,Abouzaid:2010ny}
have established that $\epsp$ is much smaller than $\eps$:
\begin{equation}
  Re(\epsp/\eps) =(1.66\pm 0.23)\times 10^{-3},
\end{equation}
and its phase, which is specified by low-energy $\pi\pi$ scattering phase shifts to be
$\phi_{\epsp}=\,(42.3\pm1.7)^\circ$~\cite{Colangelo:2001df} and equal, within errors, to $\phi_{\rm SW}$. This rather
remarkable coincidence\footnote{$\phi_{\epsp}$ is a long-distance QCD effect while
  $\phi_{\rm SW}$ is due to short-distance EW physics.}
means that $\epsp$ and $\eps$ are very nearly parallel, and the combined effect of this and the small
magnitude of $\epsp$ reduces its impact on $\delta\phi^{CPT}$ measurements to negligible, $\mathcal{O}(0.01^\circ$), levels.
The arrangement of the various quantities discussed above on the complex plane is indicated in Fig.~\ref{fig:k0kpi-Wu-Yang}(b).

\subsection{Comment on the Bell-Steinberger relation}
\label{sec:bell-steinberger}

A general procedure for determining $Im(\delta)$ that is based on unitarity and independent of any phase convention was
devised by John Bell and Jack Steinberger in 1965 and is  briefly summarized here~\cite{Bell:1965mn}.

In terms of the $\KS$ and $\KL$ eigenfunctions, a general solution to the Schr\"{o}dinger equation
can be written as
\begin{equation}
  \label{eqn:general-wave-function}
  \psi(\tau)=\alpha_1 e^{-i\lambda_S\tau}\ket{\KS} +\alpha_2 e^{-i\lambda_L\tau}\ket{\KL},
\end{equation}
where $\alpha_1$ and $\alpha_2$ can be any complex constants subject only to the normalization
condition $|\alpha_1|^2+|\alpha_2|^2=1$. The time-dependent probability associated with this wave function is 
\begin{equation}
  |\psi(\tau)|^2=|\alpha_1|^2e^{-\Gamma_S\tau}+|\alpha_2|^2e^{-\Gamma_L\tau}
  +2 Re\big(\alpha^*_1\alpha_2 e^{-{\scriptstyle \frac{1}{2}}(\Gamma_S+\Gamma_L+2i\Delta M)\tau}\braket{\KS}{\KL}\big),
\end{equation}
and the negative of its derivative at $\tau$\,=0 is
\begin{equation}
  \label{eqn:bell-steinberger-lhs}
  -\frac{d|\psi(\tau)|^2}{d\tau}\bigg|_{\tau=0}=|\alpha_1|^2\Gamma_S + |\alpha_2|^2\Gamma_L
  + Re\big(\alpha^*_1\alpha_2(\Gamma_S+\Gamma_L+2i\Delta M)\braket{\KS}{\KL}\big).
\end{equation}

In the Weisskopf-Wigner formalism, the Schr\"{o}dinger equation is not hermitian and the solutions
do not conserve probability. Instead, the overall normalization has a time dependence given by
\begin{equation}
|\psi(\tau)|^2=|\psi(0)|^2e^{-\Gamma_{\rm tot}\tau};
  ~~\frac{d|\psi(\tau)|^2}{d\tau}=-\Gamma_{\rm tot}e^{-\Gamma_{\rm tot}\tau},
\end{equation}

\noindent
and unitarity requires that
\begin{equation}
 -\frac{d|\psi(\tau)|^2}{d\tau}\bigg|_{\tau=0}=\Gamma_{\rm tot}=\sum_j\Gamma_j=\sum_j |\braket{f_j}{\psi(0)}|^2,
\end{equation}

\noindent
where the summation index $j$ runs over all of the accessible final states $\ket{f_j}$.
Applying this unitarity condition to the Eq.~(\ref{eqn:general-wave-function}) wave function, gives an independent
expression for the derivative of $|\psi(\tau)|^2$ at $\tau$\,=0
\begin{eqnarray}
    \label{eqn:bell-steinberger-rhs}
  -\frac{d|\psi(\tau)|^2}{d\tau}\bigg|_{\tau=0}&=&\sum_j|\alpha_1|^2|\braket{f_j}{\KS}|^2 + |\alpha_2|^2\braket{f_j}{\KL}|^2\\
    \nonumber
        &&~~~~~~~~+2 Re\big(\alpha^*_1\alpha_2\braket{f_j}{\KS}^*\braket{f_j}{\KL}\big).
 \end{eqnarray}

\noindent
Equations~(\ref{eqn:bell-steinberger-lhs}) and~(\ref{eqn:bell-steinberger-rhs}) have to apply for any values of
$\alpha_1$ and $\alpha_2$ (subject to $|\alpha_1|^2+|\alpha_2|^2=1$), therefore the terms multiplying $\alpha^*_1\alpha_2$ in
each expression have to be equal
\begin{equation}
  (\Gamma_S+\Gamma_L+2i\Delta M)\braket{\KS|\KL} = 2\sum_j\braket{f_j}{\KS}^*\braket{f_j}{\KL},
\end{equation}

\noindent
that, together with Eq.~(\ref{eqn:KLKS-CPT}), becomes

\begin{equation}
  \label{eqn:bell-steinberger}
        {\textstyle \frac{1}{2}}\braket{\KS|\KL}=
        \frac{Re(\eps)}{1+|\eps|^2} -i Im(\delta)
        =\frac{\sum_j\braket{f_j}{\KS}^*\braket{f_j}{\KL}}{ \Gamma_S+\Gamma_L+2i\Delta M }.     
\end{equation}

\noindent

This is known as the {\it Bell-Steinberger relation}. Here the second-order term in $|\eps|$ is retained in order to
emphasize that, with it, this equation is exact. A non-zero value of $Im(\delta)$ could only occur if $\CP T$ is violated, or
if unitarity is invalid, or if the $M_{ij}$ and/or $\Gamma_{ij}$ elements of the Hamiltonian were time-dependent. Because of
the small number of $\KS$ decay modes, the sum on the right-hand side of this equation only contains a few terms and is
manageable. In contrast, the neutral $B$-mesons have hundreds of established decay modes and probably a similar number of
modes that have yet to be detected; any attempt to apply a Bell-Steinberger-like unitarity constraint would be a hopeless task. 

A Bell-Steinberger-relation-based evaluation of  $Im(\delta)$ and $Re(\eps)$ by Giancarlo D'Ambrosio and Gino Isidori in
collaboration with the KLOE team, that used all the latest relevant measurements, was originally presented in
2006~\cite{DAmbrosio:2006hes} and recently updated in the PDG2020 review~\cite{DAmbrosio:2022pth}. Their tabulation of relevant
measurements that were available in 2022 is:

\begin{center}
  \begin{tabular}{lc}
  \hline
  Channel           &  $\braket{f_j}{\KS}^*\braket{f_j}{\KL}$~($10^{-5}\Gamma_S$)\\
  \hline
  $\pipi(\gamma)$   &    $112.1\pm 1.0~+~i(106.1\pm 1.0)$  \\
  $\piz\piz$        &     $49.3\pm 0.5~+~i(47.1\pm 0.5)$   \\
  $\pipi\gamma_{E1}$ &    $<0.01$\\
  $\pipi\piz$       &    $0.0\pm 0.2~+~i(0.0\pm 0.2)$  \\
  $\piz\piz\piz$    &    $<0.15$   \\
  $\pi^\pm\ell^\mp\nu$ &  $-0.1\pm 0.2~+~i(-0.1\pm 0.5)$,\\
  \hline
  \end{tabular}
\end{center}
\noindent
where: the $\pi\pi(\gamma)$ values come from the PDG20 averages of many  $\eta_{+-}$ and $\eta_{00}$ measurements;
the $\pipi\gamma_{E1}$ limit comes from theory~~\cite{Cheng:1993xd,DAmbrosio:1996lam}; the $\pi\pi\pi$  results are
based on thirty-year-old $\BR(\KS\rt\pipi\piz)$ measurements~\cite{Zou:1996ks,Adler:1997pp} and a
2013 KLOE upper bound on $\BR(\KS\rt\piz\piz\piz)$~\cite{Babusci:2013tr};  and the $\pi^\pm\ell^\mp\nu_\ell$ values come from
twenty-five-year-old measurements of SL charge asymmetries for tagged $\Kz(\tau)$ and $\Kzbar(\tau)$ decays from
CPLEAR~\cite{CPLEAR:1998ohr} and $\KL$~decays from KTeV~\cite{KTeV:2002kqg} plus a 2018 measurement for tagged $\KS$~decays
from KLOE-2~\cite{Anastasi:2018qqf}.

\begin{figure}[!bp]
\begin{center}
     \includegraphics[height=0.4\textwidth,width=0.8\textwidth]{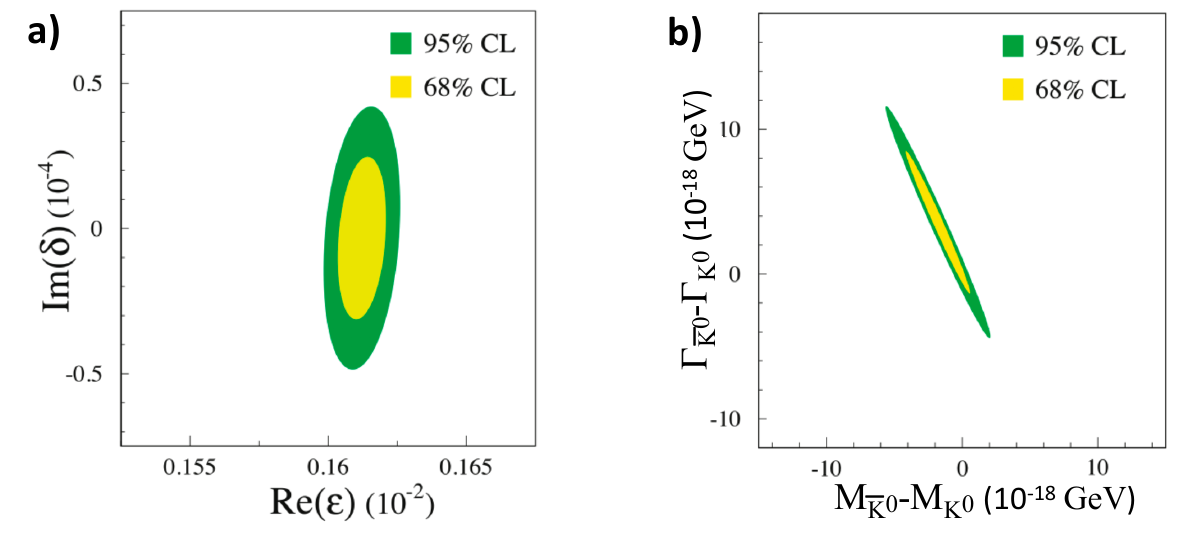}
     \caption{\footnotesize  {\bf a)} The 68\% and 95\% confidence level allowed region for
       $\Im m\delta$ and $\Re e\eps$ from a Bell-Steinberger analysis.
       {\bf b)} The corresponding allowed regions in $\Delta\Gamma$=\,$\Gamma_{\Kzbar}$$-$$\Gamma_{\Kz}$ and
       $\Delta M= M_{\Kzbar}-M_{\Kz}$. From Ref.~\cite{Zyla:2020zbs}.
     }
     \label{fig:bell-steinberger}
\end{center}
\end{figure}

The bottom line of this analysis is
 \begin{equation}
   \label{eqn:BS-results}
   Im(\delta) = (-0.3\pm 1.4)\times 10^{-5}~~~\Rightarrow~~~|M_{\Kzbar}-M_{\Kz}|<4\times 10^{-16}\,{\rm MeV},
 \end{equation}
 and the $Im(\delta)$\,{\it vs.}\,$Re(\eps)$ and $\Gamma_{\Kzbar}$$-$$\Gamma_{\Kz}$\,{\it vs.}\,$M_{\Kzbar}$$-$$M_{\Kz}$
 allowed regions are shown in Figs.\,\ref{fig:bell-steinberger}\,(a)\,\&\,(b), respectively.
 The limit on $|M_{\Kzbar}$$-$$M_{|Kz}|$ is nearly the same as the Eq.~(\ref{eqn:DeltaM_veltaPhi}) bound, that comes from simply
 comparing the PDG20 world average values for $\phi_{+-}$ and $\phi_{\rm SW}$. This is not surprising, since in both
 analyses the $\pi\pi$ channels dominate, and the sensitivities are limited by the same $\eta_{+-}$ and $\eta_{00}$ measurement
 errors. Moreover, contributions from the uncertainties associated with the other modes are smaller than those from the
 $\pi\pi$ modes.  However, these contributions are not so small and future progress in  Bell-Steinberger-relation
 analyses will require that improvements in the $\eta_{+-}$ and $\eta_{00}$ phase measurements are accompanied by
 similar improvements in the precision of the other terms.

The $\pipi\piz$ entry in the above table comes from statistics limited time-dependent Dalitz-plot analyses of
$\KS$-$\KL$ interference effects in strangeness-tagged $\Kz(\tau)$ and $\Kzbar(\tau)$ decays to $\pipi\piz$, and will be
improved with the higher statistics data samples that will be available at STCF. On the other hand, the $\piz\piz\piz$
entry is based on a $\BR(\KS$$\rt$3$\piz)$$<$2.6$\times$10$^{-8}$ upper limit from KLOE that used tagged-$\KS$ mesons produced
in a sample of 1.7\,billion $\ee$$\rt$$\phi$$\rt$$\KS\KL$ events~\cite{Babusci:2013tr}. Likewise, the precision of the
$\pi\ell\nu_\ell$ entry is limited by the measured value of the lepton charge asymmetry in a $\sim$70\,K-event sample of reconstructed
$\pi e\nu_e$ decays of tagged $\KS$ mesons in the same KLOE event sample~\cite{Anastasi:2018qqf}. Further improvements of these
measurements would need substantially larger numbers of tagged $\KS$ decays and will not be easy. The possibilities for
improving these values using $\jpsi$$\rt$$\phi\KS\KS$ and $\omega\KS\KS$ decays as sources of tagged $\KS$
decays\footnote{$\jpsi$$\rt\phi\KS\KL$
   and $\omega\KS\KL$ decays are forbidden by $\CP$ invariance.}
at STCF are currently being investigated.

\subsection{Prospects of kaon $\CP T$ study at STCF}

Based on the data sample with $10^{12}$ collected by STCF per year, the expected yields of $J/\psi \rightarrow K^{-}\pi^{+}K^{0} (K^{0} \rightarrow \pi^{+}\pi^{-}) + c.c.$ will be more than $3.9 \times 10^{9}$ events. Based on this data sample, a significant improvement in the precision of $\CP T$ measurements in the kaon system can be achieved. The following sections will introduce the event reconstruction and the expected precision of measured parameters.

%\subsubsection{MC simulation of $J/\psi \rightarrow K^{-}\pi^{+}K^{0} + c.c.$}

A Monte Carlo simulation sample with $3.9 \times 10^{9}$ $J/\psi \rightarrow K^{-}\pi^{+}K^{0} (K^{0} \rightarrow \pi^{+}\pi^{-}) + c.c.$  total events is generated by the phase space model. The time-dependent decay rates of $K^{0}(\bar{K}^{0}) \rightarrow \pi^{+}\pi^{-}$ are shown at Eq.~(\ref{eq:k0_decayrate}), and the input parameters are fixed to the PDG average values~\cite{ParticleDataGroup:2022pth}. 

%\subsubsection{Event selection procedure}
For the decay channels $J/\psi \rightarrow K^{-}\pi^{+}K^{0}$ and $J/\psi \rightarrow K^{+}\pi^{-}\bar{K}^{0}$, the flavor of the neutral $K$ meson produced at time $t=0$ is tagged by the charge sign of the charged kaon. Subsequently, the neutral $K$ meson is reconstructed to the final state with $\CP$ of +1 through the decay channel $K^{0}/\bar{K}^{0} \rightarrow \pi^{+}\pi^{-}$. 

For charged tracks not originating from the decay of neutral 
$K$ mesons, the distance of closest approach to the collision point in the direction of the beam should be within $\pm 10$ cm, and the projection onto the plane perpendicular to the beam direction should be less than 1 cm. A particle identification algorithm is performed on the selected charged tracks with information from sub-detectors, to get the probability ($Prob$) that a charged particle is a certain type of particle. The kaon candidate is required that $\mathrm{Prob}(K) > \mathrm{Prob}(p), \mathrm{Prob}(K) > \mathrm{Prob}(\pi)$ and for the pion candidate: $\mathrm{Prob}(\pi) > \mathrm{Prob}(K)$.

A vertex-constrained fit is carried out on the $\pi^{+}\pi^{-}$ to obtain the decay vertex of $K^{0}$. The two charged pions are required to originate from a common vertex, and the corresponding fitting $\chi^{2}$ should be less than 100. Subsequently, the production vertex of $K^{0}$ is obtained by performing another vertex-constrained fit on $K^{-}\pi^{+}$. The invariant mass of the $\pi^{+}\pi^{-}$ after vertex fitting is required to be within [482, 514] MeV. In order to reduce the background from random combinations of charged tracks in $K^{0}$ and from $\pi^{+}\pi^{-}$ combinations not originating from $K^{0}$ decays, the secondary vertex fitting algorithm package is utilized to obtain the decay distance $L_{K^{0}}$ and its corresponding error $\sigma(L_{K^{0}})$ for $K^{0}$.  The ratio of the decay distance to its error is required to be $L_{K^{0}}/\sigma(L_{K^{0}}) > 2$. 

In each event, the sum of the charges of the charged tracks should be zero. A four-momentum constrained kinematic fit is applied to the $\pi^{+}\pi^{-} K^{-}\pi^{+}$ hypothesis, and events with a minimum $\chi^{2} < 60$ are selected as best $J/\psi$ candidates.

After all the event selections described above, the expected background ratio is less than $0.5\%$, resulting in a high-purity data sample.

%\subsubsection{Expected sensitivity at STCF}
%Please see detailed information at Sec.~\ref{sec:kaon_CPT_sensitivity}

%\subsubsection{Estimated measurement sensitivity with ${10^{12}~\jpsi}$-decays}
\label{sec:kaon_CPT_sensitivity}

Decay curves for $\Kz(\tau)$$\rt$$\pipi$ and $\Kzbar(\tau)$$\rt$$\pipi$ for simulated BESIII events with 100 billion $J/\psi$ that were generated
with PDG values of the  $\eta_{+-}$ magnitude and phase are shown in Fig.~\ref{fig:A-prime-simul}(a), where there are strong
interference effects between the steeply falling $\KS$$\rt$$\pipi$ and nearly flat $\KL$$\rt$$\pipi$ amplitudes that
have opposite signs for $\Kz(\tau)$- and $\Kzbar(\tau)$-tagged decays. The difference between the interference terms is
displayed in a plot of the reduced asymmetry~\cite{Apostolakis:1999zw} $\mathcal{A}^\prime_{\pi\pi}$,
where\footnote{In the unmodified asymmetry,
  $\mathcal{A}_{\pi\pi}$=$[\bar{N}(\Kzbar(\tau))$$-$$N(\Kz(\tau))]/[\bar{N}(\Kzbar(\tau))$+$N(\Kz(\tau))]$,
  the $\cos(\Delta M\tau-\phi_j)$ oscillation term is multiplied by a factor $e^{\smallonehalf\Delta\Gamma\tau}$
    that, when displayed, emphasizes low-statistics, long-decay-time events that, in fact, have
    little influence on the $\phi_j$ determination.}
\begin{eqnarray}
  \label{eqn:KKbar-asymm-CPT} 
        {\mathcal A}^\prime_{\pi\pi}&\equiv&
        \frac{\bar{N}(\Kzbar(\tau))-N(\Kz(\tau))}{\bar{N}(\Kzbar(\tau))+N(\Kz(\tau))}e^{-\smallonehalf\Delta\Gamma\tau}\\
    \nonumber
    &~&~~~~~~=2 Re(\eps_L) e^{-\smallonehalf\Delta\Gamma\tau}-2\frac{|\eta_{+-}|\cos(\Delta M\tau-\phi_{+-})}
              {1+|\eta_{+-}|^2e^{\Delta\Gamma\tau}},
\end{eqnarray}
and is shown for the simulated data in  Fig.~\ref{fig:A-prime-simul}~(b), where $\phi_{+-}$ shows up as a phase shift in the
interference-generated cosine-like oscillation. 

\begin{figure}[htbp!]
\centering
\includegraphics[width=0.95\textwidth]{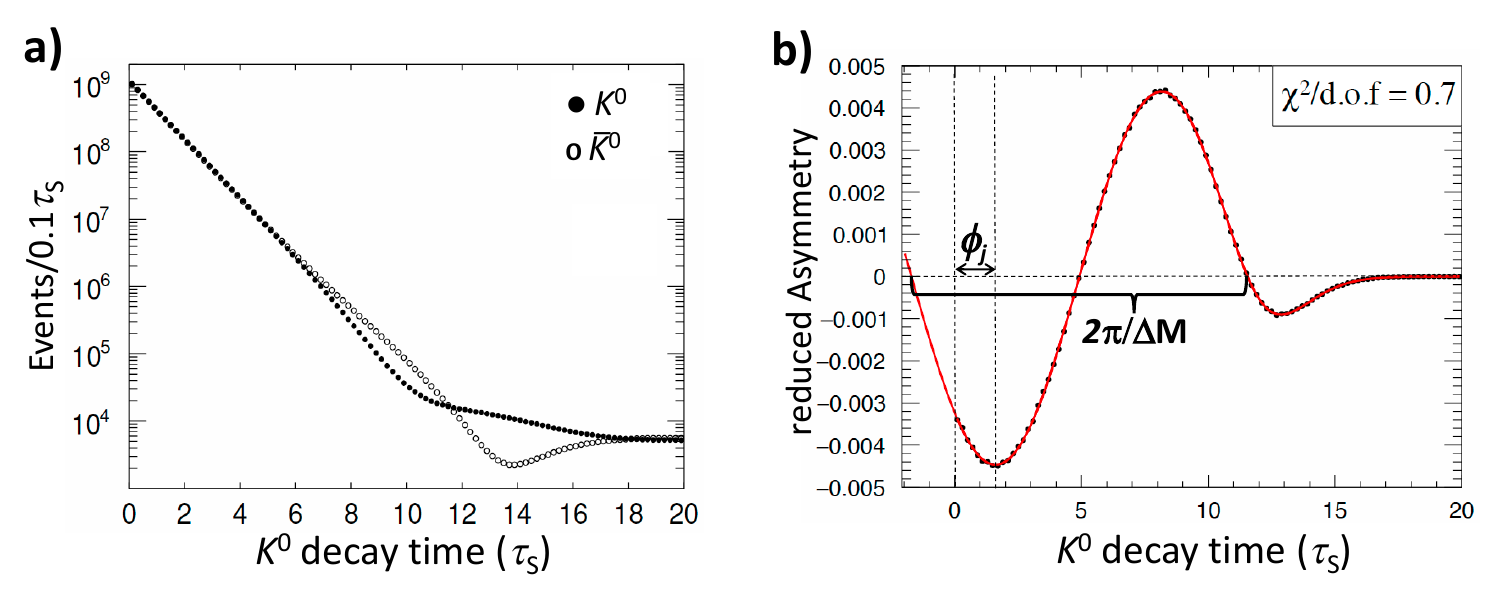}
\caption{\footnotesize {\bf a)} The solid circles show the proper time distribution for simulated strangeness-tagged
  $\Kz(\tau)\rt\pipi$ decays (the open circles are $\Kzbar(\tau)$$\rt$$\pipi$ decays). 
  {\bf b)} The reduced asymmetry, ${\mathcal A}^{\prime}_{\pipi}$, for the events shown in panel~\textbf{a)}.
 % Here $\phi=\phi_{\rm SW}+\delta\phi^{CPT}_{\eta}$. (Figures provided by Jian-Yu Zhang.) 
  }
 \label{fig:A-prime-simul}
\end{figure}

The  simulated data shown in Figs.\,\ref{fig:A-prime-simul}\,(a) and (b) correspond to 3.8 billion~tagged $K$$\rt$$\pipi$ decays
that are almost equally split between $\jpsi$$\rt$$\Km\pip\Kz$ and $\jpsi$$\rt$$\Kp\pim\Kzbar$, which were generated 
with $\Delta\phi^{CPT}$=0\, and $\phi_{\rm SW}$=\,43.4$^\circ$. This corresponds to what one would expect for a
total of $10^{12}$~$\jpsi$ decays in a detector that covered a $|\cos\theta|\le 0.85$ solid angle and was
otherwise almost perfect.  The red curve in the figure is the result of a fit to the data that determined
\begin{equation}
  \phi_{\rm SW}+\Delta\phi^{CPT} = (43.51\pm 0.05)^\circ,
\end{equation}
where the errors are statistical only. According to Eq.~(\ref{eqn:DeltaM_veltaPhi}), this $\mathcal{O}(0.1^\circ)$
precision corresponds to $|M_{\Kzbar}$$-$$M_{\Kz}|$\,$\approx$\,4$\times$10$^{-17}$\,MeV. This is an order of magnitude
improvement over the existing limit, which is based on the result in 1999 from the
CPLEAR~$\bar{p}_{\rm stop}p$ experiment at CERN~\cite{Apostolakis:1999zw} that used a sample of $\sim$70 million tagged
$\Kz(\tau)$$\rt$$\pipi$ and $\Kzbar(\tau)$$\rt$$\pipi$ events, and the result in 1995 from a Fermilab experiment
E773~that used $\sim$2 million~$K$$\rt$$\pipi$ and $\sim$0.5~million~$K$$\rt$$\piz\piz$ decays produced downstream of a regenerator
  located in a high-energy $\KL$ beam~\cite{Schwingenheuer:1995uf}:
\begin{eqnarray}
  {\rm CPLEAR:}~~~\phi_{\rm SW}+\Delta\phi^{CPT}&=&42.91^\circ\pm 0.53^\circ {\rm (stat)}\pm 0.28^\circ  {\rm (syst)}\,,\\
  \nonumber
  {\rm E773:~~}~~~\phi_{\rm SW}+\Delta\phi^{CPT}&=&42.94^\circ\pm 0.58^\circ {\rm (stat)}\pm 0.49^\circ  {\rm (syst)}.
\end{eqnarray}
The CPLEAR systematic error includes a $0.19^\circ$ component associated with the effects of regeneration in the
high-pressure hydrogen gas in the stopping target and the walls of its containment vessel. In the E773 measurement,
the phase off-set that is introduced by the regenerator is unavoidable and large, $\phi_{\rm regen}$$\sim$-130$^\circ$, and
is evaluated from a dispersion relation with an assigned error of $0.35^\circ$~\cite{Briere:1995tw}.
In the STCF detector, the production and a significant fraction of the $K$$\rt$$\pipi$ decays will occur in a high vacuum,
and the material traversed by rest of the decaying neutral kaons is very small. This will substantially reduce the
effects of regeneration  and its associated systematic error to be well below their level in  these previous experiments.

%\subsubsection{Systematic uncertainty discussion}

The systematic uncertainty of this measurement includes the following sources: the decay time resolution of $K^{0}$, background contributions, regeneration effect of $K^{0}$, and uncertainty associated with parameters fixed in the fit. The decay time resolution depends on the $K^{-}\pi^{+}$ vertex resolution, and the $\pi^{+}\pi^{-}$ vertex resolution. The high performance of track detector of STCF will reconstruct the vertex of $K^{0}$
efficiently and accurately. And its excellent position resolution and high resolution in the energy loss $dE/dx$ will help to suppress the background of miscombinations and particle misidentification. Due to the high purity of the sample at $99.5 \%$, the impact of the background can be neglected, and no decay-rate asymmetry is observed in the background. At the STCF, the flight of $K^{0}$ will occur in the low-material detectors, therefore the effects of regeneration will be much lower compared to other experiments, resulting in a controllable systematic uncertainty.

\subsection{Further comments}

Even though the $\CP T$ theorem is a fundamental element of the SM, experimental constraints on its validity
have not been significantly improved during the last three decades.  This is not because these are uninteresting and not 
important but, instead, is a consequence of the fact that none of the world's active particle physics
facilities have the capability of producing enough strangeness-tagged neutral kaons to match, much less improve on, the
CPLEAR and E773 results. Although currently operating experiments with huge numbers of charmed and beauty particles can address
many interesting subjects, they will never be able to test the $\CP T$ theorem with the exquisite sensitivity that is uniquely
provided by the neutral $K$-meson system.  As noted in this report, the STCF project will provide the first practical opportunity
to make an order of magnitude sensitivity improvement over earlier experiments. Note that in addition to probing $CP T$ invariance,
the CPLEAR experiment at also produced the world's most sensitive search  for violations of the $\Delta S$\,=\,$\Delta Q$ rule,
which can only occur in the SM as a second-order weak interaction.  The CPLEAR result only limits $\Delta S$\,=$-\Delta Q$ 
amplitudes to be less than a percent of the allowed $\Delta S$\,=\,$\Delta Q$ ones. SCTF will test rule with  an order of magnitude
higher sensitivity. 

The special requirements
such  measurements should be carefully considered during the design and implementation of the collider and its
associated detector.

\newpage

%\section{The STCF project}
%\label{sec:stcf}
%\input_sec7_stcf}

%\section{Probe of CP violation at STCF}
%\label{sec:stcfcpv}
%\input{sec8_stcfcpv}

\section{Summary}
\label{sec:sum}
%\subsection{CP violation in $J/\psi\to Y\bar{Y}$}
%\subsection{Baryon decays}

%The $CP$ violation is one of the necessary condition for the matter-antimatter asymmetry in universe.

%Searching for new source of $CP$ violation is crucial for understanding the matter-antimatter asymmetry in the universe.
%Currently, experiments have not yet to uncover evidence of $CP$ violation in strange and charmed baryons, 
%or in leptons.  The STCF provides great potential to contribute to this field through direct measurements of $CP$ violation and EDM measurements in various hadrons and tau lepton sectors. 
%The STCF is expected to produce billions of quantum-correlated baryon-antibaryon or lepton pairs. 
%The clean background and threshold production characteristics further ensure high-precision measurements with excellent systematic uncertainty control. 

We have reviewed perspectives of using unprecedented large numbers of entangled light baryon-antibaryon, meson-antimeson, and tauon-antitauon pairs will be produced at the STCF to carry out  high precision tests for $CP$ violation in these systems.

With the implementation of STCF , studies of $CP$ violation using spin-entangled and polarized baryon–antibaryon pairs, a program that was initiated by the BESIII experiment, will reach levels of precision that are compatible with SM predictions for hyperon non leptonic decays reaching $\mathcal{O}(10^{-4})$. Studies of the production process will provide stringent limits for electric dipole form factors that test $C$ and $P$ symmetry violation improve previous limit on Lambda EDM by five orders of magnitudes reaching $\mathcal{O}(10^{-21})$~$e$ cm, and to put on EDM limits on $\Sigma$, $\Xi$ to the level of $\mathcal{O}(10^{-20})$~$e$ cm. In principle, similar $CP$ violation studies are possible for charmed baryons, however, the prospects for the attainable uncertainties will soon be known with the BESIII data after the completion of the BEPCII energy and luminosity upgrade. Although no vector resonances have been found that could enhance the production of charmed baryon-antibaryon pairs today, and the data samples for the charmed baryon measurements are not competitive with that at LHCb, the simplicity and complete kinematic of the production process of the charmed baryons at STCF will drive for several unique and complementary measurements.
 
%At STCF , studies of $\CP$ violation using spin-entangled and polarized baryon-antibaryon pairs, a program that was initiated by the BESIII experiment, will reach levels of precision that are compatible with SM predictions for hyperon non leptonic decays. Studies of the production process will provide stringent limits for electric dipole form factors that test $C$ and $P$ symmetry violation.
%The simple experimental environment  will allow also for a competitive complementary to the determination of the $\CKM$ $|V_{us}|$ element based on kaon measurements. This would add a new experimental perspective on the present discrepancy known as the Cabbibo angle anomaly that has persisted at the $\sim$3$\sigma$ level since 1970~\cite{Blin-Stoyle:1970lxd}.
%In principle, similar $\CP$ violation studies are possible for charmed baryons, however, the prospects for the attainable uncertainties will soon be known with the BESIII data after the completion of the BEPCII energy and luminosity upgrade.
%Although no vector resonances have been found that could enhance the production of charmed baryon-antibaryon pairs today, and the data samples for the charmed baryon measurements are not competitive with that at LHCb, the simplicity and complete kinematic of the production process of the charmed baryons at STCF will drive for several unique and complementary measurements. 

%\subsection{CP violation in $\tau\to K_{s}\pi\nu$}

The large data samples of $\tau$ lepton pairs that are produced near  threshold at STCF, with low background and well-controlled systematic uncertainties,  will provide unique ways to make a number of precision and comprehensive studies of the $\tau$-lepton properties. Hadronic $\tau$ decays offer rich perspective to probe possible $\CP$ violation phenomena, and the relevant form factors and structure functions are important inputs to construct $\CP$ violation observables in the $\tau$ sector. Improved measurements of various hadronic $\tau$ decay that can be performed at STCF will provide a solid foundation to search for diverse $\CP$ violation signals in $\tau$ decays. These include a variety of differential observables, such as the integrated rate asymmetry, triple products and forward-backward asymmetries, that could potentially enhance $\CP$ violation features in multi-meson $\tau$-lepton decay channels. Specifically, measurements on the rate and forward-backward  asymmetries of $\tau \to K_S^0 \pi\nu_\tau$ decay at STCF will clear up the hint of a $\CP$ violation anomaly that has been reported by Babar, and shed light on BSM. Information on the $\tau$ EDM can be indirectly extracted from processes involving $\tau$ pair productions at lepton colliders. Together with dedicated studies with carefully optimized observables, STCF can provide more data on relevant processes and improve significantly the measurements of the $\tau$ EDM at the level of $|d_\tau| < 10^{-18}$ $e$ cm. 

%\subsection{CP violation in $D^{0}$-$\bar{D}^{0}$ mixing}

%Mixing and $\CP$ violation in the charm sector have long been one of the least-known areas in flavor physics because of their tiny size in the SM. Meanwhile, 
Contrary to $B$-meson and kaon systems, non-perturbative long-distance effects have been found to be the prerequisites to explain the observed charm mixing and $\CP$ violation effects. Update to date, only direct $\CP$ violation has been measured in the singly Cabbibo-suppressed decays $K^+K^-$ and $\pi^+\pi^-$. %Among the three kinds of $\CP$ violation in charmed meson system, only direct $\CP$ violation has been measured in the singly Cabbibo-suppressed decays $K^+K^-$ and $\pi^+\pi^-$. %, whereas the indirect $\CP$ violations arising from charm mixing and interference between mixing and decay amplitudes have yet to be observed. 
%In the LHCb experiment, 
%The huge charm cross section along with long flight distance of the charmed mesons in the LHCb and Belle II experiments will provide promising ways to study the charm mixing and indirect $\CP$ violation effects through time-dependent measurements. Such studies would probably enter a precision era in the next one or two decades. 
  At STCF, despite the fact that the statistics of the charmed-meson samples will be orders of magnitudes lower than that in the LHCb experiment, and time-dependent measurements are not applicable, % because the charm mesons are produced nearly at rest,
  a unique time-integrated methodology can be adopted to access the charm mixing parameters and indirect $\CP$ violation by exploiting the quantum correlations in neutral $D\bar{D}$ decays. %This is because the neutral $D\bar{D}$ mesons are produced in a coherent state with well defined $C$ parity at electron-positron colliders near threshold, known as quantum correlation. 
 Sensitivity studies have %been reported in Sec.~\ref{sec:charm} 
 demonstrated that STCF has the potential to measure charm mixing parameters and the magnitude of $\CP$ violation at $10^{-4}$ levels of precision. %with the annually produced $C$-even quantum-correlated $D\bar{D}$ decays. 
 Thes are complementary to time-dependent measurements at the LHCb and Belle II experiments. 
%It is also shown that sizable improvements can be achieved with the un-correlated and $C$-odd correlated $D\bar{D}$ decays produced along with the $C$-even ones. %Interestingly, the $\CP$-conserving phase differences in $D^0/\bar{D}^0$ decays can be obtained with the $C$-odd correlated $D/\bar{D}$ decays. These are essential inputs to $\CP$ violation studies both in bottom and charm decays and have been determined by BESIII and CLEO experiments with great success. In STCF, they could be determined with ultimate uncertainties. 
Interestingly, the $\CP$-conserving phase differences in $D^0$ and $\bar{D}^0$ decays, which are essential inputs to $\CP$ violation studies both in bottom and charm decays and have been determined by BESIII and CLEO experiments with great success, can be obtained with the correlated $D\bar{D}$ decays in STCF with ultimate uncertainties. 
%On one hand, STCF would play an important role in the forthcoming $\CP$ violation studies by providing these inputs, on the other hand, by investigating the correlated $D\bar{D}$ decays simultaneously, experimental uncertainties should be reduced significantly.

During a year of operation at the peak of the $\jpsi$ resonance, it is estimated that STCF will record
3.4~billion $\pipi$ decays of neutral $K$-produced via $\jpsi\rt K^\mp\pi^\pm K^0$ processes that have well-defined strangeness at the time of production. These events will support studies of $\CP$ violation and
tests of the $\CP T$ theorem with unprecedented precision. For comparison, the most stringent and theoretically robust
current limit on the validity of the $\CP T$ theorem, a foundational principle of the SM, is the
$|M_{\bar{K}^0}-M_{K^0}|<5\times 10^{-16}$~MeV result, that is based on a Bell-Steinberger analysis
of 25 year-old CPLEAR, KLOE and KTeV measurements with $\sim$70~million event data
samples~\cite{DAmbrosio:2022pth}. The 3.4~billion STCF $J/\psi$ event sample would advance the level of sensitivity on
this most fundamental of all fundamental SM quantities by about an order of magnitude.

%\subsection{CPT violation in $K^{0}-\bar{K}^{0}$ mixing}

\newpage
\clearpage

\section*{Acknowledgements}
\addcontentsline{toc}{section}{Acknowledgements}
We thank the Hefei Comprehensive National Science Center for their strong support on the STCF key technology research project. 
This work is supported in part by National Key R\&D Program of China under Contracts No.~2022YFA1602200; International partnership program of the Chinese Academy of Sciences Grant No.~211134KYSB20200057; National Natural Science Foundation of China~(NSFC) under Contracts Nos. 12375088, 12335003, 12305107, 12275023, 12475078, 12122509, 12221005, 12375086; National Research Foundation of Korea under Contract No.~NRF-2022R1A2C1092335; Chinese Academy of Sciences President’s International Fellowship Initiative; {Polish National Science Centre through the grants 2019/35/O/ST2/02907 and 2024/53/B/ST2/00975}; Swedish Research Council under grant No. 2021-04567. P.~R.~ acknowledges Conahcyt funding (Mexico), through project CBF2023-2024-3226 and the support of MCIN/AEI/10.13039/ 501100011033, grant PID2020-114473GB-I00 and PID2023-146220NB-I00,
and by Generalitat Valenciana, grant PROMETEO/2021 /071 (Spain, during his sabbatical).

 \bibliographystyle{elsarticle-num} 
 \bibliography{refs}





% \end{thebibliography}
%\end{sloppypar}
\end{document}